\documentclass[12pt]{JHEP3}

\usepackage{amsmath,epsfig}
\usepackage{amssymb,amsfonts}
\usepackage{latexsym}
\usepackage[latin1]{inputenc}
\usepackage{slashed}
\usepackage{empheq}
\numberwithin{equation}{section}
\usepackage{dsfont}
\usepackage{cancel}
\usepackage{overpic}
\usepackage{subcaption}
\usepackage{subeqnarray}
\usepackage{xcolor}
\let\normalcolor\relax
\usepackage{graphicx}
\usepackage{amsmath}
\usepackage{longtable}
\usepackage{color}
\usepackage{bbm}

\newcommand{\exclude}[1]{}
\def\a#1{\alpha_{#1}}

\def\beq{\begin{equation}}
\def\eeq{\end{equation}}
\def\be{\begin{equation}}
\def\ee{\end{equation}}
\def\bea{\begin{eqnarray}}
\def\eea{\end{eqnarray}}
\def\bal{\begin{align}}
\def\eal{\end{align}}

\def\2b2[#1,#2][#3,#4]{\left( \begin{array}{cc} #1 & #2 \\ #3 & #4 \end{array}
\right)}
\def\3b3[#1,#2,#3][#4,#5,#6][#7,#8,#9]{\left( \begin{array}{ccc} #1 & #2 #3 \\
#4 & #5 & #6\\#7&#8&#9\end{array} \right)}

\newcommand\fverb{\setbox\pippobox=\hbox\bgroup\verb}
\newcommand\fverbdo{\egroup\medskip\noindent%
\fbox{\unhbox\pippobox}\ }
\newcommand\fverbit{\egroup\item[\fbox{\unhbox\pippobox}]}

\newcommand{\bear}{\begin{eqnarray}}

\newcommand{\eear}{\end{eqnarray}}

\newcommand{\bsea}{\begin{subeqnarray}}
\newcommand{\esea}{\end{subeqnarray}}
\newbox\pippobox

\def\f{\varphi}

\def\6{\partial}
\def\a{\alpha}

\def\pa{\partial}

\def\m{\mu}
\def\n{\nu}
\def\r{\rho}
\def\s{\sigma}

\def\sp{\;\;\;,\;\;\;}

\def\sq
\def\a{\alpha}

\def\hri#1#2{\href{http://arxiv.org/abs/#1}{[ArXiv:#1]#2}}
\def\hre#1#2{\href{http://arxiv.org/abs/#1/#2}{[ArXiv:#1/#2]}}

\def\hrj#1#2{\href{www.doi.org/#1}{#2}}

\title{Holographic RG flows on Squashed $S^3$}
\author{
E. Kiritsis$^\natural$$^\flat$,C. Litos$^\flat$$^*$,
~\\
~\\
$^\natural$ \href{http://www.apc.univ-paris7.fr}{Universit\'e Paris Cit\' e, CNRS, Astroparticule et Cosmologie}, F-75013 Paris, France\\
~\\
$^\flat$ \href{http://hep.physics.uoc.gr}{Crete Center for Theoretical Physics}, Institute for Theoretical and Computational Physics,
Department of Physics, P.O. Box 2208,\\
University of Crete, 70013, Heraklion, Greece\\
~\\
$^*$
Department of Physics, University of Florida,
Gainesville, FL 32611, USA
}

\preprint{CCTP-2022-6\\ITCP-2022/6}
\abstract{Holographic RG flows dual to QFTs on a squashed $S^3$ are considered in the framework of Einstein dilaton gravity in four dimensions. A general dilaton potential is used and flows are driven by a scalar relevant operator. The general
properties of such flows are analysed and the UV and IR asymptotics are computed.
Exotic asymptotics  are found,  that are different from the  standard Fefferman-Graham asymptotics.}

\begin{document}
\maketitle

\section{Introduction, summary of results and outlook}

Renormalization group (RG) flows in Quantum Field Theory (QFT) are usually studied in flat space.
There are, however, many reasons to consider QFTs on curved manifolds
and study the associated RG flows.

One reason is that curved manifolds are considered in order to render QFTs well defined or well controlled in the IR, by taming IR divergences. There are many facets of this idea, going back to \cite{Adler,CW}, and to \cite{KK} for a similar approach of regulating IR divergences in string theory. On the holographic side of QFTs, this is the role played by global AdS space. There, the QFT lives on $R\times S^d$, where the spatial part is a sphere.

Curved manifolds,  provide IR modifications to supersymmetric QFTs in a way that the supersymmetric indices or any other supersymmetric observable are well-defined,
\cite{Komar, Marte}. They also provide control parameters on which supersymmetric observables can depend upon.
Non-trivial backgrounds can be useful in deriving non-trivial results like exact $\beta$-functions, \cite{NSVZ}. Spheres have been  used to regulate otherwise singular holographic solutions in \cite{Bu1}. Holographic flows with de Sitter or other cosmological slices were analyzed in \cite{Bu2}-\cite{CdL}.

Partition functions on curved manifolds are further objects of interest. The partition function on spheres, in particular,   was argued to serve as an analogue of the c-function in odd dimensions. The case of three dimensions is known, \cite{J},  but from holographic arguments the case can also be made for other dimensions, \cite{M,F}.
The dynamics of QFTs on curved manifolds may have a different structure from that on flat manifolds, especially in the case of QFTs on AdS manifolds, \cite{ABTY}.

Cosmology is another important context where quantum  field
theories on curved space-times are central.  The  non-perturbative renormalization group flow   on de Sitter backgrounds  was
studied for large-$N$ scalar field theories in \cite{Tsamis:1992xa, Tsamis:1994ca, Tsamis:1996qm,Ramsey:1997qc,Burgess:2009bs,Serreau:2011fu}.

A folk theorem says that RG flows of QFTs on curved
manifolds are very similar to those on flat manifolds. The argument
for this  is that $\beta$-functions are determined by short-distance divergences, and the UV structure of a given QFT is independent of the curvature.
Although the leading intuition of such statements is correct, the folk theorem fails on several grounds. Indeed, the leading UV divergences are independent of curvature. However, subleading divergences  do depend on curvature.
 A further observation is that already for
CFTs, curvature is a source of breaking of scale invariance via the
conformal anomaly, \cite{Duff}. For generic QFTs, driven by relevant
couplings, the $\beta$-functions do  depend on curvature,
\cite{Osborn}. Vacuum expectation values of
operators also depend on curvature. It is expected that curvature becomes very
important in the IR and this expectation is generically correct.

\vspace{0.3cm}
In this work we shall study the RG behavior of holographic QFTs on the squashed-$S^3$ manifold  using the framework of holography.

The holographic correspondence, \cite{M1,M2,M3}, provides a map between QFT and
gravity/string-theories in higher dimensions, at least in the limit of large $N$ and strong coupling.
 In this context, the holographic dimension serves as an effective RG scale in the
  dual QFT, thus geometrizing the notion of RG flow. In essence, RG flows can be
  understood as bulk evolution in the holographic dimensions \cite{R1}-\cite{axion2}, \cite{C}.

Holography also provides a means of calculating the induced action for the background
metric of a  QFT. Integrating out a QFT coupled to a background metric, provides
a Schwinger functional for the metric that can be turned into an effective action
 for the expectation value of the stress tensor. This is the starting point for
  many cosmology setups, like that of Starobinsky, \cite{staro}, and its
  generalizations, \cite{kcosmo}. It is also the formalism relevant for
  emergent gravity, \cite{e1}-\cite{e5}.

The metric on the squashed $S^3$ manifold is
\be
d\Omega^2_a = \frac{L^2}{4} \left( a^2 \left( d\psi + \cos\theta d\phi\right)^2 + d\Omega^2 \right)
\label{ga4n}\ee
where $L$ is the radius and $a \in \mathbb{R}^+$ is a dimensionless parameter characterising the deformation of $S^3$. The round three-sphere is obtained when we substitute $a=1$, with $L$ corresponding to its radius.
The three-dimensional quantum field theories on a squashed $S^3$ have been considered in the past with emphasis on N=2 supersymmetric theories, \cite{Ima,Hama}.
With a judicious choice of a background gauge field, the theory preserves supersymmetry. Localization techniques can be used for the computation of the partition function. This  depends non-trivially on the squashing parameter which  provides a nontrivial dimensionless control parameter for the theory.
Holographic duals have been discussed both in the vector case, \cite{Hartnoll} and in the standard adjoint case, \cite{Myers}-\cite{B2}.

For CFT$_3$ on squashed $S^3$, the holographic solution, \cite{Myers} made contact with Nut\footnote{A Nut is a point-like zero of a Killing vector, while a Bolt is an $S^2$ famiy of fixed points.} and Bolt structures found earlier in gravitational solutions.
Two families of solutions where found in \cite{Myers}. Both solutions can be thought as conifolds of the $S^1$ and $S^2$ factors of the squashed three-sphere.
The AdS-Taub-Nut solution has the property that both $S^1$ and $S^2$ shrink together in the interior and this is how the manifold ends.
The solution exists for all values of $a\in {\mathbb{R}}^+$.

The AdS-Taub-Bolt solution has the property that $S^2$ shrinks to zero in the bulk while $S^1$ remains finite. It exists only if the squashing factor is sufficiently small, $a\lesssim 0.3$.
Therefore, for $a\gtrsim 0.3$ there is only one saddle point for the ground-state of a  holographic CFT$_3$ on squashed-$S^3$.
However, for $a\lesssim 0.3$ there are two competing saddle points, the Nut and the Bolt solution. The Nut solution has lower free energy as long as $ 0.08\lesssim a\lesssim 0.3$
while the Bolt is the dominant saddle for smaller values of $a$.

\subsection{Summary of results}

In this paper we  start a systematic analysis of RG flows associated to three-dimensional holographic QFTs on squashed-$S^3$.
We shall use a general four-dimensional  Einstein-dilaton action. as in (\ref{ga1}), that is capable of describing QFTs with a single relevant coupling.
Overall, such theories depend on this relevant coupling, the radius of curvature of the squashed-$S^3$ manifold as well as the dimensionless squashing parameter $a$.
Therefore, such holographic QFTs  have two dimensionless parameters that can be chosen at will.

 We use a metric ansatz with two independent scale factors, as in (\ref{ga3}), that control the size of the $S^1$ and the $S^2$ factors of the metric. In total, with the dilaton we have three unknown functions that control the RG flow and will be determined by solving the equations of motion.

 We derive the first order equations, appropriate for the RG interpretation,
 analyze their singular points and interpret their parameters in terms of the dimensionless sources of the theory as well as the dimensionless vevs of the scalar and the stress tensor.

 We then find analytically the solutions near the singular points. These points include, extrema of the potential as well as points where $\dot \f$ vanishes.
 We also analyze the bulk curvature invariants, and determine all the ways the geometry can end regularly in the bulk.
 We confirm that   there are only two ways this can happen, at a Nut or a Bolt as in the conformal solution, \cite{Myers}.

 For all such asymptotics, analytic expansions are found, which include the non-analytic parts of the solutions, and all relevant integration constants that are mapped
 to physical dual QFT parameters (sources and vevs).
 Such analytic expansions are crucial in implementing numerical solutions for the bulk equations.
 In particular, the two dimensionless sources are given by the squashing parameter $a$ and the dimensionless curvature ${\cal R}$ that is the physical curvature divided by the appropriate power of the relevant coupling constant.
 Moreover, the scalar vev as well as the stress tensor vevs can be also written in terms of the solution parameters, and this is done in  equations (\ref{ga93})-(\ref{ga99}).

We then solve numerically the bulk equations for a generic bulk $\f^4$  scalar potential and find the families of regular solutions.
From them, we compute and plot the two-dimensionless source parameters $a$ and ${\cal R}$ as defined in
(\ref{ga93})-(\ref{ga99}) in terms of the solution parameters, and determine the vevs as functions of the sources.

 Finally, we have discovered new classes of asymptotic solutions, near  extrema of the potential, where they have a boundary, that  does not fit the Fefferman-Graham asymptotics.
 They exist only in the presence of relevant couplings (running scalars). For such asymptotics, the boundary metric is degenerate, not unlike the cases discussed in
 \cite{S,T}.
 We do not know if such asymptotics can be part of global regular solutions. We plan to investigate this in future work.

\subsection{Open questions and outlook}

The present papers sets the basics for a systematic study of holographic  RG Flows on squashed-$S^3$, There are however several issues that remain open and
require further study.

\begin{itemize}

\item The landscape of solution in the presence of several competing extrema needs to be studied in more detail. For each end-point of the flow in field space, $\f_0$, there are two solutions, the Nut-like and the Bolt-like. Which UV fixed point they end-up, depends on the scalar landscape.

\item The on-shell action as a function of the flow and source data needs to be computed. In particular,
its asymptotics for $\f_0$ near maxima or minima of the potential, must be worked-out analytically. The structure of Nut-Bolt phase transitions must be studied in the QFT case.
We do expect that for a given dimensionless curvature $\mathcal{ R}$ there is a maximum squashing parameter $a^2_{max}(\mathcal{R})$ below which bolt-like solutions exist.
We have indeed verified this in the present paper. We expect that at above some squashing parameter $a^2_*(\mathcal{R})$ the Nut-like solution dominate while below it is the Bolt-like solution that dominates. This a smooth continuation of the what we know already happens in CFTs.
This would probably persist for very large and very small values of $\mathcal{R}$ as in these two limits we approach the UV and the IR CFTs.

\item The novel, non-Fefferman-Graham asymptotics found in appendix \ref{exot} must be studied further. In particular we need to know whether they can appear as part of a global regular
flow solution, and under what conditions. Moreover, we need to understand the holography rules in that case, as they transcend the Fefferman-Graham paradigm.

\item The qualitative dependence of the on-shell action on the squashing parameter must be elucidated from the previous elements of the study.
This may be usufull for applications of the wave-function of the universe idea, \cite{B2}.

 \item Last but not least, other manifolds with parameters may play the role of the squashed-$S^3$ in this study.

\end{itemize}

The structure of this paper is as follows.
In section \ref{s2} we present the general setup and equations for the flows. We review the Nut and Bolt solutions of \cite{Myers} for the CFT$_3$ case,
and by comparing their free energies we demonstrate the quantum phase transition as a function of the squashing parameter.
In section \ref{thefof} we develop the first order formalism. In section \ref{potextrema} we derive the asymptotics of solutions near extrema of the potential.
In section \ref{interiorgeo} we study the interior geometry, the end-points of the flow as well as $\f$-bounces.
Finally, in section \ref{numerics} we study complete flows and their data based on numerical solution of the equations of motion.

In appendix \ref{squashedsphere} we review the geometry of the squashed-$S^3$. In appendix \ref{PGexpansion} we study in detail the Fefferman-Graham expansion of the conformal
Nut and Bolt solutions. In appendix \ref{Curvatureinvariants} we analyze the bulk curvature invariants to be used as litmus test of the regularity of the flow solutions.
In appendix \ref{extremalpoints} we find all possible end points of the flows.
In appendix \ref{IR-endpoints} we develop the perturbative expansion around end-points of the flow that are not extrema of the potential.
In appendix \ref{Extremaexpansions} we develop the perturbative expansion of the solutions near endpoints that are extrema of the potential.
Finally in appendix \ref{expval} we review some formulae for expectation values of the dual field theory and their relation to the holographic flow solutions.

\section{Holographic space-times with squashed $S^3$ slicing and RG flows\label{s2}}

Our object of study are holographic RG flows of QFTs on squashed $S^3$. We shall be using for simplification, an  Einstein-dilaton theory in (3+1)-dimensions with Euclidean  metric.
 The scalar is expected to be dual to the relevant operator of the dual QFT that is driving the flow. Although this does not include effects of operator mixing during the RG flow, it sufficient to show the most important aspects of the RG dynamics. The action we are using is

\be
S[g,\f] = M^2 \int d^3x du \sqrt{\vert g \vert} \left( R^{(g)} - \frac{1}{2} \partial_\mu \f \partial^\mu \f - V(\f ) \right) + S_{GHY}
\label{ga1}\ee
where $S_{GHY}$ is the Gibbons-Hawking-York boundary term and can be written as

\be
S_{GHY} = 2M^2 \left[\int d^3x \sqrt{\gamma} K \right]_{UV}
\label{ga2}\ee
with $\gamma$ being the determinant of the induced metric defined on fixed $u$ slices and $K$ the associated trace of the extrinsic curvature. The Euclidean action can be obtained by setting $S_E = -S$ and changing the metric to positive signature.

In this work we shall be interested in boundary field theories defined on three-dimensional squashed spheres. Without loss of generality, we can employ domain wall coordinates and choose the following ansatz for $\f$ and the (3+1)-dimensional metric:

\be
\f = \f(u) \ , \  ds^2 = du^2 + L^2 \left( e^{2A_1(u)} \left( d\psi + \cos\theta d\phi\right)^2 + e^{2A_2(u)} \left( d\theta^2 + \sin^2\theta d\phi^2 \right) \right)
\label{ga3}\ee
where $L$ has units of length, $\theta,\phi,\psi$ are the Euler angles, taking the values $\theta \in [0,\pi]$,$\phi \in [0,2\pi]$,$\psi \in [0,4\pi]$ and $A_1,A_2$ are scale factors depending only on the domain wall coordinate $u$. We may parametrise the three-dimensional squashed sphere metric as

\be
d\Omega^2_a = \frac{L^2}{4} \left( a^2 \left( d\psi + \cos\theta d\phi\right)^2 + d\Omega^2 \right)
\label{ga4}\ee
where $a \in \mathbb{R}^+$ is a dimensionless parameter characterising the deformation of $S^3$.
This metric has $SU(2)\times U(1)$ isometries compared to the $SU(2)\times SU(2)$ isometry of the round S$^3$. There are further squashed classes of metrics with lower symmetry,
like $SU(2)$ or only U(1) symmetry but we do not consider them here.
The round three-sphere is obtained when we substitute $a=1$, with $L$ corresponding to its radius. For completeness, in Appendix \ref{squashedsphere} we calculate the curvature invariants of the squashed 3-sphere.

In the following, we shall also adhere to the following convention for derivatives, ie. primes are derivatives with respect to the scalar $\f$ while dots are derivatives with respect to the holographic domain wall coordinate $u$

\be
f'(\f) \equiv \frac{df}{d\f} \sp \dot{g}(u) \equiv \frac{dg}{du}
\label{ga5}\ee

Varying the action (\ref{ga1}) with respect to the metric $g_{\mu\nu}$ found in (\ref{ga3}) and the scalar field $\f$ gives rise to the following equations of motion:

\be
\ddot{\f} + \left( \dot{A}_1 + 2 \dot{A}_2 \right) \dot{\f} - V'(\f) = 0
\label{ga6}\ee
\be
4\dot{A}_2^2 + 8 \dot{A}_1 \dot{A}_2 + \frac{1}{L^2 }e^{2A_1 -4 A_2} - \frac{4}{L^2} e^{-2A_2} - \dot{\f}^2 + 2 V = 0
\label{ga7}\ee
\be
4 \ddot{A}_2 + 4 \dot{A}_2 \left( \dot{A}_2 - \dot{A}_1 \right) + \frac{1}{L^2} e^{2A_1- 4A_2} + \dot{\f}^2 = 0
\label{ga8}\ee
\be
\ddot{A}_1 - \ddot{A}_2 + \left(\dot{A}_1 - \dot{A}_2 \right)\left( \dot{A}_1 + 2 \dot{A}_2 \right) + \frac{1}{L^2}e^{-2A_2} - \frac{1}{L^2} e^{2A_1-4A_2} = 0
\label{ga9}\ee
where equation (\ref{ga6}) arises from variation with respect to $\f$, and can be derived from equations (\ref{ga7}) - (\ref{ga9}).
Holographic RG flow are in one-to-one correspondence with regular solutions of the equations (\ref{ga6}) - (\ref{ga9}),with Dirichlet boundary conditions at the AdS boundary. Hence, in the following we shall be interested in solutions to these equations for various choices of the potential of $V(\f)$. We note, however, that $V(\f)$ must be negative-definite.

The potential may have maxima and/or minima representing distinct UV or IR fixed points of the dual CFT. Our first goal is  to study the solutions \text{at} the extrema of the potential, that are dual to CFT$_3$s on the squashed $S^3$. This was already studied in \cite{Myers,Hartnoll} but we briefly present it for setting up our notation.

\subsection{Conformal Fixed Points: Nuts and Bolts}\label{conffixedpoints}

We are interested in cases in which $\f = \text{const.}$, i.e. we have $\dot{\f} = \ddot{\f} = 0$. Such solutions are associated with extrema $\f_{ext}$ of the potential, ie. we have $V'(\f_{ext}) = 0$. The value of the potential at this point is defined to be $V(\f_{ext}) = - \frac{6}{\ell^2}$, where $\ell$ is the AdS length scale.

Before we continue, we note that for this subsection only we are going to use the following convention:

\be
\dot{f}(r) \equiv  \frac{df}{dr}.
\label{ga10}\ee

We are also going to change our coordinate system from (\ref{ga3})  to the following coordinates as they are more convenient in describing the associated solutions:

\be
ds^2 = \frac{dr^2}{f(r)} + \frac{L^2}{4}f(r) \left( d\psi + \cos\theta d\phi \right)^2 + \frac{L^2}{4} g(r) d\Omega^2.
\label{ga11}\ee
In these coordinates, the AdS boundary is at $r \to \infty$. By comparing the previous metric with the one in (\ref{ga3}) we observe that we have the following relations:

\be
du = \frac{dr}{\sqrt{f(r)}} \ , \ e^{2A_1} = \frac{1}{4} f(r) \ , \ e^{2A_2} = \frac{1}{4} g(r).
\label{ga12}\ee

Substituting the previous equation into the equations  (\ref{ga6})  - (\ref{ga7}) and setting the
scalar $\f$ to one of the extrema of the potential $V(\f)$ yields the following equations:

\be
2 \frac{\dot{f}(r)\dot{g}(r)}{f(r)g(r)} - \frac{16}{L^2f(r)g(r)} - \frac{12}{f(r)\ell^2} + \left( \frac{\dot{g}(r)}{g(r)} \right)^2 + \frac{4}{L^2 g(r)^2} = 0,
\label{ga13}\ee
\be
2 \frac{\dot{f}(r)\dot{g}(r)}{f(r)g(r)} - \frac{16}{L^2f(r)g(r)} - \frac{12}{f(r)\ell^2} + 4 \frac{\ddot{g}(r)}{g(r)}- \left( \frac{\dot{g}(r)}{g(r)} \right)^2 + \frac{12}{L^2 g(r)^2} = 0,
\label{ga14}\ee
\be
\frac{\ddot{g}(r)}{g(r)} + \frac{8}{L^2g(r)^2} - \frac{8}{L^2f(r)g(r)} - \frac{\ddot{f}(r)}{f(r)} = 0.
\label{ga15}\ee
The general solution of equations (\ref{ga13})-(\ref{ga15}) can be written as,
\be
g(r) = b \left( r^2 - \frac{1}{b^2L^2} \right) \ , \ f(r)=\frac{b^4L^4r^4+b^2L^2(4b\ell^2-6)r^2+(4b\ell^2-3 )-C_1b^2L^2\ell^2 r}{ b^2L^2\ell^2(b^2L^2 r^2-1)},
\label{ga16}\ee
and it contains the following integration constants: $b,C_1$.
The metric (\ref{ga11}) has the following behavior near the boundary:
\be
ds^2 \approx \frac{\ell^2}{r^2} dr^2 + \frac{L^2 r^2}{4\ell^2} \left( d\psi + \cos\theta d\phi \right)^2 + \frac{L^2b r^2}{4} d\Omega^2,
\label{ga18}\ee
which, after substituting
\be
\tilde{r} = \frac{\ell^2}{r} \sp L = \hat{L} a \sp b = \frac{1}{a^2\ell^2}
\label{ga19}\ee
becomes
\be
ds^2 \approx \frac{\ell^2}{\tilde{r}^2} \left( d\tilde{r}^2 + \frac{\hat L^2}{4} \left( a^2    \left( d\psi + \cos\theta d\phi \right)^2 + d\Omega^2 \right) \right).
\label{ga20}\ee
This indicates that the solution has as a source the squashed sphere metric with radius $\hat{L}$ and deformation parameter $a$.

It shall be useful to rewrite the functions $g(r),f(r)$ in a different form that shall make it easier to compare our notation with the one presented in \cite{Myers} and \cite{Hartnoll}. For that, we change our metric (\ref{ga11}) as follows:
\be
ds^2 = \frac{dv^2}{U(v)} + 4n^2 U(v)\left(d\psi + \cos\theta d\phi \right)^2 + \left(v^2-n^2 \right)d\Omega^2
\label{ga21}\ee
with
\be
U(v) = \frac{v^2 +n^2 -2mv + \ell^{-2}\left( v^4 -6n^2v^2-3n^4\right)}{v^2-n^2},
\label{ga22}\ee
where $m$,$n$ are parameters with dimensions of length. The latter one is commonly known as \textit{the Nut parameter}. Upon comparing the two metrics (\ref{ga21}) , (\ref{ga11}) we observe that the functions $f,g,V$ are related as follows:

\be
f(r) = \frac{16n^2}{L^2} U(\frac{L}{4n}r) \sp g(r) = \frac{1}{4n^2}\left( r^2 - \frac{16n^4}{L^2} \right).
\label{ga23}\ee

From the previous relation and from (\ref{ga19}) we observe that we can write the Nut parameter in terms of $a,\ell$ as follows:
\be
n = \frac{a\ell}{2 }.
\label{ga24}\ee

The form of the metric (\ref{ga3}) indicates that the geometry ends when either of the functions $f(r),g(r)$ vanish. The point at which this happens is called the IR end point. We shall call the case in which $g(r) \to 0$ the \textit{Taub-Nut} solution, and abbreviate the root of $g$ with $r^n_{IR}$. The second solution, ie. the one in which the function $f(r) \to 0$, shall be called the \textit{Taub-Bolt} solution. We shall abbreviate this root of $f$ with $r_0$. We note that in since the geometry ends first for the Taub-Nut solution, we must demand that $r_0 > r^n_{IR}$.

\subsubsection{The Taub-Nut } \label{taubnut}

In this case the IR point coincides with the root of $g(r)$, ie. we have

\be
r^{n}_{IR} = \frac{4n^2 }{L},
\label{ga26}\ee
where the index $n$ stands for "Nut". In order to determine the uknown parameter $m$, we are going to demand that the curvature invariants are regular at the endpoint of the geometry. As was discussed in Appendix \ref{Curvatureinvariants}, the only curvature invariant that gives a non-trivial condition is the Kretschmann scalar. Demading that the Kretschmann scalar of the metric (\ref{ga11}) does not diverge yields for the parameter $m$ the following:

\be
m_n = n - \frac{4n^3}{\ell^2}.
\label{ga27}\ee

The function $f$ now takes the form

\be
f_n = \frac{(Lr - 4n^2)(L^2 r^2 + 8n^2(Lr + 2\ell^2) - 48n^4)}{L^2\ell^2(Lr + 4n^2)}.
\label{ga28}\ee

We observe that in order for the Kretschmann scalar to be regular, we have to demand $f$ to vanish as we approach the IR. The action (\ref{ga1}) now becomes

\be
S_{on-shell}^{nut} = 2\pi^2 N^2 \left(S_{div} + 32 \frac{n^2}{\ell^2} \left( 3 - \frac{14n^2}{\ell^2}\right) + \mathcal{O}(\epsilon)\right),
\label{ga29}\ee
where  $\epsilon$ is a dimensionless parameter that approaches $0$ as we approach the UV, and is related to the cutoff R of the radial coordinate through the following equation:
\be
\epsilon = \frac{\ell^2}{R L}.
\label{ga30}\ee

We have also denoted with $N = M \cdot \ell$  the number that represents the degrees of freedom, and we have denoted with
\be
S_{div} =  - \frac{16(1 - 7 \frac{n^2}{\ell^2})}{\epsilon}  - \left(\frac{n}{\ell}\right)^{-2} \frac{2}{\epsilon^3}
\label{ga30.a}\ee
the diverging part of the action. Near the IR endpoint, the metric (\ref{ga3}) becomes

\be
ds^2 \approx \frac{L}{2} \left( \frac{dr^2}{r-r_{IR}^n} + (r -r_{IR}^n) d\Omega_3^2 \right).
\label{ga30.b}\ee

Therefore, this solution amounts to a shrinking of the three-dimensional sphere.
\subsubsection{The Taub-Bolt} \label{taubbolt}

In this case the IR endpoint coincides with the root of $f$: $f(r_0) = 0$. This relation can be solved for the parameter $m$ as follows

\be
m_b = \frac{L^3 r_0^3}{128n^3 \ell^2} + \frac{2n^3\ell^2 - 6n^5}{Lr_0 \ell^2} + L \left(  \frac{r_0}{8n} - \frac{3nr_0}{4\ell^2} \right).
\label{ga31}\ee

The function $f$ now becomes

\be
f_b=\frac{(r-{r_0}) \left(L^4 r
   {r_0} \left(r^2+r
   {r_0}+{r_0}^2\right)-96
   L^2 n^4 r {r_0}+16 L^2 n^2 r
   {r_0} \ell ^2+768 n^8-256 n^6
   \ell ^2\right)}{L^2 {r_0}
   \ell ^2 \left(L^2 r^2-16
   n^4\right)}.
\label{ga32}\ee

We must demand regularity of the curvature at $r_0$ and for this we expand the
metric (\ref{ga11}) in the vicinity of $r_0$:

\be
ds^2 \approx \frac{dr^2}{f'(r_0)(r-r_0)} + \frac{L^2}{4} f'(r_0)(r-r_0)\left(d\psi + \cos\theta d\phi\right)^2 + \frac{L^2}{4}g(r_0) d\Omega^2.
\label{ga33}\ee

After defining $ u \equiv \frac{2\sqrt{r-r_0}}{\sqrt{f'(r_0)}} $ becomes
\be
ds^2 \approx du^2 + \frac{L^2}{16} f'(r_0)^2u^2  \left(d\psi + \cos\theta d\phi\right)^2 + \frac{L^2}{4}g(r_0) d\Omega^2,
\label{ga34}\ee
which has a topology $\mathbb{R}^2 \times S^2$. Since $\psi \in [0,4\pi]$, in order to avoid having a conical singularity we must choose
\be
f'(r_0) = \frac{2}{L},
\label{ga35}\ee
which implies
\be
r_0 = \frac{\ell^2}{3L} \left( 1 + \sqrt{144 \left( \frac{n}{\ell} \right)^4 - 48 \left( \frac{n}{\ell} \right)^2 + 1}\right).
\label{ga36}\ee

Demading that the previous root is smaller that $r^n_{IR}$, we obtain the following condition:
\be
\frac{n}{\ell} \leq \sqrt{\frac{2-\sqrt{3}}{12}} \implies a \leq a_{crit} \approx 0.3,
\label{ga37}\ee
therefore the Bolt solution exists only for the previous values of the squashing parameter. In contrast, the Nut solution exists for all possible values.

The on-shell action now becomes

\be
S_{on-shell}^{bolt} = 2\pi^2 N^2 \left( S_{div} +  \frac{16 n \left(3 \ell ^2 \left(n^2+r_b^2\right)-24 n^2 r_b^2-9 n^4+5 r_b^4\right)}{r_b \ell ^4} + \mathcal{O}(\epsilon) \right),
\label{ga38}\ee
where  $\epsilon$ is a dimensionless parameter that approaches $0$ as we approach the UV, and is related to the cutoff R of the radial coordinate through the following equation:

\be
\epsilon = \frac{\ell^2}{R L},
\label{ga39}\ee
and
\be
r_b = \frac{L}{4n} r_0 = \frac{\ell^2}{12n}\left( 1 + \sqrt{144 \left( \frac{n}{\ell} \right)^4 - 48 \left( \frac{n}{\ell} \right)^2 + 1}\right)
\label{ga40}\ee
is a rescaling of the root $r_0$ and
\be
S_{div} = - \frac{16(1 - 7 \frac{n^2}{\ell^2})}{\epsilon} - \left( \frac{n}{\ell} \right)^{-2} \frac{2}{\epsilon^3}
\label{ga40.a}\ee
is the diverging part of the action. Finally, near the IR endpoint $r_0$ the metric (\ref{ga3}) becomes

\be
ds^2 \approx \frac{L^2}{16n^2} \frac{dr^2}{\frac{r}{r_0}-1} + 4n^2 \left( \left( \frac{r}{r_0} - 1 \right) \left( d\psi + \cos\theta d\phi \right)^2 +  (\frac{L^2 r^2}{16n^4} - 1) d\Omega_2^2 \right).
\label{ga40.b}\ee

Therefore, this solution amounts to a shrinking of the circle $S^1$.
\subsubsection{A comparison of the two theories\label{alpha}}

We compare the free energies of the two solution in order to determine which one dominates the Euclidean gravitational path integral. For this we shall take the difference of their on-shell actions. Defining
\be
S_{on-shell}^{diff} = \lim_{\epsilon \to 0 }\left( S_{nut} - S_{bolt} \right)
\label{ga41}\ee
we obtain, after substituting the values of $r_b,n$ and converting  to Fefferman-Graham (FG) coordinates (see Appendix \ref{PGexpansion} for the relevant calculations), the following expression:

$$
S_{on-shell}^{diff}=  - 8\pi^2 N^2\left(-\frac{\left(9 a^4-3 \left(\sqrt{9 a^4-12 a^2+1}+6\right) a^2+4 \sqrt{9 a^4-12 a^2+1}+6\right)}{9} \right.$$
\be
\left. + \frac{\sqrt{9 a^4-12 a^2+1}+1}{27a^2} \right).\label{ga42}\ee

We redefine

\be
a^2 = \frac{1}{1+\alpha}.
\label{ga43}\ee

Substituting the previous redefinition into the action (\ref{ga42}) yields the following expression:
$$
S_{on-shell}^{diff} =- \frac{8}{27} \pi^2 N^2 \left(\frac{9}{(\alpha+1)^2} \left( \sqrt{\alpha(\alpha-10)-2} -3 \right)   \right.
$$
\be\left. + \frac{6}{\alpha+1}\left( 9 - 2 \sqrt{\alpha(\alpha-10)-2}\right) + \sqrt{\alpha(\alpha-10)-2} -17 + \alpha \right).\label{ga44}\ee

We note that the previous action is defined only for the values of $\alpha$ for which both the Nut and Bolt solution exist, which as can be found from (\ref{ga37}) is

\be
\alpha \geq \alpha_{crit} \approx 10.2 .
\label{ga45}\ee

The Nut solution, however, exists $\forall \ \alpha > -1$. A plot of the action $S_{on-shell}^{diff}$ can be seen in Figure 1. We note the following:

\begin{itemize}
   \item[$\bullet$] When $-1 < \alpha < 12.3$, the on-shell action is positive. This implies that $S_{nut} > S_{bolt}$, i.e. the Nut solution is dominant.
   \item[$\bullet$] The action $S^{diff}_{on-shell}$ changes sign at $\alpha = 12.3$. This means that we have a first order phase transition from the Nut to the Bolt.
   \item[$\bullet$] When $\alpha > 12.3$ the Bolt solution is dominant.
\end{itemize}

The work presented in this subsection has been studied extensively by \cite{Myers} and \cite{Hartnoll}.
\FIGURE{
      \includegraphics[scale = 0.53]{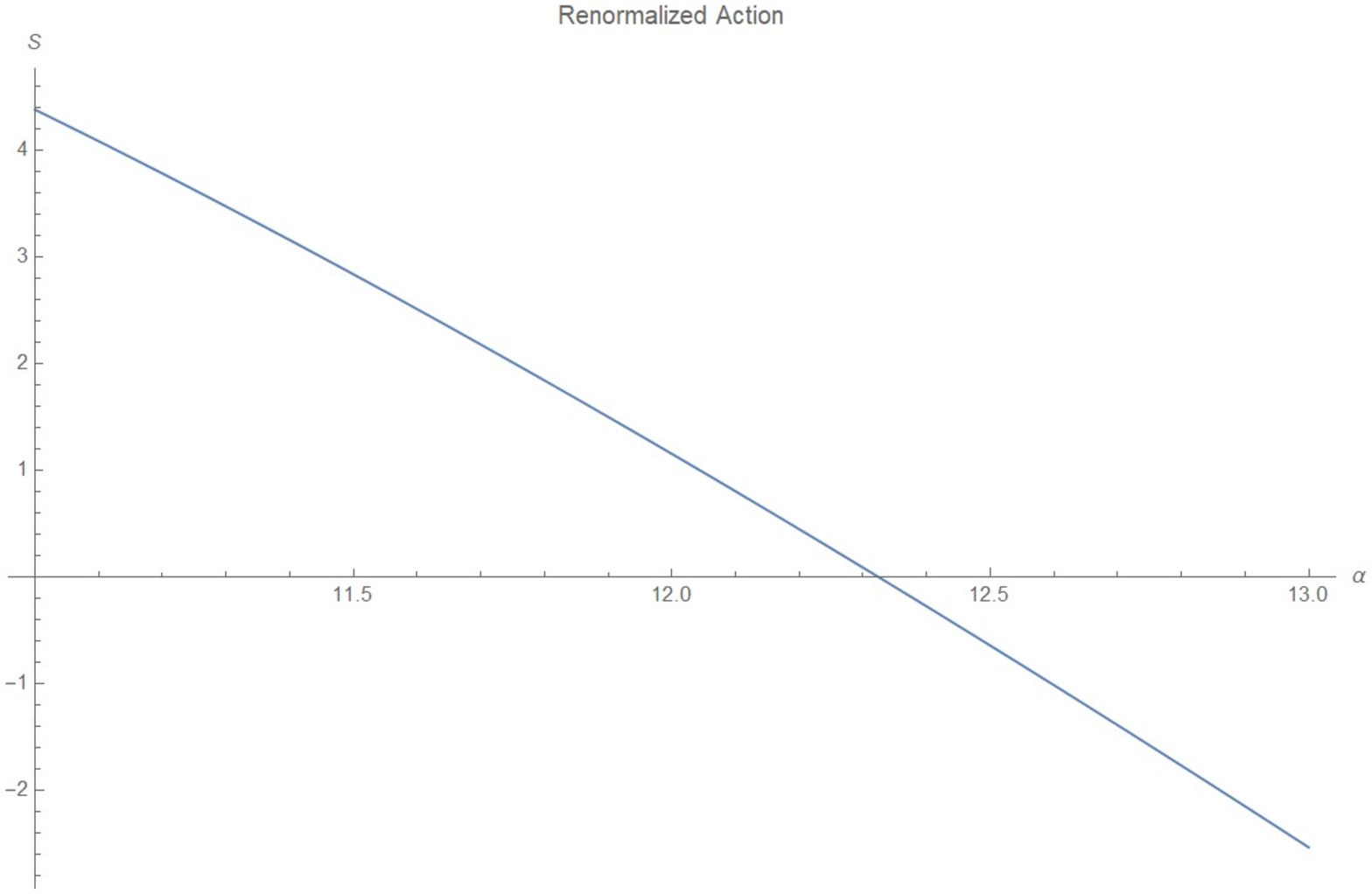}
      \caption{ The action $S_{on-shell}^{diff}$ plotted as a function of the scaling parameter $\alpha$.}
      \label{fig1}
}

\section{The First Order Formalism} \label{thefof}

We now turn our attention to solutions in which the scalar field has a non trivial dependence on the coordinate u. Our goal is to interpret the solutions to the equations (\ref{ga7}) - (\ref{ga9}) in terms of RG flows.
To this end, it shall be convenient to rewrite the second order Einstein equations as a set of first-order equations, which shall allow an interpretation as gradient RG flows. This is locally always possible, except at special points where $\dot{\f} = 0$. Given a solution, as long as $\dot{\f}(u) \neq 0$, we can invert the relation between $u$ and $\f$ and define the following scalar functions:

\be
W_1(\f ) \equiv-4 \dot{A}_1,
\label{ga46}\ee
\be
W_2(\f )\equiv- 4 \dot{A}_2,
\label{ga47}\ee
\be
T_1 \equiv \frac{4}{L^2}e^{2A_1-4A_2},
\label{ga48}\ee
\be
T_2 \equiv \frac{4}{L^2} e^{-2A_2},
\label{ga49}\ee
\be
S \equiv \dot{\f}.
\label{ga50}\ee

Equations (\ref{ga48}),(\ref{ga49}) can be inverted and written in terms of the exponentials $e^{2A_1},e^{2A_2}$ as follows:

\be
e^{2A_1} = \frac{4T_1}{L^2 T_{2}^2} \ , \ e^{2A_2} = \frac{4}{L^2 T_2}.
\label{ga48.a}\ee

The equations of motion (\ref{ga7}) - (\ref{ga9}) and the scalar one (\ref{ga6}) now become

\be
W_2^2+2W_1W_2+T_1-4T_2-4S^2+8V=0,
\label{ga51}\ee
\be
4(W_2'-W_1')S+(W_1-W_2)(W_1+2W_2)+4(T_2-T_1)=0,
\label{ga53}\ee
\be
SS' - \frac{1}{4} S\left( W_1 + 2 W_2 \right) - V' = 0,
\label{ga55}\ee
\be
-4W_2'S+W_2^2-W_1W_2+T_1+4S^2=0.
\label{ga52}\ee

We also have the defining relations,
\be
ST_1'=-\frac{1}{2}(W_1-2W_2)T_1\sp ST_2'= \frac{1}{2} W_2 T_2 ,
\label{ga54}\ee
obtained by differentiating the definitions in (\ref{ga48}) and (\ref{ga49}).
The previous set of equations is comprised of four first order differential equations and an algebraic one. Therefore, we have a total of 4 integration constants. In the analysis that shall follow, we are going to make use of equations (\ref{ga51}),(\ref{ga53}) - (\ref{ga54}). We note, however, that equation (\ref{ga52}) is not  independent, as we shall see in the sequel.

The original system of equations (\ref{ga7})-(\ref{ga9}) has 5 integration constants. These five are given by the four integration constants of the previous equations plus one more , that emerges when solving $\frac{df}{du}=S(\f)$. Upon examining equations (\ref{ga46}),(\ref{ga47}) we observe that there are two extra integration constants introduced when we integrate with respect to $u$. However, from the definitions of the functions $T_1,T_2$ in (\ref{ga48}),(\ref{ga49}) it becomes clear that there are also two constraints that completely define the previously mentioned integration constants.
The dual QFT has three sources, $\hat L, a$ and $\f_0$ (the scalar coupling) and the respective three vevs. Out of the six one can make 5 dimensionless parameters and these are the integration constants of the original system.
Out of these $\hat L, a$ in units of $\f_0$ are arbitrary, while the other three vevs must be determined by regularity conditions in the IR.

In this work, we shall be concerned with solutions that start from a UV point and end in an IR point. For regularity, we shall be interested in flows in which S is always finite and approaches 0 as we approach one of the fixed points. Before we present our results, we shall manipulate equations (\ref{ga51}),(\ref{ga53}) - (\ref{ga54}) and bring them to a more convinient form. To that end, we solve (\ref{ga51}),(\ref{ga53}) with respect to $T_1,T_2$:

\be
T_1 = \frac{1}{3} \left(-4 S W_1'+4 S W_2'-4 S^2+8 V+3 W_1 W_2+W_1^2-W_2^2\right),
\label{ga158}\ee
\be
T_2 = \frac{1}{12} \left(-4 S W_1'+4 S W_2'-16 S^2+32 V+9W_1W_2+W_1^2+2 W_2^2\right),
\label{ga159}\ee
and substitute the expressions for the functions in equations (\ref{ga54}):

$$
-8 S^2 \left(W_1''+W_1-W_2''\right)-2 S (W_1-5 W_2) W_1'+4 S (3 W_1-2 W_2) W_2'+8 V' \left(W_2'-W_1'\right)+
$$
\be  +(W_1-2 W_2) \left(8 V+3 W_1 W_2+W_1^2-W_2^2\right) = 0 , \label{ga160}\ee
$$
2 S \left((W_1+9 W_2) W_1'+2 (5 W_1+2 W_2) W_2'\right)-8 S^2 \left(W_1''+2 W_1-W_2''+2 W_2\right)+
$$
\be+8 V' \left(W_2'-W_1'\right) -W_2 \left(32 V+9 W_1 W_2+W_1^2+2 W_2^2\right) = 0.  \label{ga161}\ee

Upon subtracting the two previous equations and using (\ref{ga55}) we obtain
\be
8S^2 + 8V + W_1^2 + 2 W_2^2 - 4 S\left( W_1' + 2 W_2' \right) = 0.
\label{ga162}\ee

Using (\ref{ga162}) and (\ref{ga53}), equation (\ref{ga52}) can be shown to be equivalent to (\ref{ga51}), as advertised.

The independent equations satisfied by the functions $W_{1,2}$ and $S$ are therefore,
\be
SS' - \frac{1}{4}S(W_1+2W_2) - V' = 0,
\label{num1}\ee
\be
8S^2 + 8V + W_1^2 + 2 W_2^2 - 4 S\left( W_1' + 2 W_2' \right) = 0,
\label{num3}\ee
$$
-8 S^2 \left(W_1''+W_1-W_2''\right)-2 S (W_1-5 W_2) W_1'+4 S (3 W_1-2 W_2) W_2'+8 V' \left(W_2'-W_1'\right)+
$$
\be
+(W_1-2 W_2) \left(8 V+3 W_1 W_2+W_1^2-W_2^2\right) = 0.
\label{num2}\ee

These equations have the same number of unknown integration constants as the original ones (namely four).
The analysis of these equations  is performed in  Appendices \ref{extremalpoints} - \ref{Extremaexpansions}.

\section{Perturbative analysis near  extrema of the potential}\label{potextrema}

We shall now examine solutions of our system in the vicinity of extremal points of $V(\f)$. Near the extremum the potential can be parametrized as

\be
V(\f) = - \frac{6}{\ell^2} - \frac{\Delta \Delta_-}{2\ell^2} x^2 + \mathcal{O}\left( x^3 \right),
\label{ga56}\ee
where we have defined for brevity
\be
x \equiv \f - \f_0 \ , \ x > 0,
\label{ga57}\ee
with  $\f_0$ being a generic point, and
\be
\Delta_- \equiv 3 - \Delta,
\label{ga58}\ee
with $\Delta$ being a dimensionless parameter that takes the values $\frac{3}{2} \leq \Delta < 3$ when we are at a maxima, and $\Delta > 3$ when we are at a minima\footnote{
When ${3\over 2}\Delta<2$ there is the possibility of alternative quantization, which corresponds to the scalar operator having dimension $3-\Delta$.
In this case the role of sources and vevs are interchanged, and the on-shell action of alternative quantization is the Legendre transform of the standard on-shell action with respect to the source.
The regular bulk solutions however remain the same.}. Moreover, in equation (\ref{ga57}) we tacitly assumed that $\f > \f_0$, ie. we approach the extremal points from the right. This is not always the case. However, we can easily find the expansions in the case $\f < \f_0$ by making the following substitutions:

$$ W_1 \to - W_1 \ , \ W_2 \to - W_2.$$

In the following, we are going to solve equations (\ref{ga51}),(\ref{ga53}) - (\ref{ga55}) for $W_1,W_2$,$T_1,T_2,S$ near $\f = \f_0$. The relevant calculations are performed in Appendix \ref{Extremaexpansions}. Here we present and discuss the results.

\subsection{Expansion near maxima of the potential}\label{expansionnearmaxima}

We work in an expansion in $\f$ about the maximum at $\f = \f_0$. Overall, there are 2 solutions to the equations (\ref{ga51}),(\ref{ga53}) ,(\ref{ga55}),(\ref{ga54}), that have a 3d near-boundary metric, which can be identified as the $-$ and $+$ branches that were studied in the context of maximally symmetric spheres in \cite{C}. The $-$ solution is:

\be
S_- =  \frac{\Delta_-}{\ell} x \left( 1 + \mathcal{O}(x) \right) + \frac{\Delta_-(\mathcal{R}_1 + 2\mathcal{R}_2)}{(2\Delta_- - 1)32\ell} x^{\frac{2}{\Delta_-}+1} \left(1 + \mathcal{O}(x) \right) + \frac{C_1 +2 C_2}{\Delta_- \ell} x^{\frac{\Delta}{\Delta_-}}  \left(1 + \mathcal{O}(x) \right),
\label{ga59}\ee

\be
W_{1-} = \frac{4}{\ell} + \frac{\Delta_-}{2\ell}x^2 + \mathcal{O}(x^3) + \frac{\mathcal{R}_1}{8\ell} x^{\frac{2}{\Delta_-}}  \left(1 + \mathcal{O}(x) \right) + \frac{C_1}{\ell} x^{\frac{3}{\Delta_-}} \left(1 + \mathcal{O}(x) \right),
\label{ga60}\ee

\be
W_{2-} = \frac{4}{\ell} + \frac{\Delta_-}{2\ell}x^2 + \mathcal{O}(x^3) + \frac{\mathcal{R}_2}{8\ell} x^{\frac{2}{\Delta_-}}  \left(1 + \mathcal{O}(x) \right) + \frac{C_2}{\ell} x^{\frac{3}{\Delta_-}} \left(1 + \mathcal{O}(x) \right),
\label{ga61}\ee

\be
T_{1-} = \frac{\mathcal{R}_1 + \mathcal{R}_2}{2\ell^2} x^{\frac{2}{\Delta_-}}  \left(1 + \mathcal{O}(x) \right),
\label{ga62}\ee

\be
T_{2-} = \frac{5\mathcal{R}_2 + 3\mathcal{R}_1}{8\ell^2} x^{\frac{2}{\Delta_-}}  \left(1 + \mathcal{O}(x) \right),
\label{ga63}\ee
where $C_1,C_2,\mathcal{R}_1,\mathcal{R}_2$ are integration constants, and $\frac{3}{2} \leq \Delta < 3$.

The $+$ solution is given by:

\be
S_+ =  \frac{\Delta}{\ell} x \left( 1 + \mathcal{O}(x) \right) + \frac{\Delta(\mathcal{R}_1 + 2\mathcal{R}_2)}{(2\Delta - 1)32\ell} x^{\frac{2}{\Delta}+1} \left(1 + \mathcal{O}(x) \right),
\label{ga64}\ee

\be
W_{1+} =  \frac{4}{\ell} + \frac{\Delta}{2\ell}x^2 + \mathcal{O}(x^3) + \frac{\mathcal{R}_1}{8\ell} x^{\frac{2}{\Delta}}  \left(1 + \mathcal{O}(x) \right) - \frac{2C_2}{\ell} x^{\frac{3}{\Delta}} \left(1 + \mathcal{O}(x) \right),
\label{ga65}\ee

\be
W_{2+} =  \frac{4}{\ell} + \frac{\Delta}{2\ell}x^2 + \mathcal{O}(x^3) + \frac{\mathcal{R}_2}{8\ell} x^{\frac{2}{\Delta}}  \left(1 + \mathcal{O}(x) \right) + \frac{C_2}{\ell} x^{\frac{3}{\Delta}} \left(1 + \mathcal{O}(x) \right),
\label{ga66}\ee

\be
T_{1+} = \frac{\mathcal{R}_1 + \mathcal{R}_2}{2\ell^2} x^{\frac{2}{\Delta}}  \left(1 + \mathcal{O}(x) \right),
\label{ga67}\ee

\be
T_{2+} = \frac{5\mathcal{R}_2 + 3\mathcal{R}_1}{8\ell^2} x^{\frac{2}{\Delta}}  \left(1 + \mathcal{O}(x) \right).
\label{ga68}\ee

The above expressions describe four continuous families of solutions, whose structure is an analytic expansion in integer powers of $\f$, plus a series of non-analytic, subleading terms. However, each solution has a different number of integration constants. As the reader can observe in Appendix \ref{Extremaexpansions}, each solution initially had 4 integration constants. However, some of them were set to 0, since they introduced non-subleading terms. Finally, both these solutions describe the shrinking of the three dimensional squashed sphere, given by the metric (\ref{ga1}).

Given our results for the functions $W_1,W_2,S,T_1,T_2$, we are now in a position to solve for $\f(u)$ and $A_1,A_2$.
For the $-$ branch we solve (\ref{ga50}) and (\ref{ga46}),(\ref{ga47}) subsject to (\ref{ga59}) - (\ref{ga63}) to obtain

\be
\f - \f_0 = \f_- \ell^{\Delta_-} e^{\Delta_-\frac{u}{\ell}} \left(1+\mathcal{O}\left(\left(\mathcal{R}_{1}+2 \mathcal{R}_{2}\right)\left|\varphi_{-}\right|^{\frac{2}{\Delta_-}} e^{2 \frac{u}{\ell}}\right)\right)
\label{ga81}\ee
$$ + \frac{C_1 + 2 C_2}{\Delta_-(3-2\Delta_-)} \vert \f_- \vert^{\frac{\Delta}{\Delta_-}} \ell^{\Delta}  e^{\Delta \frac{u}{\ell}} \left(1+\mathcal{O}\left(\left(\mathcal{R}_{1}+2 \mathcal{R}_{2}\right)\left|\varphi_{-}\right|^{\frac{2}{\Delta_-}} e^{2 \frac{u}{\ell}}\right)\right) + \ldots ,$$

\be
A_{1-}=A_{1-}^{c}-\frac{u}{\ell}-\frac{\varphi_{-}^{2} \ell^{2\Delta_-}}{16} e^{2\Delta_-\frac{u}{\ell}}-\frac{\mathcal{R}_{1}\left|\varphi_{-}\right|^{\frac{2}{\Delta_-}} \ell^{2}}{196} e^{2 \frac{u}{\ell}}
\label{ga82}\ee
$$-\frac{\Delta C_{1}\left|\varphi_{-}\right|^{\frac{3}{\Delta_-}} \ell^{3}}{6(3-2\Delta_-)} e^{3 \frac{u}{\ell}}+\ldots,$$
\be
A_{2-}=A_{2-}^{c}-\frac{u}{\ell}-\frac{\varphi_{-}^{2} \ell^{2\Delta_-}}{16} e^{2\Delta_- \frac{u}{\ell}}-\frac{\mathcal{R}_{2}\left|\varphi_{-}\right|^{\frac{2}{\Delta_-}} \ell^{2}}{196} e^{2 \frac{u}{\ell}}+\ldots
\label{ga83}\ee
$$\quad-\frac{\Delta C_{2}\left|\varphi_{-}\right|^{\frac{3}{\Delta_-}} \ell^{3}}{6(3-2\Delta_-)} e^{3 \frac{u}{\ell}}+\ldots,$$
where we introduced the integration constants $\f_-,A^c_{1-}, A^c_{2-}$. We can repeat the same analysis for the other solution as well. For the $+$ branch we have instead

\be
\f - \f_0 = \f_+ \ell^\Delta e^{\Delta \frac{u}{\ell}} \left( 1 + \mathcal{O} \left( (\mathcal{R}_1 + 2 \mathcal{R}_2) \vert \f_+ \vert^{\frac{2}{\Delta}} e^{2 \frac{u}{\ell}} \right) \right) + \ldots,
\label{ga84}\ee

\be
A_{1+} = A_{1+}^c - \frac{u}{\ell} - \frac{\f_+^2 \ell^{2\Delta}}{16} e^{2\Delta \frac{u}{\ell}} - \frac{\mathcal{R}_1 \vert \f_+ \vert^{\frac{2}{\Delta}} \ell^2}{64} e^{2 \frac{u}{\ell}} + \frac{C_2 \ell^3 \vert \f_+ \vert^{\frac{3}{\Delta}}}{6} e^{3 \frac{u}{\ell}} + \ldots,
\label{ga85}\ee

\be
A_{2+} = A_{2+}^c - \frac{u}{\ell} - \frac{\f_+^2 \ell^{2\Delta}}{16} e^{2\Delta \frac{u}{\ell}} - \frac{\mathcal{R}_2 \vert \f_+ \vert^{\frac{2}{\Delta}} \ell^2}{64} e^{2 \frac{u}{\ell}} -  \frac{C_2 \ell^3 \vert \f_+ \vert^{\frac{3}{\Delta}}}{12} e^{3 \frac{u}{\ell}} + \ldots,
\label{ga86}\ee
where we introduced the integration constants $\f_+,A^c_{1+}, A^c_{2+}$.

A few comments are in order.

\begin{enumerate}
   \item[$\bullet$] Since our solutions for the functions $W_{1,2}, T_{1,2}, S$ are valid for $\f$ near $\f_0$, the above results are the leading terms in $\f(u)$ and $A_{1,2}$ for $u \to - \infty$.
   \item[$\bullet$] For both the $+$ and $-$ branch, the two scale factors $A_1,A_2$ have the same behavior, which is the one expected in the near-boundary region of an $AdS_{3+1}$ spacetime with length scale $\ell$.
   \item[$\bullet$] For the $-$ branch of the solution, we can identify $\f_-$ as the source for the scalar operator $\mathcal{O}$ in the boundary field theory associated with $\f$. The vacuum expectation value of $\mathcal{O}$ depends on the integration constants $C_1,C_2$ and is given by
   \be
   \left\langle \mathcal{O} \right\rangle = \frac{C_1 + 2 C_2}{\Delta_-} \vert \f_- \vert^{\frac{\Delta}{\Delta_-}}.
   \label{ga93}\ee
   \item[$\bullet$] For the $+$ branch of solutions, the bulk field $\f$ is also associated with a scalar operator $\mathcal{O}$. However, in this case the source is identically zero, yet there is a non zero vev given by
   \be
   \left\langle \mathcal{O} \right\rangle = (2\Delta-3) \f_+.
   \label{ga94}\ee
   \item[$\bullet$] For the $+$ and $-$ branches we can associate the integration constants $\mathcal{R}_1,\mathcal{R}_2$ with the UV curvature and the scaling parameter $a$. Specifically, by looking at the near boundary metric one finds
   \be
   R^{uv} = \left\{ \begin{array}{ll}
      \frac{\mathcal{R}_1+ 2 \mathcal{R}_2}{8} \vert \f_- \vert^{\frac{2}{\Delta_-}} \ , \ (-)  \text{branch} \\
      \frac{\mathcal{R}_1+ 2 \mathcal{R}_2}{8} \vert \f_+ \vert^{\frac{2}{\Delta}} \ , \ (+)  \text{branch}
   \end{array} \right .
   \label{ga95}\ee
   \be
   a^2 = \frac{4 \left( \mathcal{R}_1 + \mathcal{R}_2 \right)}{3\mathcal{R}_1 + 5 \mathcal{R}_2}.
   \label{ga96}\ee
   We observe that the integration constants $\mathcal{R}_1,\mathcal{R}_2$ are related to the curvature $R^{uv}$ of the manifold on which the UV QFT is defined. For future use, we can define the dimensionless curvature, $\mathcal{R}$, as
   \be
   \mathcal{R} = \frac{\mathcal{R}_1 + 2 \mathcal{R}_2}{8}.
   \label{ga96.b}\ee
   \item[$\bullet$] We can also calculate the expectation value of the stress-energy tensor that is dual to the metric (\ref{ga3}). It is given by
   \be
   \left\langle T_{ij} \right\rangle = 3(M \ell)^2 g_{ij}^{(3)},
   \label{ga97}\ee
   where the metric tensor $g_{ij}^{(3)}$ can be found by writing the metric (\ref{ga3}) in FG coordinates. In each case we have
   \be
   \left\langle T_{ij} \right\rangle_+ = \frac{C_2 \ell^3 \vert \f_+ \vert^{\frac{1}{\Delta}}}{2\pi G_N (3\mathcal{R}_1 + 5 \mathcal{R}_2)} \left( 2a^2 (d\psi + \cos\theta d\phi)^2 - d\Omega^2 \right),
   \label{ga98}\ee
   \be
   \left\langle T_{ij} \right\rangle_- = \frac{-\Delta \ell^3 \vert \f_- \vert^{\frac{1}{\Delta_-}}}{\pi G_N(3-2\Delta_-)(3\mathcal{R}_1 + 5\mathcal{R}_2)}  \left( a^2 C_1(d \psi+\cos \theta d \phi)^{2}+C_{2} d \Omega^{2}\right)
   \label{ga99}\ee

   Therefore for the minus solution $C_{1,2}$ control both the scalar as well as the stress tensor vev.
\end{enumerate}

Overall, the above findings imply the following:

\begin{enumerate}
   \item[$\bullet$] For the ($+$) and ($-$) branch we observe that maxima of the potential are associated with UV fixed points. The bulk space-time asymptotes to $AdS_{3+1}$ and reaching the maximum of the potential is equivalent to reaching the boundary. Moreover, moving away from the boundary is equivalent to a flow leaving the UV. For the
   ($-$) branch, flows corresponding to solutions are driven by the existence of a non-zero source $\f_-$ for the perturbing operator $\mathcal{O}$. However, for the ($+$) branch, the source is identically zero, and the flows are driven by a non-zero vev of $\mathcal{O}$.
\end{enumerate}

\subsection{Expansion near minima of the potential}\label{expansionnearminima}

In the following, we shall describe solutions for $T_{1,2},W_{1,2},S$ corresponding to flows that end at minima of the potential.

In the following, we present the results, while detailed calculations can be found in Appendix \ref{Extremaexpansions}. Overall, there is only one solution to the equations (\ref{ga51}) , (\ref{ga53}) - (\ref{ga55}) that has a 3d near-boundary metric, which we can identified as the $+$ branch that was studied in the context of maximally symmetric spheres in \cite{C}:

\be
S_+ =  \frac{\Delta}{\ell} x \left( 1 + \mathcal{O}(x) \right) + \frac{\Delta(\mathcal{R}_1 + 2\mathcal{R}_2)}{(2\Delta - 1)32\ell} x^{\frac{2}{\Delta}+1} \left(1 + \mathcal{O}(x) \right),
\label{ga100}\ee

\be
W_{1+} =  \frac{4}{\ell} + \frac{\Delta}{2\ell}x^2 + \mathcal{O}(x^3) + \frac{\mathcal{R}_1}{8\ell} x^{\frac{2}{\Delta}}  \left(1 + \mathcal{O}(x) \right) - \frac{2C_2}{\ell} x^{\frac{3}{\Delta}} \left(1 + \mathcal{O}(x) \right),
\label{ga101}\ee

\be
W_{2+} =  \frac{4}{\ell} + \frac{\Delta}{2\ell}x^2 + \mathcal{O}(x^3) + \frac{\mathcal{R}_2}{8\ell} x^{\frac{2}{\Delta}}  \left(1 + \mathcal{O}(x) \right) + \frac{C_2}{\ell} x^{\frac{3}{\Delta}} \left(1 + \mathcal{O}(x) \right),
\label{ga102}\ee

\be
T_{1+} = \frac{\mathcal{R}_1 + \mathcal{R}_2}{2\ell^2} x^{\frac{2}{\Delta}}  \left(1 + \mathcal{O}(x) \right),
\label{ga103}\ee

\be
T_{2+} = \frac{5\mathcal{R}_2 + 3\mathcal{R}_1}{8\ell^2} x^{\frac{2}{\Delta}}  \left(1 + \mathcal{O}(x) \right),
\label{ga104}\ee
where $C_1,C_2,\mathcal{R}_1,\mathcal{R}_2$ are integration constants, and $\Delta > 3$.

The above expressions describes three continuous families of solutions, whose structure is an analytic expansion in integer powers of $\f$, plus a series of non-analytic, subleading terms. As we explained in the previous subsection, initally all solutions had 4 integration constants, however, some where intentionally set to 0.

Given our results for the functions $W_1,W_2,S,T_1,T_2$, we are now in a position to solve for $\f(u)$ and $A_1,A_2$.
For the $+$ branch we solve (\ref{ga50}) and (\ref{ga46}),(\ref{ga47}) subject to (\ref{ga59}) - (\ref{ga63}) to obtain

\be
\f - \f_0 = \f_+ \ell^\Delta e^{\Delta \frac{u}{\ell}} \left( 1 + \mathcal{O} \left( (\mathcal{R}_1 + 2 \mathcal{R}_2) \vert \f_+ \vert^{\frac{2}{\Delta}} e^{2 \frac{u}{\ell}} \right) \right) + \ldots,
\label{ga117}\ee

\be
A_{1+} = A_{1+}^c - \frac{u}{\ell} - \frac{\f_+^2 \ell^{2\Delta}}{16} e^{2\Delta \frac{u}{\ell}} - \frac{\mathcal{R}_1 \vert \f_+ \vert^{\frac{2}{\Delta}} \ell^2}{64} e^{2 \frac{u}{\ell}} + \frac{C_2 \ell^3 \vert \f_+ \vert^{\frac{3}{\Delta}}}{6} e^{3 \frac{u}{\ell}} + \ldots,
\label{ga118}\ee

\be
A_{2+} = A_{2+}^c - \frac{u}{\ell} - \frac{\f_+^2 \ell^{2\Delta}}{16} e^{2\Delta \frac{u}{\ell}} - \frac{\mathcal{R}_2 \vert \f_+ \vert^{\frac{2}{\Delta}} \ell^2}{64} e^{2 \frac{u}{\ell}} -  \frac{C_2 \ell^3 \vert \f_+ \vert^{\frac{3}{\Delta}}}{12} e^{3 \frac{u}{\ell}} + \ldots,
\label{ga119}\ee
where we introduced the integration constants $\f_+,A^c_{1+},A^c_{2+}$.

A few comments are in order.

\begin{itemize}
   \item[$\bullet$] Recall that the expansions for the functions $S,W_{1,2},T_{1,2}$ are valid only for $\f$ in the vicinity of $\f_0$. It follows from equations (\ref{ga117}), (\ref{ga118}) , (\ref{ga119}) that small values of $(\f - \f_0)$ require $u \to - \infty$. This in turn implies $e^{2A_1}.e^{2A_2} \to \infty$ as we approach a minima of the potential. This is the behavior expected when approaching a UV fixed point.
   \item[$\bullet$] As we've mentioned in the previous subsection, in the boundary QFT the bulk field $\f$ shall be associated with an operator $\mathcal{O}$. Looking at expression (\ref{ga117}) , we see that the absense of a term of the form $\sim e^{\Delta_- \frac{u}{\ell}}$ implies that the source of this operator vanishes. Its vev, however, is non zero and is given by
   \be
   \left\langle \mathcal{O} \right\rangle_+ = (2\Delta - 3) \f_+.
   \label{ga126}\ee
   \item[$\bullet$] For the $+$ branch we can associate the integration constants $\mathcal{R}_1,\mathcal{R}_2$ with the UV curvature and the scaling parameter $a$. Specifically, by looking at the near boundary metric one finds
   \be
   R^{uv} = \frac{\mathcal{R}_1 + 2 \mathcal{R}_2}{8}\vert \f_+ \vert^{\frac{2}{\Delta}} \sp a^2 = \frac{4 \left( \mathcal{R}_1 + \mathcal{R}_2 \right)}{3\mathcal{R}_1 + 5 \mathcal{R}_2}.
   \label{ga127}\ee
\end{itemize}

\section{The geometry in the interior}\label{interiorgeo}

After having analysed the behavior close to the UV boundary, we now turn our attention to the geometry in the interior. Specifically, we are interested in the way the spacetime can "end' regularly, ie. where the scale factor shrinks to 0 but the bulk curvature invariants remain finite.  As is shown in Appendix \ref{Curvatureinvariants}, the only curvature invariant that we need to worry about is the Kretschmann scalar.

The analysis of the possible ways with which $S$ can vanish is presented in appendix \ref{extremalpoints}. As is shown there, when we look near the IR the asymptotic behavior of the functions $S,W_{1,2},T_{1,2}$ always takes the general form of an expansion in half-integer powers of $(\f-\f_0)$ (which we assume to be positive, for simplicity)\footnote{As was explained in the beginning of section \ref{potextrema}, we can expand the functions in terms of $(\f_0-\f)$ by inverting the sign of the functions $W_1,W_2$.}:

\be
S = \sqrt{\f - \f_0} \left( S_0 + S_1 \sqrt{\f - \f_0} + \mathcal{O}\left( \f - \f_0\right) \right),
\label{ga128}\ee
\be
W_1 = \frac{1}{\sqrt{\f-\f_0}} \left( W_{(1)0} + W_{(1)1} \sqrt{\f -\f_0} + \mathcal{O} \left(\f - \f_0 \right) \right),
\label{ga129}\ee
\be
W_2 = \frac{1}{\sqrt{\f-\f_0}} \left( W_{(2)0} + W_{(2)1} \sqrt{\f -\f_0} + \mathcal{O} \left(\f - \f_0 \right) \right),
\label{ga130}\ee
\be
T_1 = \frac{1}{\f - \f_0} \left( T_{(1)0} + T_{(1)1} \sqrt{\f - \f_0} + \mathcal{O} \left( \f - \f_0 \right) \right),
\label{ga131}\ee
\be
T_2= \frac{1}{\f - \f_0} \left( T_{(2)0} + T_{(2)1} \sqrt{\f - \f_0} + \mathcal{O} \left( \f - \f_0 \right) \right).
\label{ga132}\ee

On the other hand, we have assumed that the potential is an analytic function of $\f$ and can be expanded in terms of $(\f - \f_0)$ as
\be
V(\f ) = V_0 + V_1( \f - \f_0 ) + V_2 (\f -\f_0)^2 + \mathcal{O}\left( ( \f -\f_0)^3 \right).
\label{ga133}\ee

Depending on the values of the coefficients appearing in (\ref{ga128}) - (\ref{ga132}), the solution around $\f = \f_0$ can be of three possible types:
\begin{enumerate}
   \item Bolt-like IR end-points.
   \item Nut-like IR end-points.
   \item $\f$-Bounces.
\end{enumerate}
We proceed to discuss each one in turn.

\subsection{Bolt-like IR end-points}\label{boltir}

In this case, the factor $e^{2A_1}$ shrinks to 0 while the factor $e^{2A_2}$ tends to a constant value. Specifically, the first order formalism functions become to leading order

\be
S \approx S_0 \sqrt{\f - \f_0} + \ldots\ , \ W_1 \approx  \frac{-2S_0}{\sqrt{\f-\f_0}} + \ldots \ , \ W_2 \approx \frac{2V_0 - T_{(2)2}}{S_0} \sqrt{\f - \f_0} + \ldots,
\label{ga134}\ee
\be
T_1 \approx \frac{T_{(2)2}^2}{4V_1} (\f - \f_0) + \ldots \sp T_2 \approx T_{(2)2} + \ldots,
\label{ga135}\ee
where

\be
S_0^2 = V_1 \sp T_{(2)2} = \text{arbitrary}.
\label{ga136}\ee

We see that this solution, obtained assuming $\f > \f_0$, makes sense only for $V_1 > 0$, ie. $V'(\f_0) > 0$ (cfr. equation (\ref{ga133})).  It is easy to show that for $V'(\f_0) <0$ we have to reach $\f_0$ from below.

With expressions (\ref{ga134}) - (\ref{ga135}) for $S,W_{1,2}$ and $T_{1,2}$, one can integrate equations (\ref{ga46}) - (\ref{ga50}) order by order in $(\f-\f_0)$ to find the expressions for the scale factors $A_1,A_2$ and the scalar field $\f(u)$.  To lowest order one finds

\be
\f \approx \f_0 + \frac{V_1}{4} (u-u_0)^2 + \ldots \ , \ A_1(u) = \ln\left( \frac{u-u_0}{2L} \right) + \ldots \ ,\ A_2(u) = \ln\left( \frac{2}{L\sqrt{ T_{(2)2}}} \right) + \ldots,
\label{ga137}\ee
where $u_0$ is an integration constant. Near $u \to u_0$, the metric (\ref{ga3}) becomes

\be
ds^2 \approx du^2 + \frac{(u-u_0)^2}{4} \left( d\psi + \cos\theta d\phi \right)^2 + \frac{4}{T_{(2)2}} d\Omega^2.
\label{ga138}\ee

\subsection{Nut-like IR end-points} \label{nutir}

In this case  the leading coefficients $S_0,W_{(1)0},W_{(2)0},T_{(1)0},T_{(2)0}$ are all non-vanishing, and both factors $e^{2A_1},e^{2A_2}$ shrink to 0. However, this does not imply a singularity.

To leading order in $(\f-\f_0)$ we have:

\be
S \approx S_0 \sqrt{\f - \f_0} + \ldots \ , \ W_1 \approx \frac{-2S_0}{\sqrt{\f-\f_0}} + W_{(1)2} \sqrt{\f -\f_0} + \ldots,
\label{ga139}\ee
\be
W_2 \approx \frac{-2S_0}{\sqrt{\f - \f_0}} + \left( \frac{7V_0 - 6V_2}{12 S_0} - \frac{W_{(1)2}}{2} \right) \sqrt{\f - \f_0} + \ldots \ , \ T_1 \approx \frac{2V_1}{\f - \f_0} + \ldots,
\label{ga140}\ee
\be
T_2 \approx \frac{2V_1}{\f - \f_0},
\label{ga141}\ee
where
\be
S_0^2 = \frac{V_1}{2} \ , \ W_{(1)2} = \text{arbitrary}.
\label{ga142}\ee

We see that this solution, obtained assuming $\f > \f_0$, makes sense only for $V_1 > 0$, i.e. $V'(\f_0) > 0$ (cfr. equation (\ref{ga142})).  It is easy to show that for $V'(\f_0) <0$ we have to reach $\f_0$ from below.

With expressions (\ref{ga139}) - (\ref{ga141}) for $S,W_{1,2}$ and $T_{1,2}$, one can integrate equations (\ref{ga46}) - (\ref{ga50}) order by order in $(\f-\f_0)$ to find the expressions for the scale factors $A_1,A_2$ and the scalar field $\f(u)$.  To lowest order one finds

\be
\f \approx \f_0 + \frac{1}{8}V_1(u-u_0)^2 + \ldots,
\label{ga143}\ee

\be
A_1 = \ln \left( \frac{u-u_0}{2L} \right) + \ldots \ , \ A_2 = \ln \left( \frac{u-u_0}{2L} \right) + \ldots,
\label{ga144}\ee
where $u_0$ is an integration constant. Near $u \to u_0$, the metric (\ref{ga3}) becomes

\be
ds^2 \approx du^2 + \frac{(u-u_0)^2}{4} \left( \left(d\psi + \cos\theta d\phi\right)^2 + d\Omega^2 \right).
\label{ga145}\ee

\subsection{$\f$-Bounces} \label{bounce}

In this case the leading coefficients of $W_1,W_2,T_1,T_2$ are vanishing, while the leading coefficients of $S$ are non-vanishing. This implies that we have finite $W_{1,2},T_{1,2}$. To leading order in $(\f - \f_0)$ we have:

\be
S \approx S_0 \sqrt{\f - \f_0} + \ldots \ , \ W_1 = \frac{4T_{(2)2} - T_{(1)2} - 8 V_0 - W_{(2)1}^2}{2W_{(2)1}} + \ldots,
\label{ga146}\ee
\be
W_2 = W_{(2)1} + \ldots \ , \ T_1 = T_{(1)2} + \ldots \ , \ T_2 = T_{(2)2} + \ldots,
\label{ga147}\ee
where

\be
S_0^2 = 2 V_1 \ , \ T_{(1)2},T_{(2)2},W_{(2)1} = \text{arbitrary}.
\label{ga148}\ee

In this case, the scalar field has a turning point, but the scale factors have a non-vanishing derivative.  The solution (\ref{ga148}) describes two branches, with $S_0 = \pm \sqrt{2V_1}$, corresponding to the two branches of the superpotential equations: at a bounce, the two branches can be glued, giving rise to a regular geometry.

With expressions (\ref{ga146}) - (\ref{ga148}) for $S,W_{1,2}$ and $T_{1,2}$, one can integrate equations (\ref{ga46}) - (\ref{ga50}) order by order in $(\f-\f_0)$ to find the expressions for the scale factors $A_1,A_2$ and the scalar field $\f(u)$.  To lowest order one finds

\be
\f \approx \f_0 + \frac{1}{2} V_1 (u-u_0)^2 +\ldots,
\label{ga149}\ee
\be
A_1 \approx \frac{1}{2} \ln\left( \frac{4T_{(1)2}}{L^2 T_{(2)2}^2} \right) -\frac{4T_{(2)2} - T_{(1)2} - 8 V_0 - W_{(2)1}^2}{8W_{(2)1}} (u-u_0 ) + \ldots,
\label{ga150}\ee
\be
A_2 \approx \frac{1}{2} \ln\left( \frac{4}{L^2 T_{(2)2}} \right) - \frac{W_{(2)1}}{4} (u-u_0) + \ldots,
\label{ga151}\ee
with $u_0$ being an integration constant. Near $u \to u_0$, the metric (\ref{ga3}) becomes

\be
ds^2 \approx du^2 + \frac{4}{T_{(2)2}} \left( \frac{T_{(1)2}}{T_{(2)2}} \left( d\psi + \cos\theta d\phi\right)^2 +d\Omega^2 \right).
\label{ga152}\ee

\section{Complete RG flows} \label{numerics}

In this section, we shall display solutions corresponding to full RG flows for a specific potential. The flows originate from a UV fixed point at a maximum of the potential and  end at an IR point at $\f=\f_0$, which does not  coincide with a minimum  of the potential.
From experience with sphere-sliced solutions and general qualitative considerations, \cite{C}, as $\f_0$ approaches the UV fixed point (maximum) the overall boundary curvature  becomes larger and larger.
When $\f_0$ approaches the nearby minimum of the potential, the boundary curvature vanishes.

We shall consider the following potential:

\be
V(\f) = - \frac{6}{\ell^2} - \frac{\Delta \Delta_-}{2\ell^2} (\f -1)^2 - \frac{\Delta \Delta_-}{2\ell^2} (\f-1)^3 - \frac{\Delta \Delta_-}{8\ell^2}(\f-1)^4,
\label{ga153}\ee
where $\Delta_- = 3-\Delta$, and $\Delta$ is a dimensionless parameter corresponding to the scaling dimension of the dual operator to the scalar $\f$ at the UV fixed point, which is located at one of the  maxima. The potential has one minimum at $\f = 0$, and two maxima at $\f = \pm 1$. $\ell$ is the AdS curvature scale at the two isomorphic maxima. Due to the symmetry of the potential around $\f = 0$, it shall be sufficient to confine ourselves to RG flows starting from the maximum $\f_{max} = 1$ (UV fixed point), and ending at a IR endpoint $\f_0$, which takes values inside the interval $[0,1]$. Our results can be easily then extended to flows starting from $\f = -1$ and ending somewhere in the interval $[-1,0]$. A graph of the potential can be seen in Figure \ref{fig2}. In the following numerical calculations , we shall use the typical value $\Delta = 1.8$ and we also set the scale to $\ell = 1$. We have checked that the qualitative nature of our results is not strongly dependent on the (generic) value of $\Delta$.

\FIGURE{
   \centering
   \includegraphics[scale = 0.8]{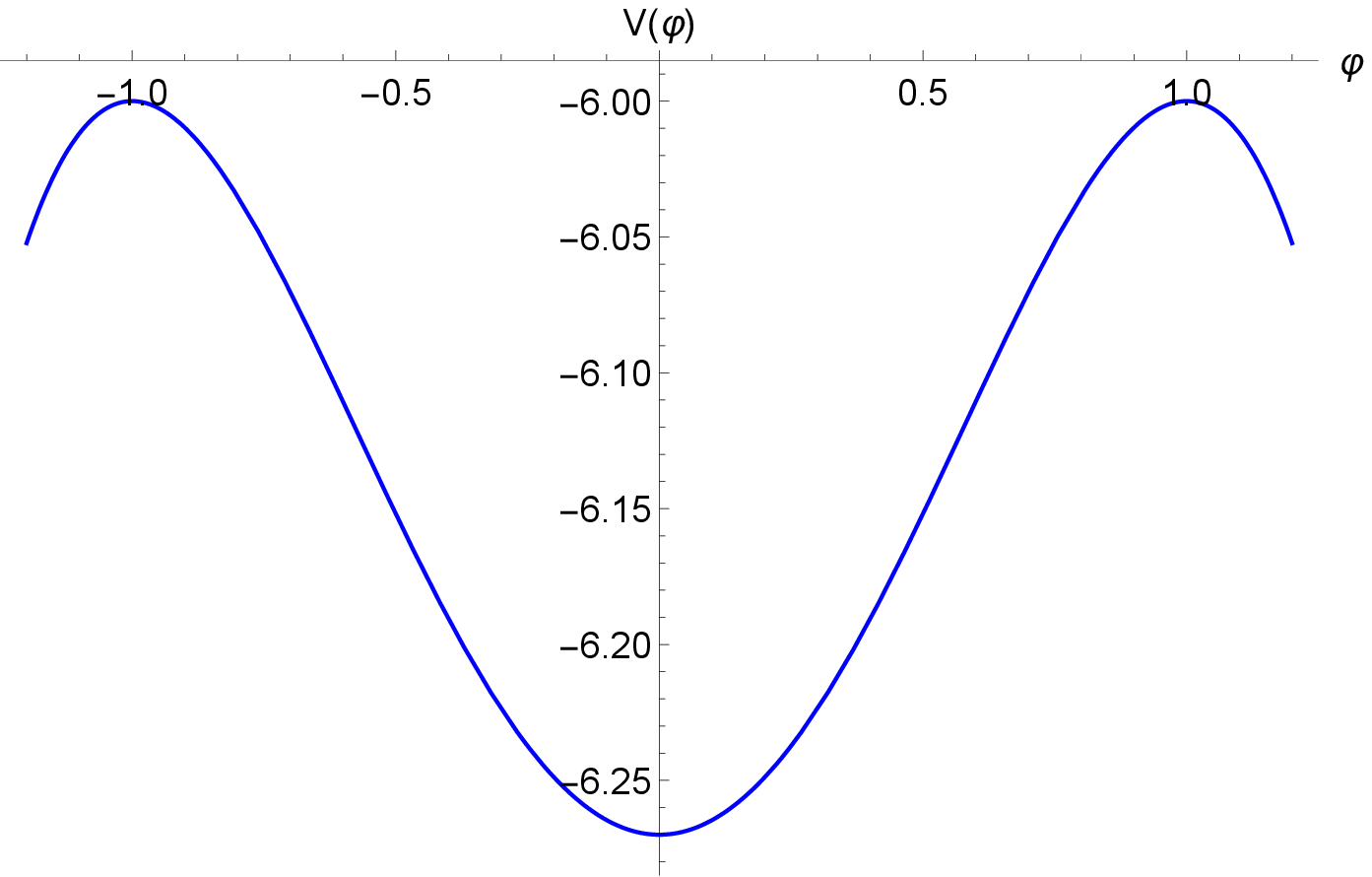}
   \caption{ Plot of the scalar potential with parameters $\Delta = 1.8$, $\ell = 1$.}
   \label{fig2}
}

We shall numerically solve equations (\ref{num1})-(\ref{num3}) and find regular solutions by applying the appropriate IR regulaity conditions we derived.
There are two ways with which a flow can regularly end at the IR, which are both described in section \ref{interiorgeo}. These are the Nut IR endpoint (section \ref{nutir}), which amounts to a shrinking of the three dimensional sphere, and the Bolt IR endpoint (section \ref{boltir}), which amounts to a shrinking of the one dimensional circle. Therefore we shall always obtain two branches of solutions.

\subsection{The case of the Nut-IR endpoint}

We start with the case of the Nut IR endpoint. For convenience, we shall rewrite here the Ir regularity  conditions associated with this case, (\ref{ga139})-(\ref{ga142}):
that we reproduce here for convenience,
\be
S \approx S_0 \sqrt{\f - \f_0} + \ldots \ , \ W_1 \approx \frac{-2S_0}{\sqrt{\f-\f_0}} + W_{(1)2} \sqrt{\f -\f_0} + \ldots,
\ee
\be
W_2 \approx \frac{-2S_0}{\sqrt{\f - \f_0}} + \left( \frac{7V_0 - 6V_2}{12 S_0} - \frac{W_{(1)2}}{2} \right) \sqrt{\f - \f_0} + \ldots \ , \ T_1 \approx \frac{2V_1}{\f - \f_0} + \ldots,
\ee
\be
T_2 \approx \frac{2V_1}{\f - \f_0},
\ee
where
\be
S_0^2 = \frac{V_1}{2} \ , \ W_{(1)2} = \text{arbitrary}.
\ee

We are interested in the behavior of the observables with respect to the free parameter $W_{(1)2}$ and the position of the scalar enpoint $\f_0$. They  are the only two parameters on which a regular solution depends.
For convenience, we  make the following definitions:

\begin{equation}
\tilde{W} \equiv \frac{7V_0-6V_2}{18S_0} = \frac{-0.148492 \varphi _0^4+1.82434 \varphi _0^2-3.95744}{\sqrt{1.08 \varphi _0-1.08 \varphi _0^3}},
\label{n1}\end{equation}
\begin{equation}
\alpha \equiv  {\tilde{W}\over W_{(1)2} }.
\label{n1a}\end{equation}

We defined two constants: $\tilde{W}$, which is negative, and $\alpha$.
Note that $\alpha$ defined above bears no relation with the parameter $\alpha$ defined in (\ref{ga43}) and used in subsection \ref{alpha}.
 For $\alpha = 1$ one can easily deduce from equations (\ref{ga146}),(\ref{ga147}) that we have $W_{(1)2} = W_{(2)2}$, i.e. there is no squashing and the slices are $S^3$.
Therefore, instead of $\f_0$ and $W_{1(2)}$ as parameters of the solution we shall use equivalently, $\f_0$ and $\alpha$.

In Figure \ref{fig3} we show the behavior of the superpotentials $W_{1}$ as well as $W_2$,  for various values of the constant $\alpha$ in Nut end-point flows.
In Figure \ref{fig3a}, we have plotted the superpotential $S$ as a function of  $\f_0$ for different choices of $\alpha$ as well.

\FIGURE[ht]{
   \includegraphics[scale = 0.5]{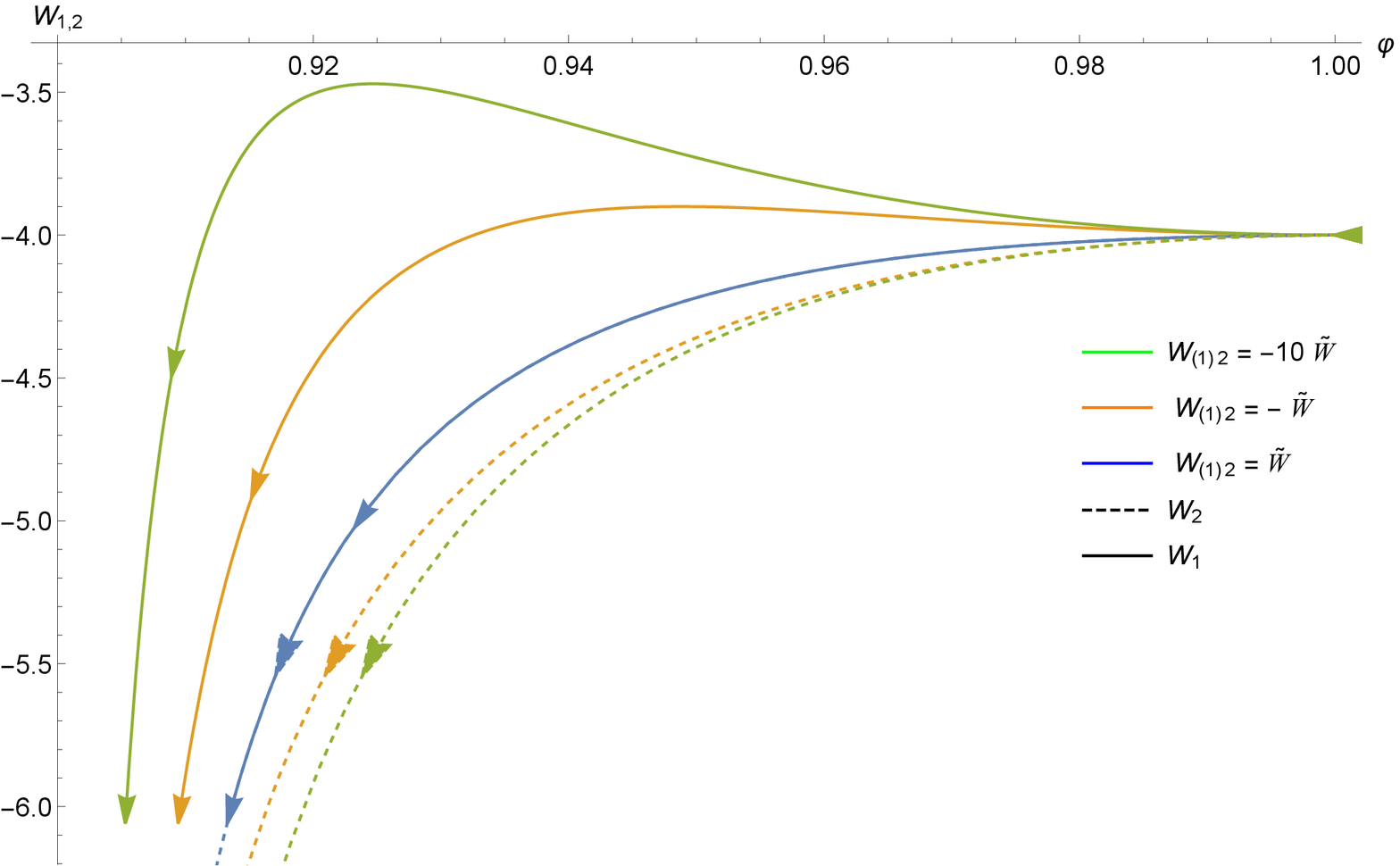}
   \caption{ Plot of the superpotentials  $W_1,W_2$ as functions of  $\f$ for different choices of $\alpha$, defined in (\protect{\ref{n1}}), in the case of a Nut IR end-point. The IR endpoint for this plot has been chosen to be at $\f_0 = 0.9$. The  direction of the flow is indicated in the figure: it starts from the UV fixed point and ends in the IR.
    The different lines correspond to different choices of the integration constant $W_{(1)2}$. The constant $\tilde W$ is defined in (\protect{\ref{n1a}}). Note that when we choose $W_{(1)2} = \tilde{W}$, the two functions coincide and the slice geometry in this case is that of a round $S^3$.}
   \label{fig3}
}
\FIGURE[ht]{
   \includegraphics[scale = 0.5]{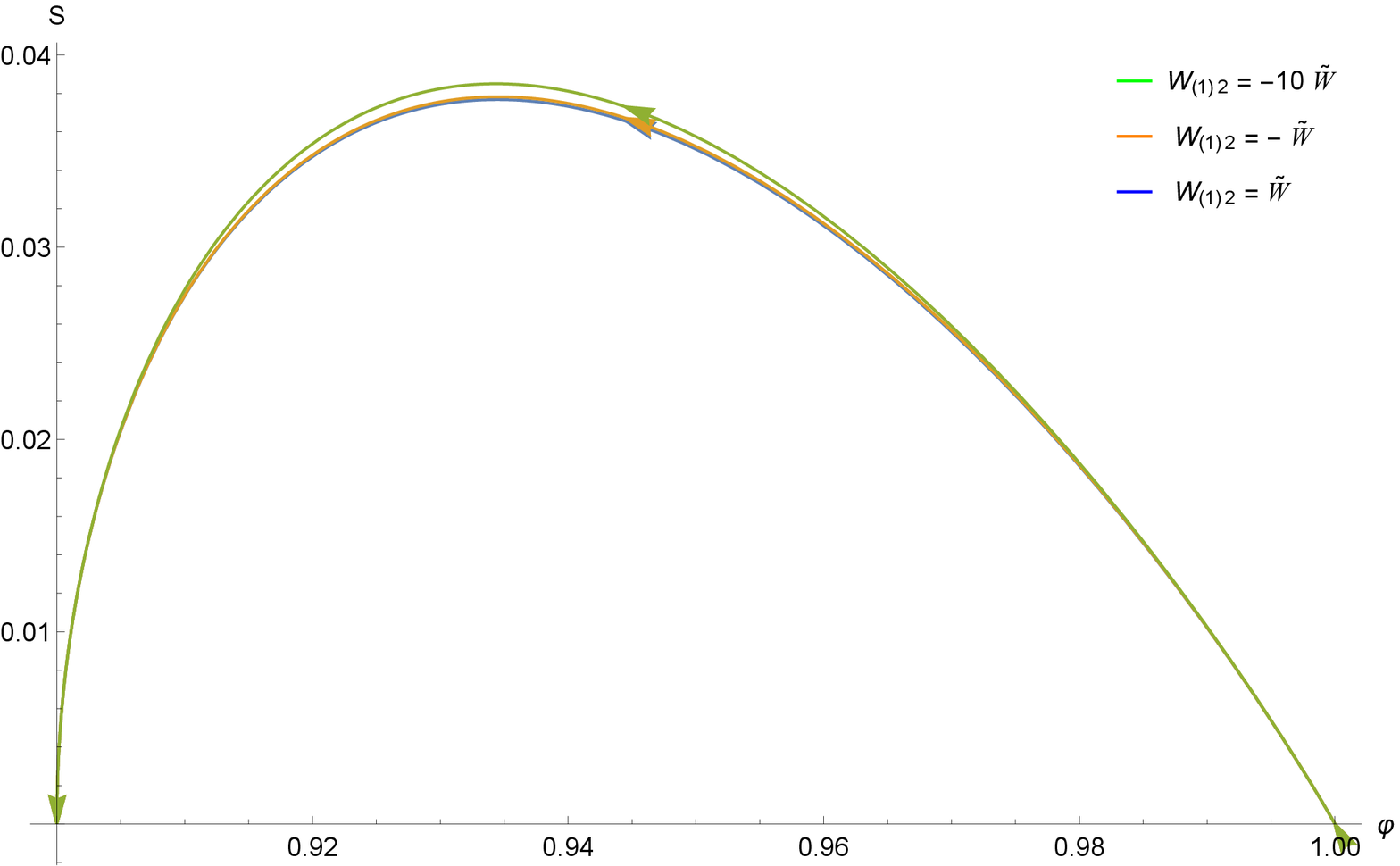}
   \caption{Plot of the superpotential  $S$ as a function of  $\f$ for different choices of $\alpha$, defined in (\protect{\ref{n1}}), in the case of a Nut IR end-point. The IR endpoint for this plot has been chosen to be at $\f_0 = 0.9$. The  direction of the flow is indicated in the figure: it starts from the UV fixed point and ends in the IR.
    The different lines correspond to different choices of the integration constant $W_{(1)2}$. The constant $\tilde W$ is defined in (\protect{\ref{n1a}}).  Note that the curves corresponding to $W_{(1)2} = \tilde{W}$ and $W_{(1)2} = -\tilde{W}$ almost coincide.}
   \label{fig3a}
}

Our results are as follows:

\begin{enumerate}
   \item[$\bullet$] The main result that is confirmed is that for every value of $\f_0$ between $\f_{min} = 0$ and $\f_{max} = 1$, there exists a unique regular solution to the superpotential equations (\ref{num1})-(\ref{num2}) corresponding to an RG flow that starts from the UV fixed point at $\f = 1$ and ends at an IR endpoint $\f_0$, located between the minimum of the potential and the maximum at $\f = 1$, with a Nut end-point.

        We note, however, that no solution exists for $\f_0 = 0$ exactly. As was discussed in \cite{C}, the solution with $\f_0 = 0$ corresponds to $\mathcal{R} = 0$, i.e. we must have a flat slice or an infinite relevant coupling.

   \item[$\bullet$] We note in Figure \ref{fig3} that as we approach the IR endpoint, the superpotentials $W_1,W_2$ diverge as $(\f - \f_0)^{-1/2}$, which is expected for an RG flow with positive curvature (for a more detailed discussion, see \cite{C}). This divergence does not imply a singularity in the bulk geometry.

   \item[$\bullet$] In the vicinity of the UV fixed points the solutions are collectively denoted by $W_{1-}(\f),W_{2-}(\f)$ in section \ref{expansionnearmaxima}. These solutions depend on four continuous parameters: $\mathcal{R}_1,\mathcal{R}_2,C_1,C_2$. In turn, these correspond to the following observables: the dimensionless curvature, $\mathcal{R}$, the vev of the dual operator, $\left\langle \mathcal{O} \right\rangle$, the vev of the energy-stress tensor, and the squashing parameter $a^2$. Picking a regular solution with the appropriate IR behavior for a RG flow fixes two  of these parameters (the vevs as a function of the couplings, $\mathcal{R}$ and $a$). The remaining freedom ($\mathcal{R},a$), is then related to the choice of IR endpoint $\f_0$ and to the choice of a value for the free parameter $W_{(1)2}$, or equivalently $\alpha$.

   \newpage
   \item[$\bullet$] In Figure \ref{fig4} we have plotted the dimensionless curvature, $\mathcal{R}$, as a function of  the two free parameters, $\alpha$ and $\f_0$. We observe that $\mathcal{R}$ increases as we increase $\f_0$ and as we increase $\alpha$ as well. The former was expected, since it was found in \cite{C} that the dimensionless curvature is increasing as the end-point $\f_0$ approaches the UV fixed point (here at $\f=1$). For completeness, we also provide slices of $\mathcal{R}$ for specific values of $\f_0$ and $\alpha$ in  the two figures, \ref{fig5} and \ref{fig5a}.
\end{enumerate}
\FIGURE[ht]{
   \includegraphics[scale = 0.59]{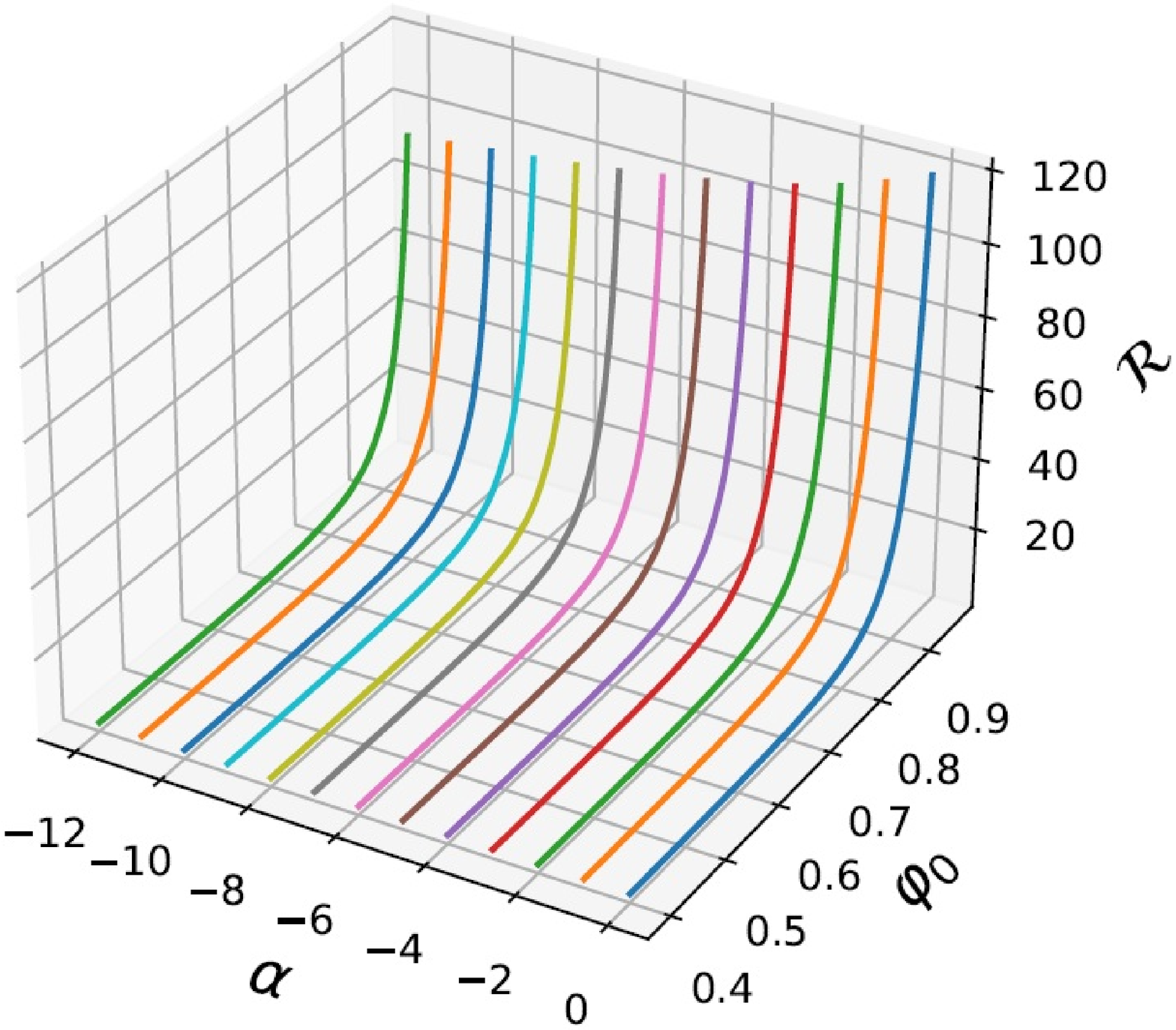}
   \caption{The dimensionless curvature $\mathcal{R}$ as a function of  the two parameters of the solution, $\f_0$ and $\alpha$, for Nut end-point solutions.}
   \label{fig4}
}
\FIGURE[ht]{
      \centering
      \includegraphics[width = 0.85\textwidth]{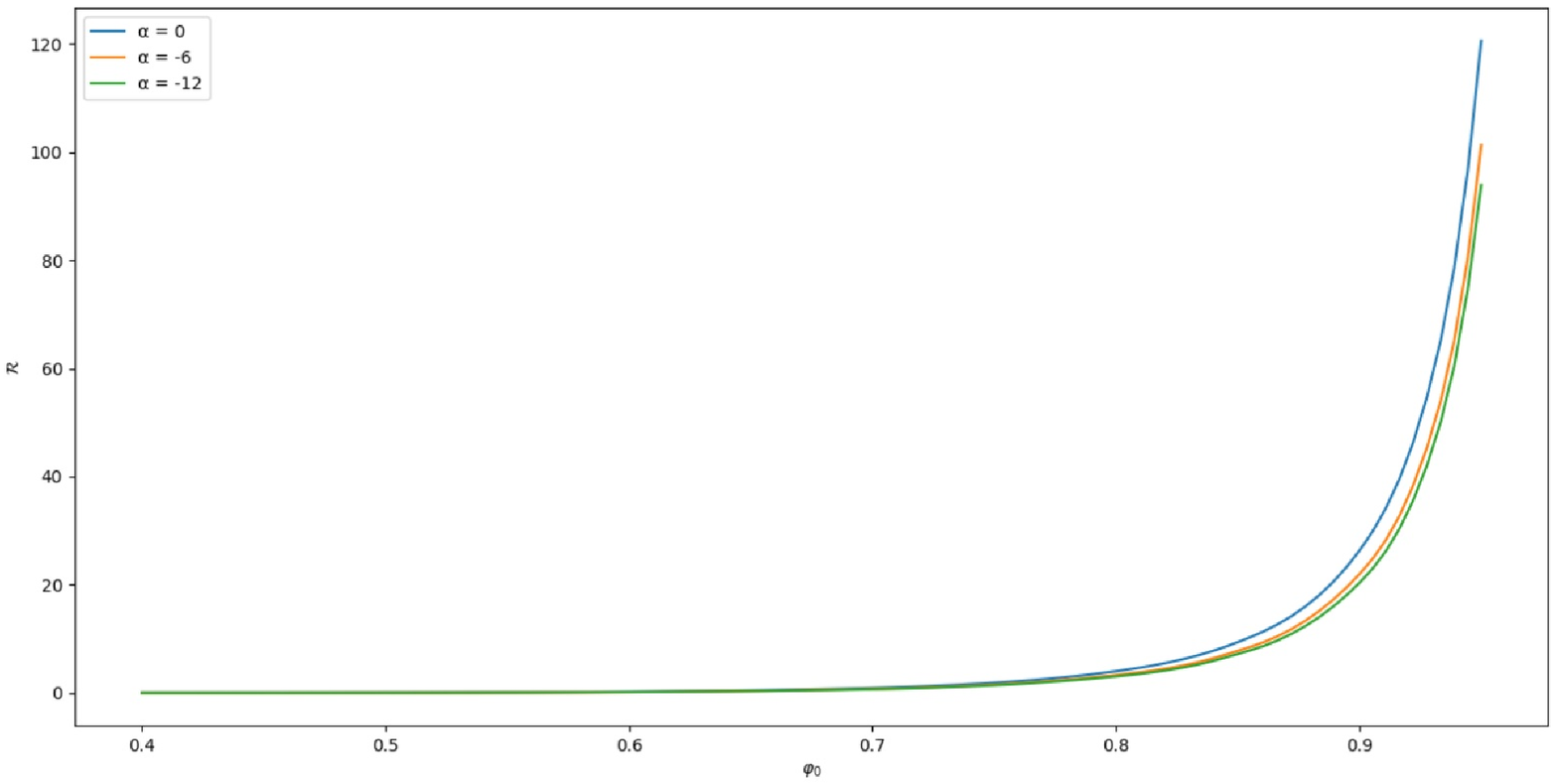}
      \caption{ The dimensionless curvature $\mathcal{R}$ as a function of $\f_0$ for fixed  values of $\alpha$, for Nut end-point solutions. }
      \label{fig5}
}
\FIGURE[ht]{
      \centering
      \includegraphics[width = 0.85\textwidth]{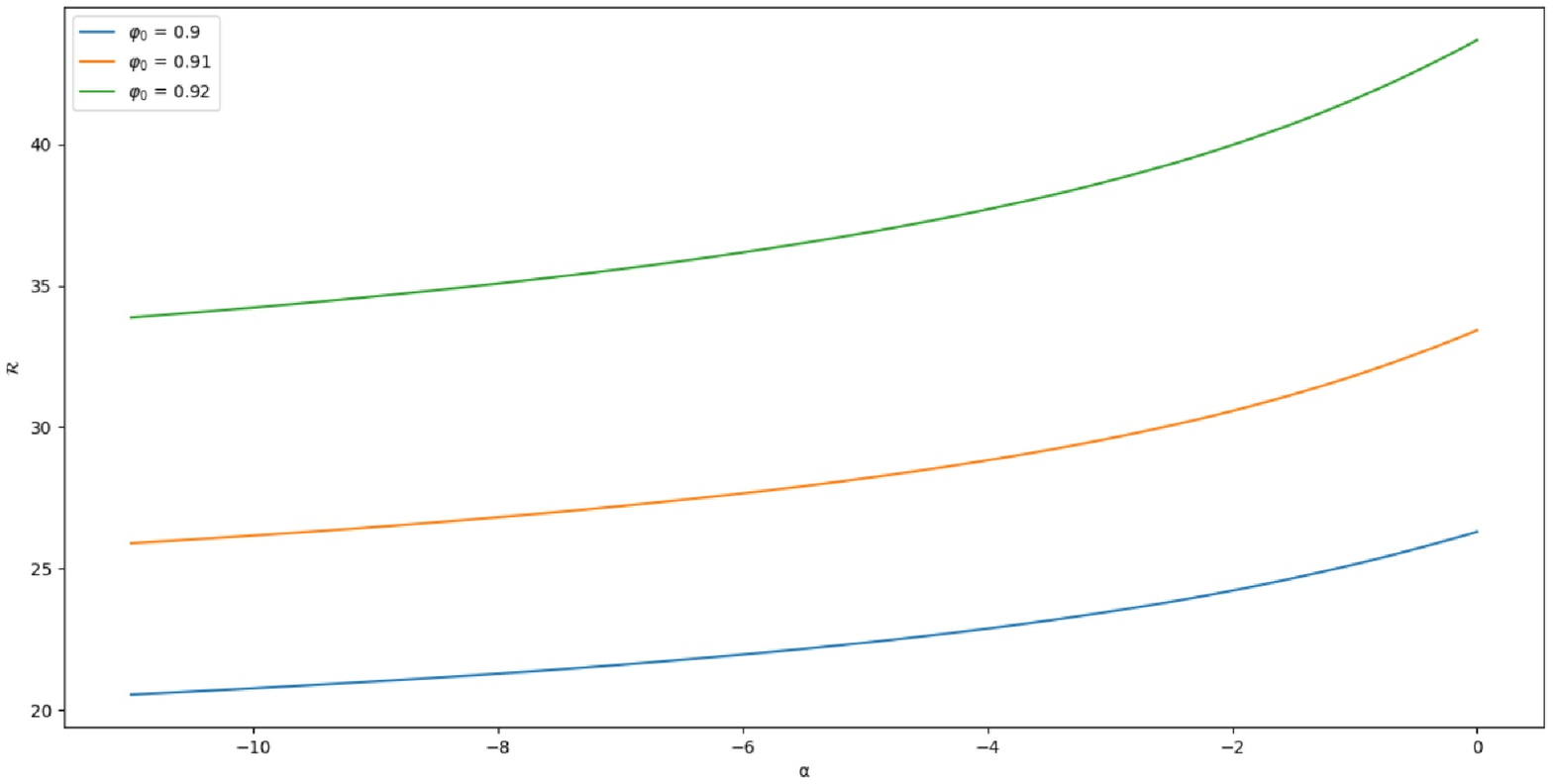}
      \caption{The dimensionless curvature $\mathcal{R}$ as a function of $\alpha$ for fixed  values of $\f_0$, for Nut end-point solutions.}
      \label{fig5a}
}

%\newpage

\begin{enumerate}
   \item[$\bullet$] Continuing, in Figure \ref{fig7} we have plotted the squashing parameter $a^2$ as a function of the two free parameters, $\alpha$ and $\f_0$. We observe that $a^2$ increases as we increase $\alpha$, and tends to 0 as $\alpha \to -\infty$. This is illustrated in the right  figure of \ref{fig8}, where $a^2$ is plotted as a function of  $\alpha$ for certain values of $\f_0$. We note that independently of the value of $\f_0$, $\alpha = -\infty$ corresponds to a vanishing $a^2$, and $\alpha = -1$ corresponds to $a^2 = 1$, i.e no squashing.
\end{enumerate}
\FIGURE[h]{
   \includegraphics[scale = 0.59]{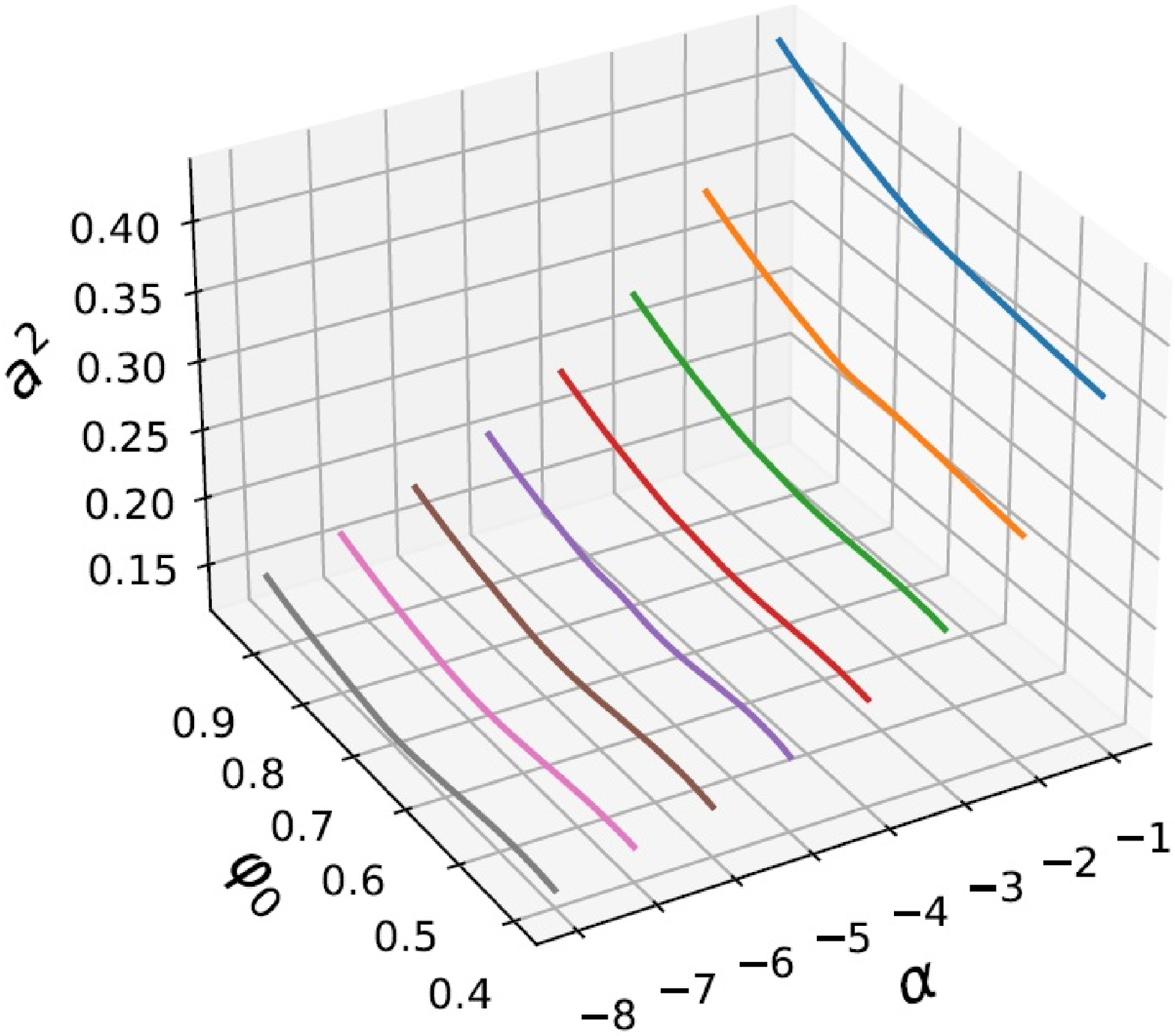}
   \caption{The squashing parameter $a^2$ as a function of the two parameters, $\f_0$ and $\alpha$, for Nut end-point solutions.}
   \label{fig7}
}
\FIGURE[h]{
   \begin{minipage}{.5\textwidth}
      \includegraphics[scale = 0.34]{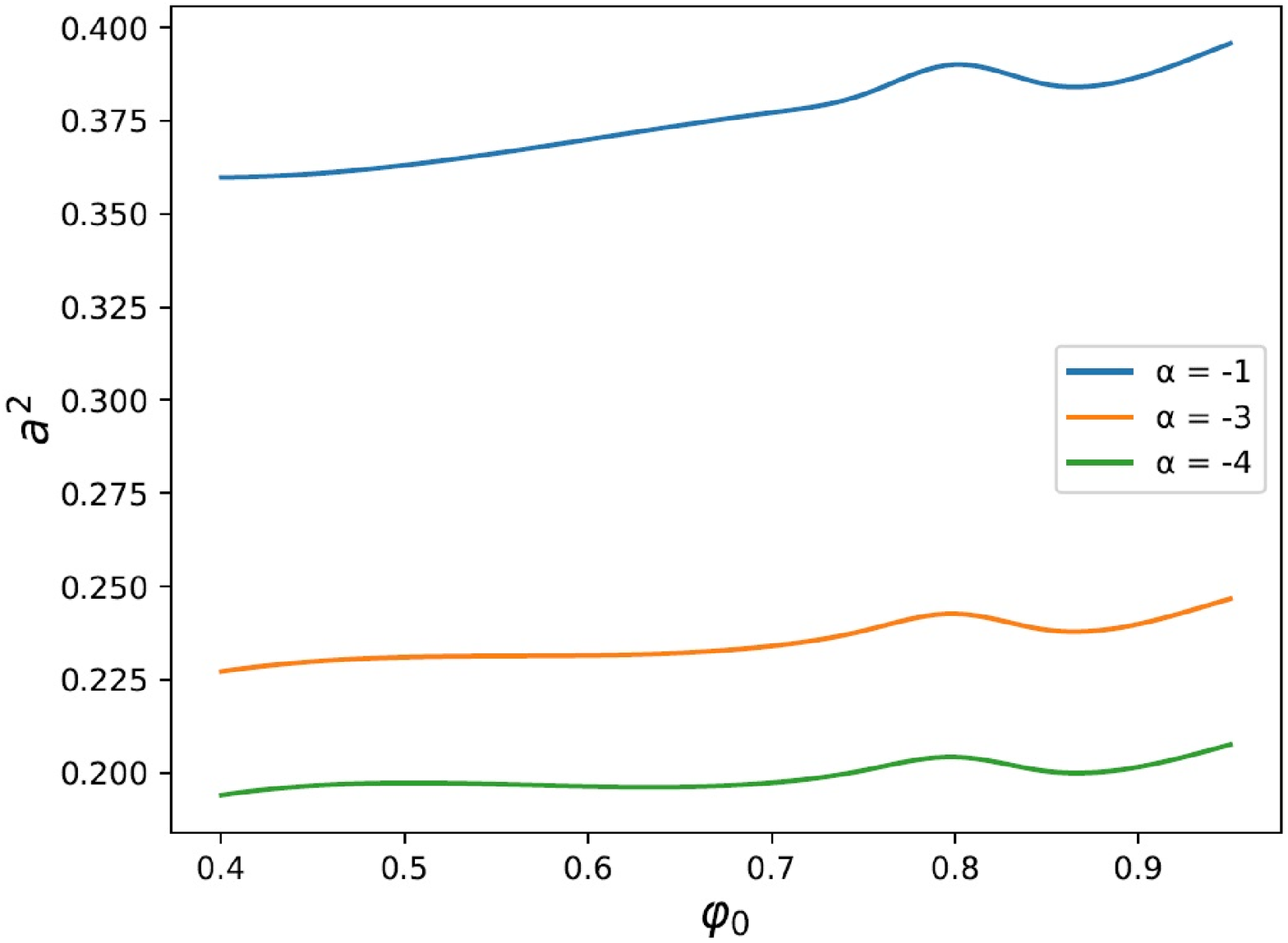}
      \caption{Left: The squashing parameter $a^2$ as a function of $\f_0$ for fixed  values of $\alpha$ for Nut end-point solutions. Right: The squashing parameter $a^2$ as a function of $\alpha$ for fixed values of $\f_0$, for Nut end-point solutions.}
      \label{fig8}
   \end{minipage}%
   \begin{minipage}{.5\textwidth}
\includegraphics[scale = 0.35]{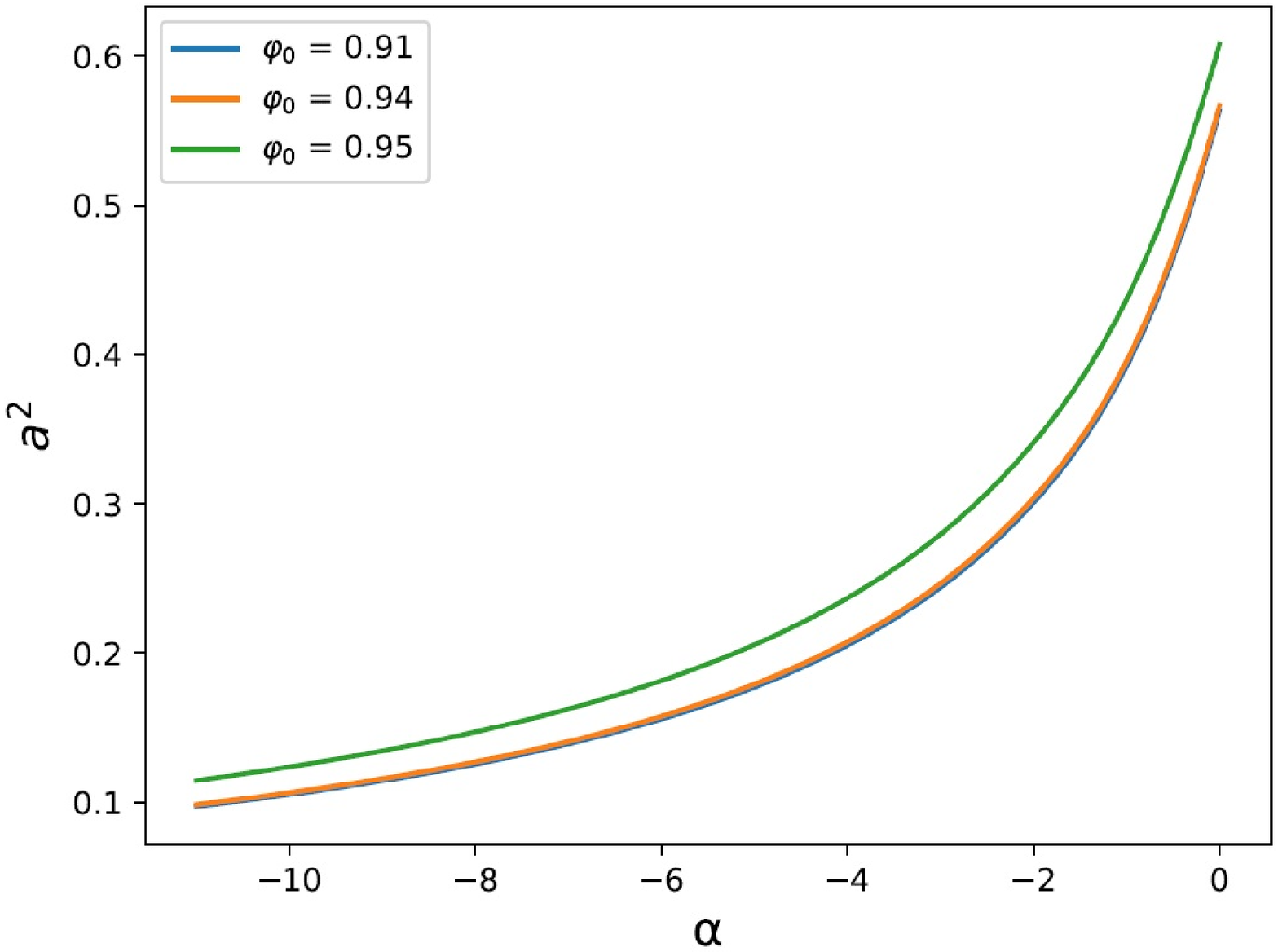}
   \end{minipage}
}
We now turn to the parameters that control the vevs, $C_{1,2}$. The relationship of the stress tensor vev as well as the scalar vev to $C_{1,2}$ is  given in (\ref{ga93}), (\ref{ga98}), (\ref{ga99}).
\newpage

\begin{enumerate}
   \item[$\bullet$] In Figure \ref{fig9} we have plotted the integration constant $C_2$ as a function of  $\f_0$ and $\alpha$, and in figures \ref{fig10}, we provide certain slices of this behavior for specific values of $\f_0$, $\alpha$. We observe that $C_2$ decreases monotonically with respect to $\f_0$ for a given value of $\alpha$. Specifically, for fixed $\alpha$, the form of $\vert C_2 \vert$ is similar to that of $\mathcal{R}$.
\end{enumerate}

\FIGURE[h]{
   \includegraphics[scale = 0.59]{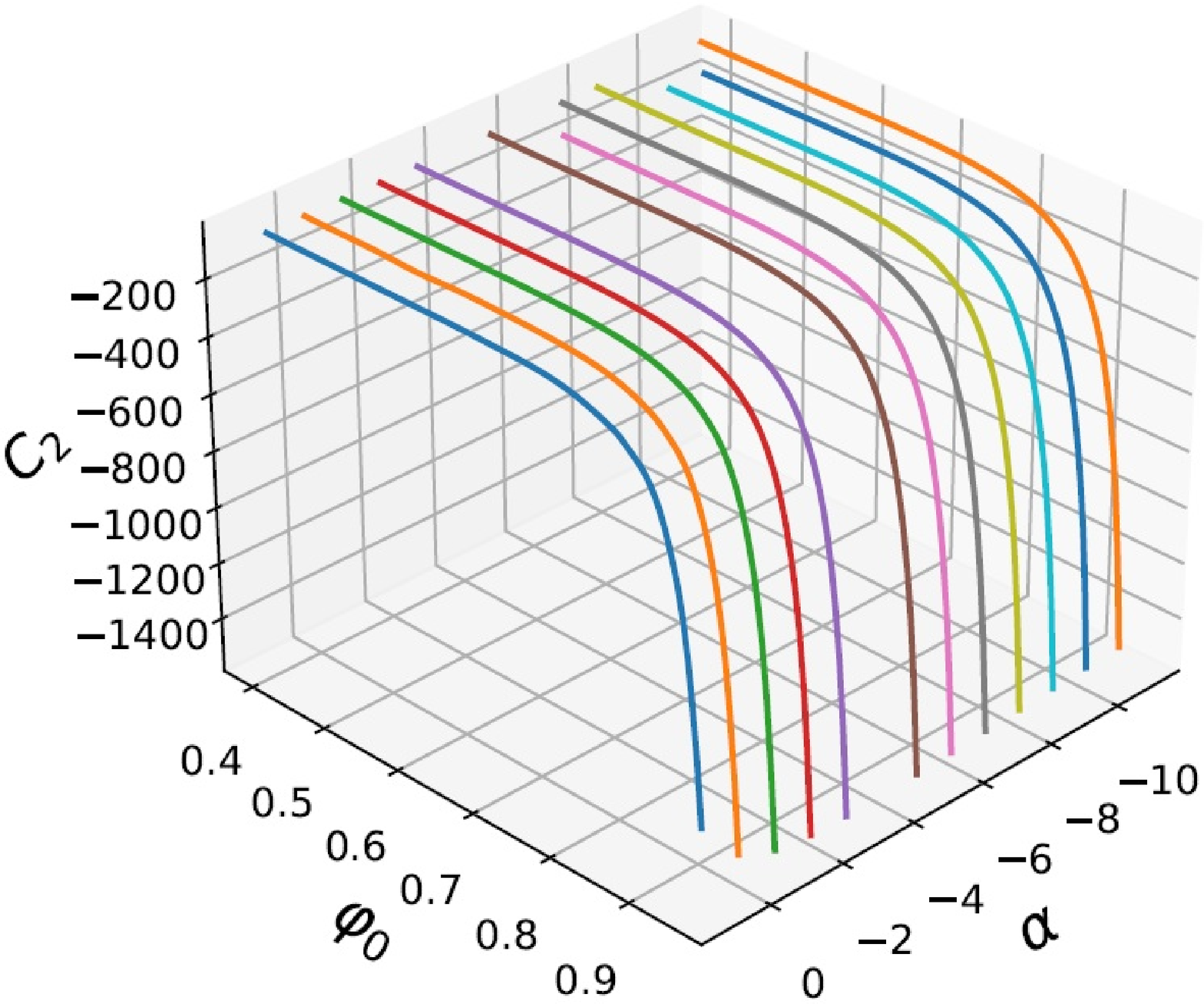}
   \caption{The integration constant $C_2$ as a function of  the two parameters, $\f_0$ and $\alpha$, for Nut end-point solutions.}
   \label{fig9}
}
\FIGURE[h]{
   \begin{minipage}{.5\textwidth}
      %\centering
      \includegraphics[scale = 0.34]{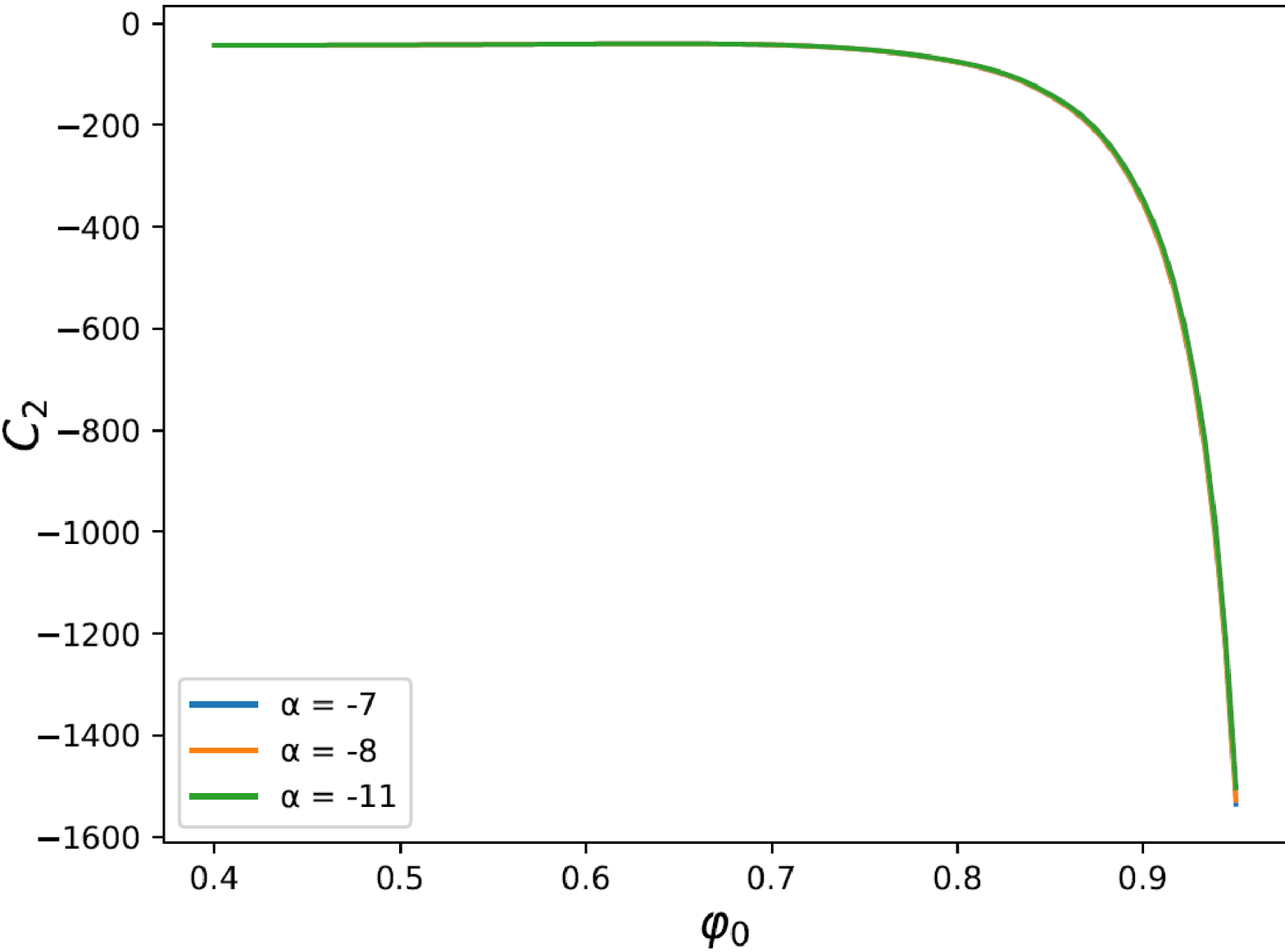}
      \caption{Left: The integration constant $C_2$ as a function of $\f_0$ for fixed values of $\alpha$ for Nut end-point solutions. Right:  The integration constant  $C_2$ as a function of  $\alpha$ for fixed values of $\f_0$, for Nut end-point solutions.}
      \label{fig10}
   \end{minipage}%
   \begin{minipage}{.5\textwidth}
      %\centering
\includegraphics[scale = 0.35]{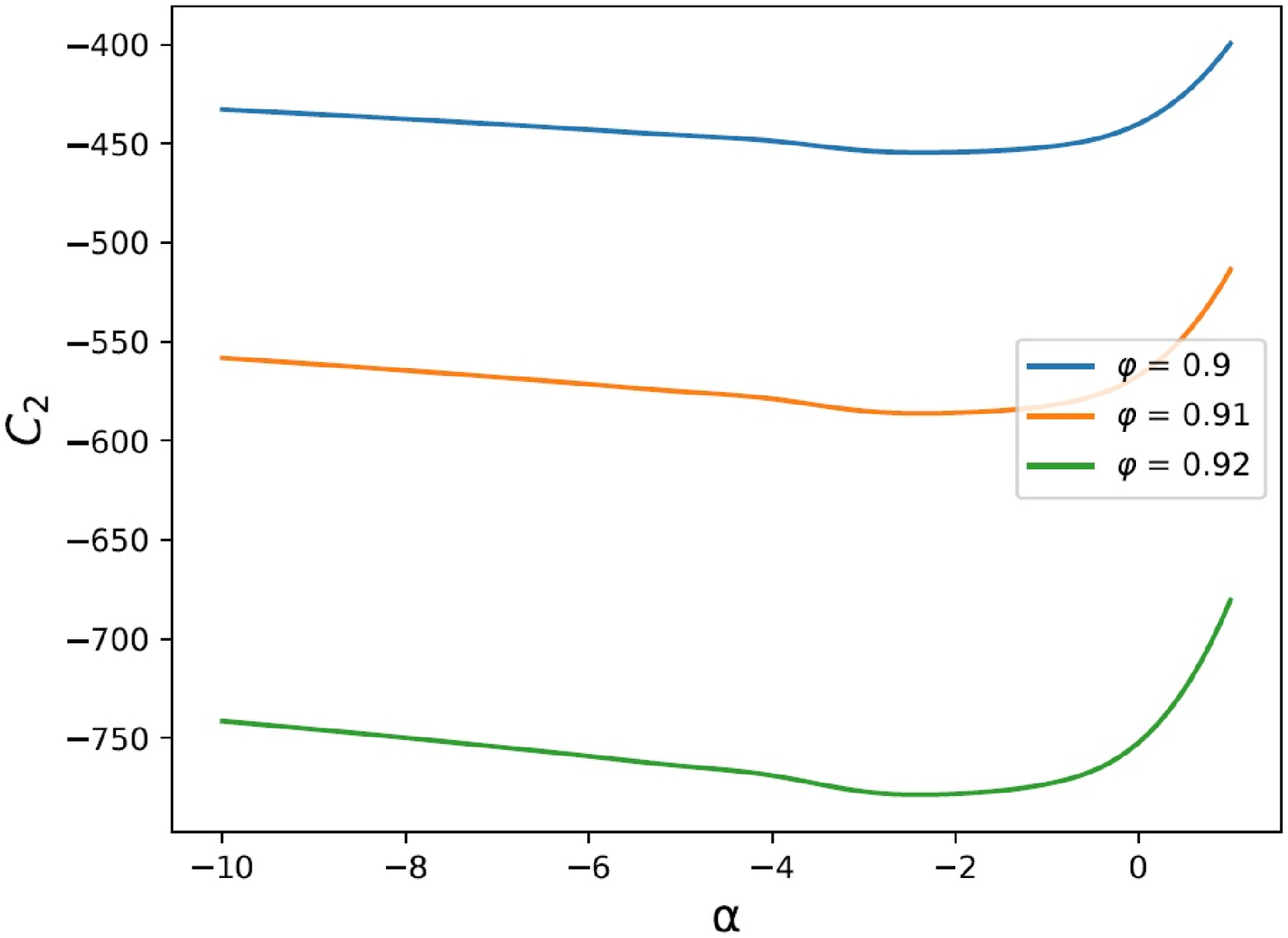}
   \end{minipage}
}

\newpage

\begin{enumerate}
   \item[$\bullet$] We also show $C_1$ as a function of  the two parameters in Figure \ref{fig11}, along with certain slices for specific values of $\f_0$, $\alpha$ in the two figures  \ref{fig12}. We observe that for fixed values of $\alpha$, the function $C_1$ has a similar behavior to that of $C_2$, with the core difference being a "small" deformation in the region $0.6 < \f_0 < 0.8$. This deformation can be attributed to the squashing parameter, which for $\alpha \neq 1$ is non-trivial.
\end{enumerate}

\FIGURE[h]{
   \includegraphics[scale = 0.59]{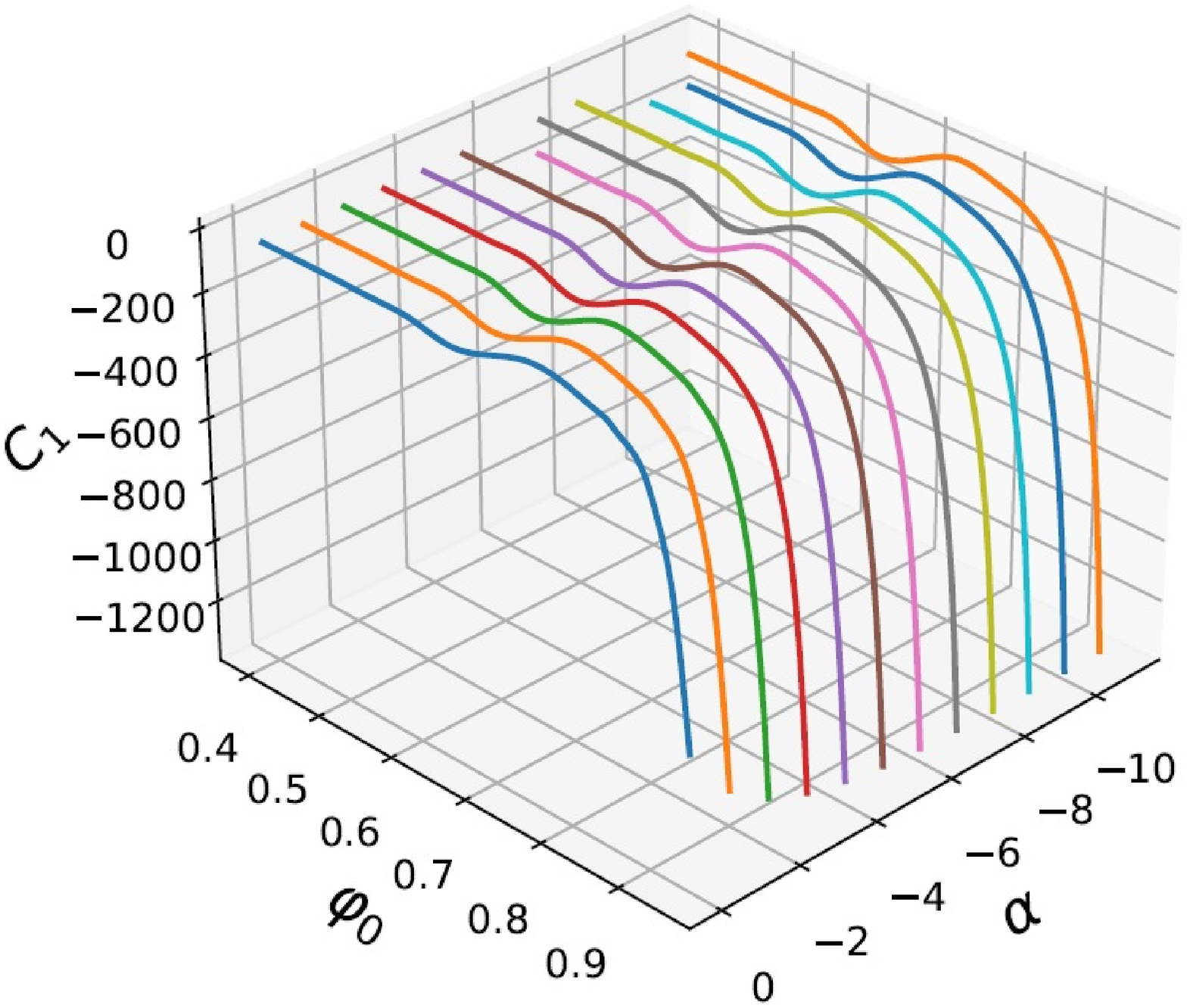}
   \caption{The integration constant $C_1$ as a function of  the two parameters, $\f_0$ and $\alpha$, for Nut end-point solutions.}
   \label{fig11}
}
\FIGURE[h]{
   \begin{minipage}{.5\textwidth}
      %\centering
      \includegraphics[scale = 0.34]{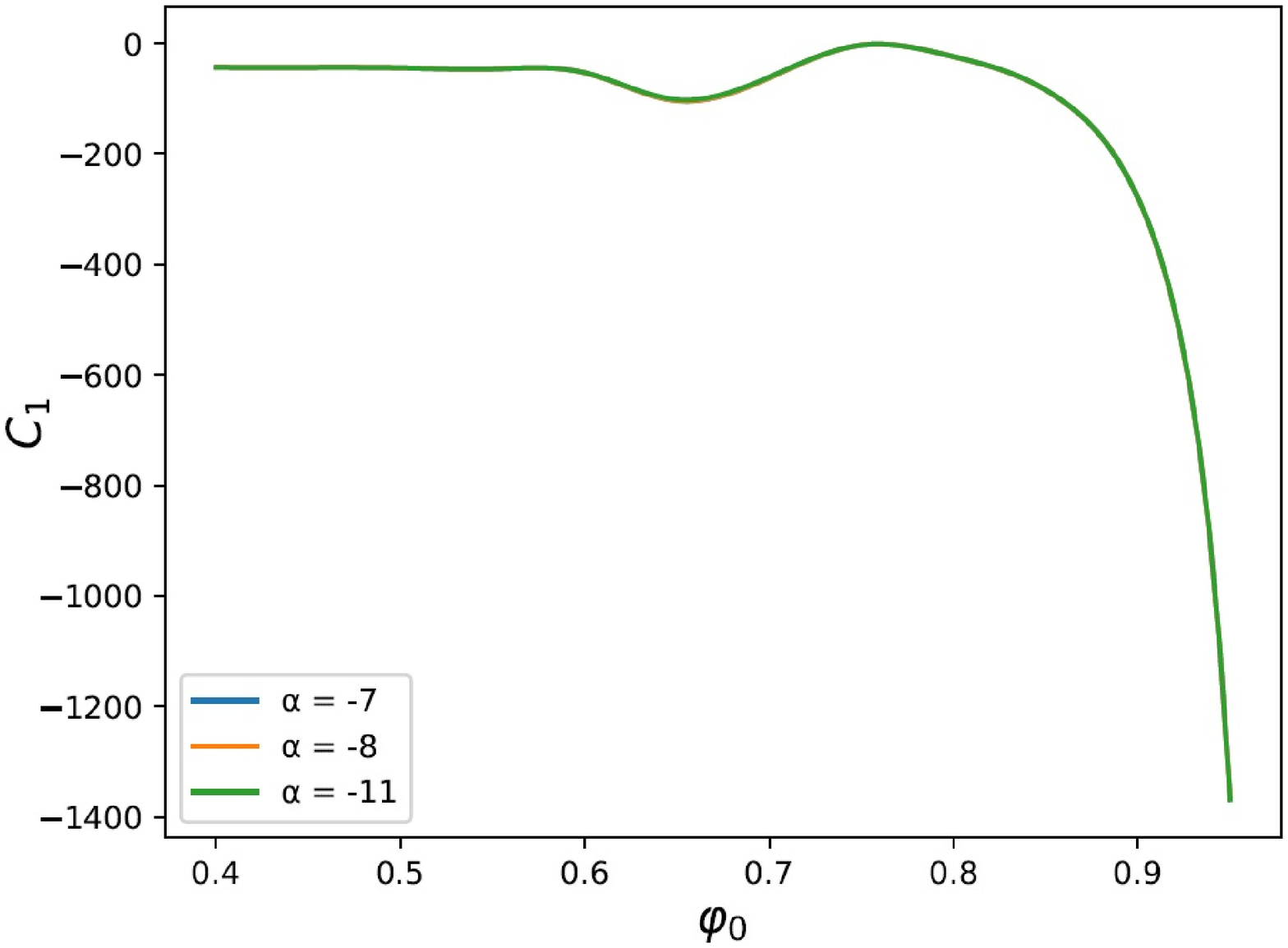}
      \caption{Left: The integration constant $C_1$ as a function of  $\f_0$ for fixed  values of $\alpha$, for Nut end-point solutions. Right: $C_1$ as a function of $\alpha$  for fixed  values of $\f_0$, for Nut end-point solutions.}
      \label{fig12}
   \end{minipage}%
   \begin{minipage}{.5\textwidth}
      %\centering
\includegraphics[scale = 0.34]{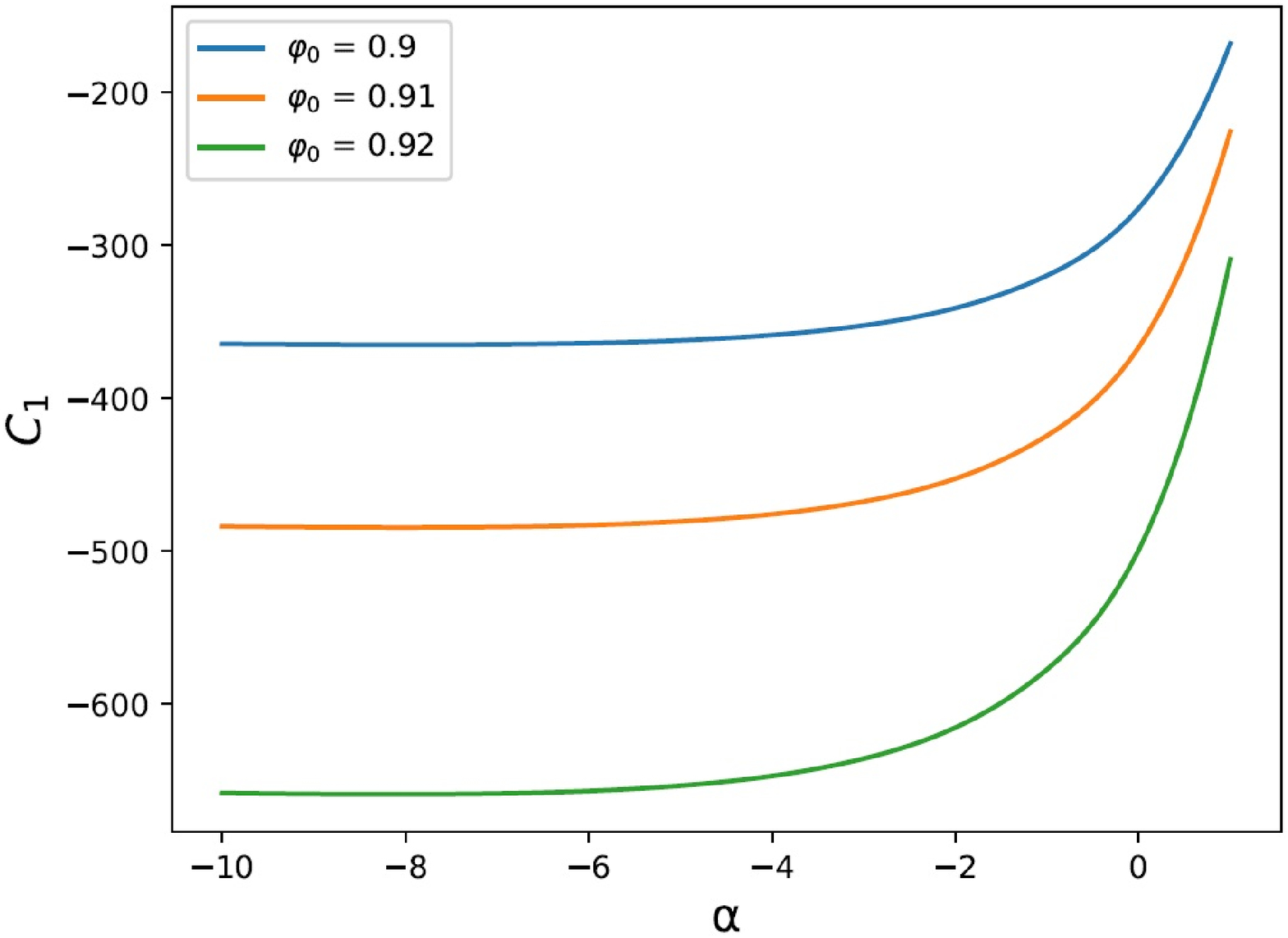}
   \end{minipage}
}

\newpage

Our findings can be summarised as follows: We consider a QFT on a squashed $S^3$, with a relevant coupling $\f_-$, a fixed UV curvature, $R^{uv}$, and a value for the squashing parameter $a^2$.
 Out of the first two scales, we construct the dimensionless curvature, $\mathcal{R} = R^{uv} \vert \f_- \vert^{-2/\Delta_-}$.
  Our regular flow solutions are controled by the two IR parameters $\f_0$ and $W_{(1)2}$.
  For fixed $W_{(1)2}$,  $\mathcal{ R}$  is in one to one correspondence with $\f_0$.

  As the value of $\mathcal{R}$ increases, the endpoint $\f_0$ approaches the maximum of the potential, located at $\f_{max} = 1$. Similarly, as the value of $\mathcal{R}$ approaches 0, the endpoint $\f_0$ approaches the minimum of the potential, situated at $\f_{min} = 0$. This is in agreement with QFT intuition.

  For fixed $\f_0$, the value of $\mathcal{R}$ decreases with increasing $W_{(1)2}$. Furthermore, having specified a value for the free parameter $W_{(1)2}$ and for $\f_0$, we can calculate the squashing parameter $a^2$. The remaining parameters, $C_1,C_2$, which control the vevs, are then completely fixed and calculable.

\subsection{The case of the Bolt IR endpoint}

Similarly with the case of the Nut IR endpoint, we shall rewrite for convenience the IR regularity  conditions for the Bolt, (\ref{ga134})-(\ref{ga136}), which we derived in subsection \ref{boltir}:

\be
S = S_0 \sqrt{\f - \f_0} + \ldots  \ , \ W_{1} = \frac{-2S_0}{\sqrt{\f - \f_0}} + \frac{3T_{(2)2} - V_0 - 2V_2}{4S_0} \sqrt{\f - \f_0} + \ldots,
\ee
\be
W_2 = \frac{2V_0-T_{(2)2}}{S_0}\sqrt{\f - \f_0} + \ldots \ , \ T_1 = \frac{T_{(2)2}^2}{4V_1} (\f - \f_0) + \ldots \ , \ T_{2} = T_{(2)2} + \ldots,
\ee
where
\be
S_0 = \sqrt{V_1} \ , \ T_{(2)2} = \text{arbitrary}.
\ee

Here we have two free parameters, namely $T_{(2)2}$, and the value of the scalar end-point, $\f_0$. For convenience, we shall make the following definitions:

\be
\alpha \equiv { T_{(2)2} \over  \tilde{W}} \ , \ \tilde{W} \equiv \frac{9 V_0 + 2 V_2}{7} = -0.347143 \varphi _0^4+0.231429 \varphi _0^2-7.90714,
\ee
where $\alpha$ and $\tilde{W}$ are constants. When $\alpha = 1$, we have to leading order

\be
W_1 + \frac{2S_0}{\sqrt{\f-\f_0}} = W_2 + \mathcal{O} \left( (\f - \f_0)^{3/2}.\right)
\ee

In Figure \ref{fig13} we have plotted the superpotentials, $W_{1,2}$, as a function of  $\f$ , for various choices of the free parameter $\a$. Moreover, we have set $\f_0 = 0.9$. We observe that as we increase $T_{(2)2}$ ($\tilde{W}$ is negative), the maximum of the function $W_1$ decreases, and its position approaches the endpoint. Similarly, the minimum of the function $W_2$ decreases, and its position approaches the endpoint as well. For completeness, we have also plotted the function $S$ as a function of  $\f$ for specific values of $\alpha$ in Figure \ref{fig13a}.

\FIGURE[ht]{
   \includegraphics[scale = 0.69]{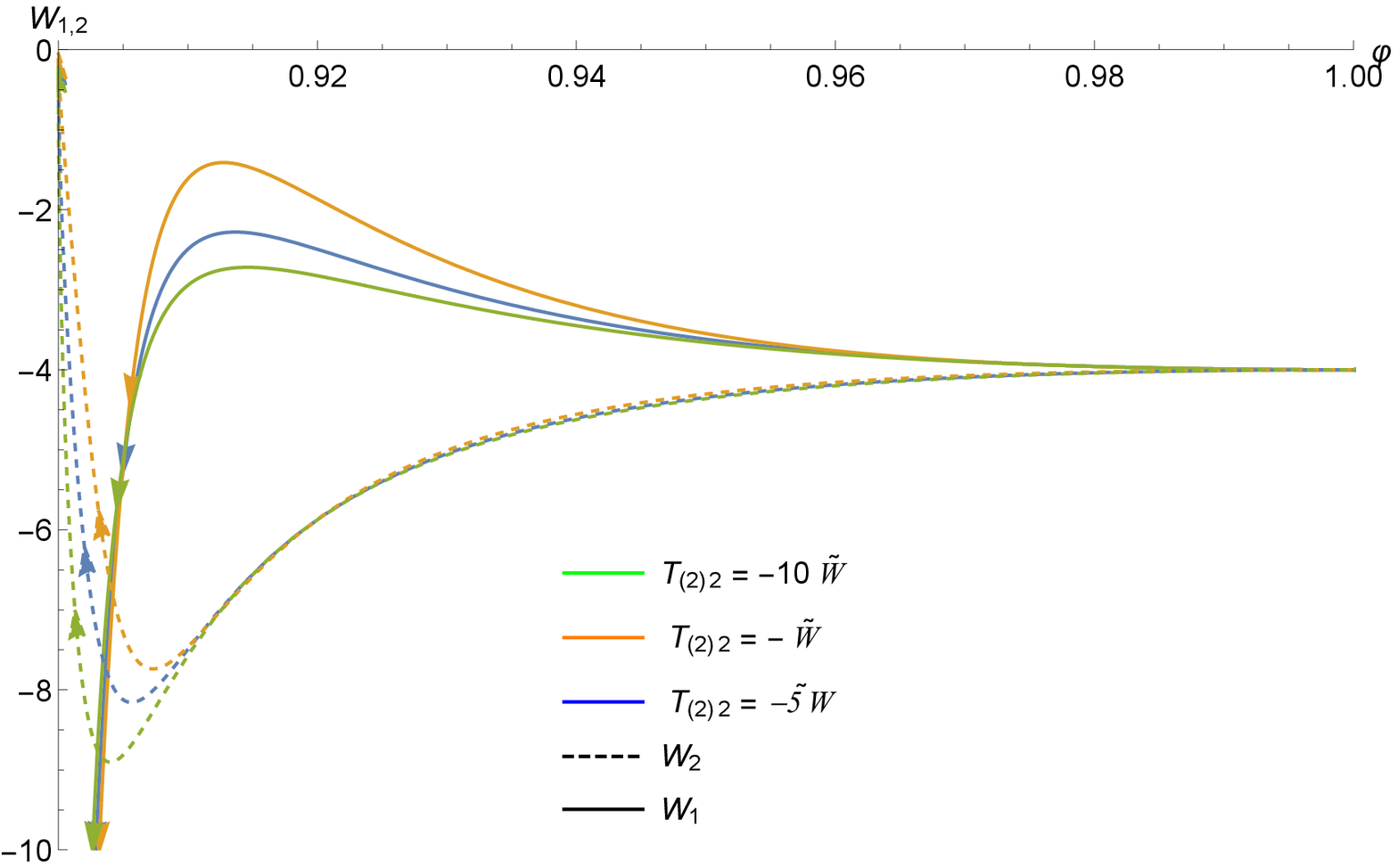}\\
   \caption{ Plot of the superpotentials  $W_1,W_2$ as functions of  $\f$ for different choices of $\alpha$, for Bolt end-point solutions. The IR endpoint for this plot has been chosen to be at $\f_0 = 0.9$. The   direction of the flow is indicated in the figure: It starts from the UV fixed point and ends at the IR.
   The different lines correspond to different choices of the integration constant $T_{(2)2}$. The constant $\tilde W$ is defined in (\protect{\ref{n1a}}).
   }
   \label{fig13}
   }
\newpage

\FIGURE[ht]{
      \includegraphics[scale = 0.61]{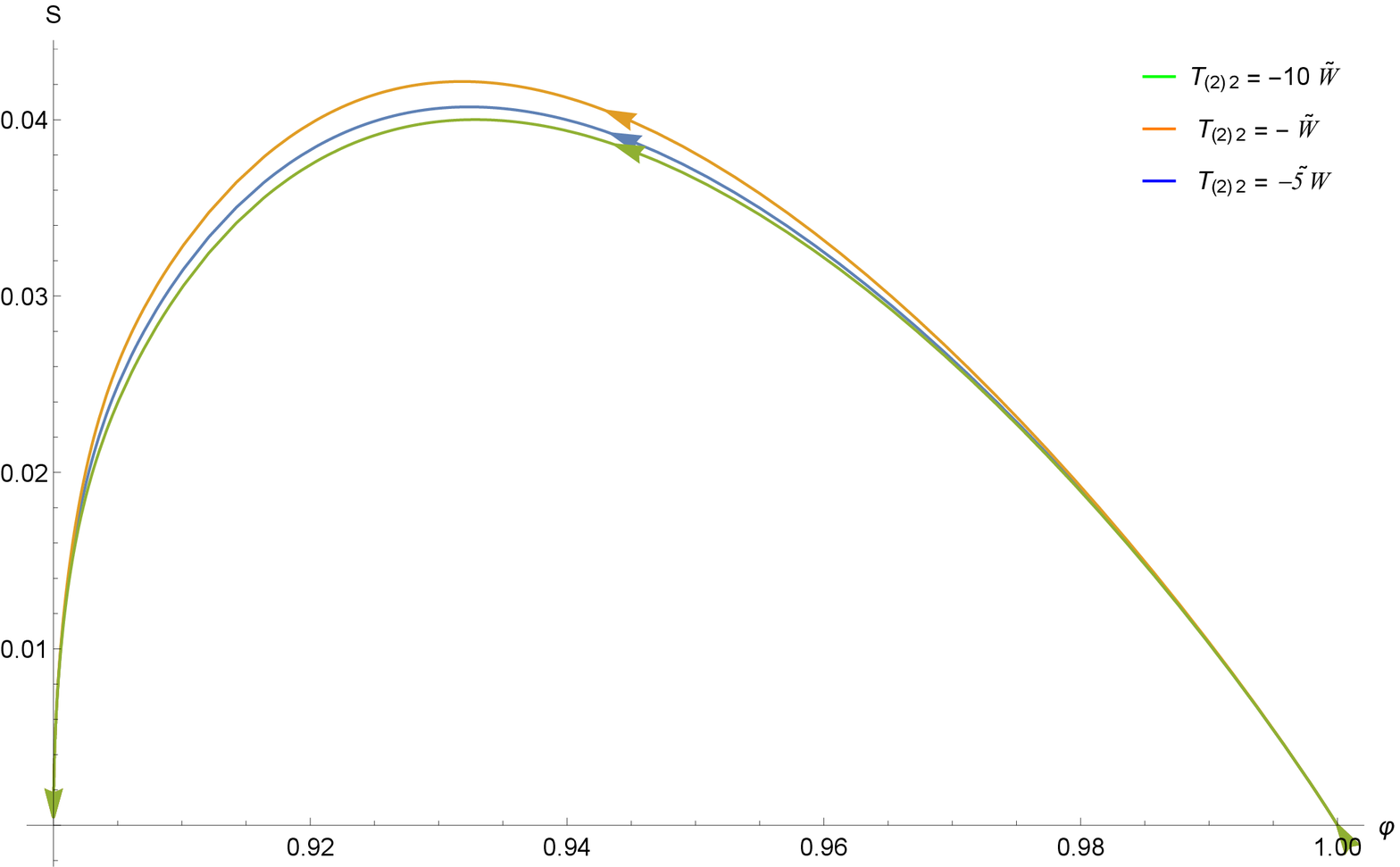}
      \caption{Plot of the superpotential  S as a function of $\f$ for different choices of $\alpha$,  for Bolt end-point solutions. The IR endpoint for this plot has been chosen to be at $\f_0 = 0.9$.  The  direction of the flow is indicated in the figure: It starts from the UV fixed point and ends at the IR.
       The different lines correspond to different choices of the integration constant $T_{(2)2}$. The constant $\tilde W$ is defined in (\protect{\ref{n1a}}).
      }
      \label{fig13a}
   }
Our remarks are as follows:
\begin{enumerate}
   \item[$\bullet$] Similarly to the case of the Nut end-point, for every value of $\f_0$ between $\f_{min} = 0$ and $\f_{max} = 1$ there exists a unique solution to the superpotential equations (\ref{num1})-(\ref{num2}) corresponding to an RG flow that starts from the UV fixed point at $\f = 1$ and ends at an IR endpoint $\f_0$, which is located between $\f_{min}$ and $\f_{max}$. Furthermore, no solution exists for $\f_0 = 0$ exactly, which corresponds to $\mathcal{R} = 0$, i.e. a flat space.
\end{enumerate}

\newpage

\begin{enumerate}
   \item[$\bullet$] The dimensionless curvature, $\mathcal{R}$, has a qualitatively similar  behavior, as in the case of the Nut,  when we fix the integration constant $\alpha$. This is illustrated in Figure \ref{fig14}. However, note that the scale of $\mathcal{R}$ is significantly less. Moreover, as $\alpha$ increases, the value of $\mathcal{R}$ decreases, in contrast to the previous case, where $\mathcal{R}$ increased as we increased $\alpha$. This is illustrated in the right figure of \ref{fig15}. From figure \ref{fig14} it becomes apparent that since the dimensionless curvature $\mathcal{R}$ is positive, the parameter $\alpha$ cannot take arbitrarily large values. Specifically, we find that its largest value is $\alpha = 0$. For values larger than that, $\mathcal{R}$ either vanishes or becomes negative and this is inconsistent with our assumptions.

   Finally, for completeness, we have also plotted in the left figure \ref{fig15}, $\mathcal{R}$ as function of  $\f_0$ for specific values of $\alpha$.
\end{enumerate}

\FIGURE[ht]{
   \includegraphics[scale = 0.6]{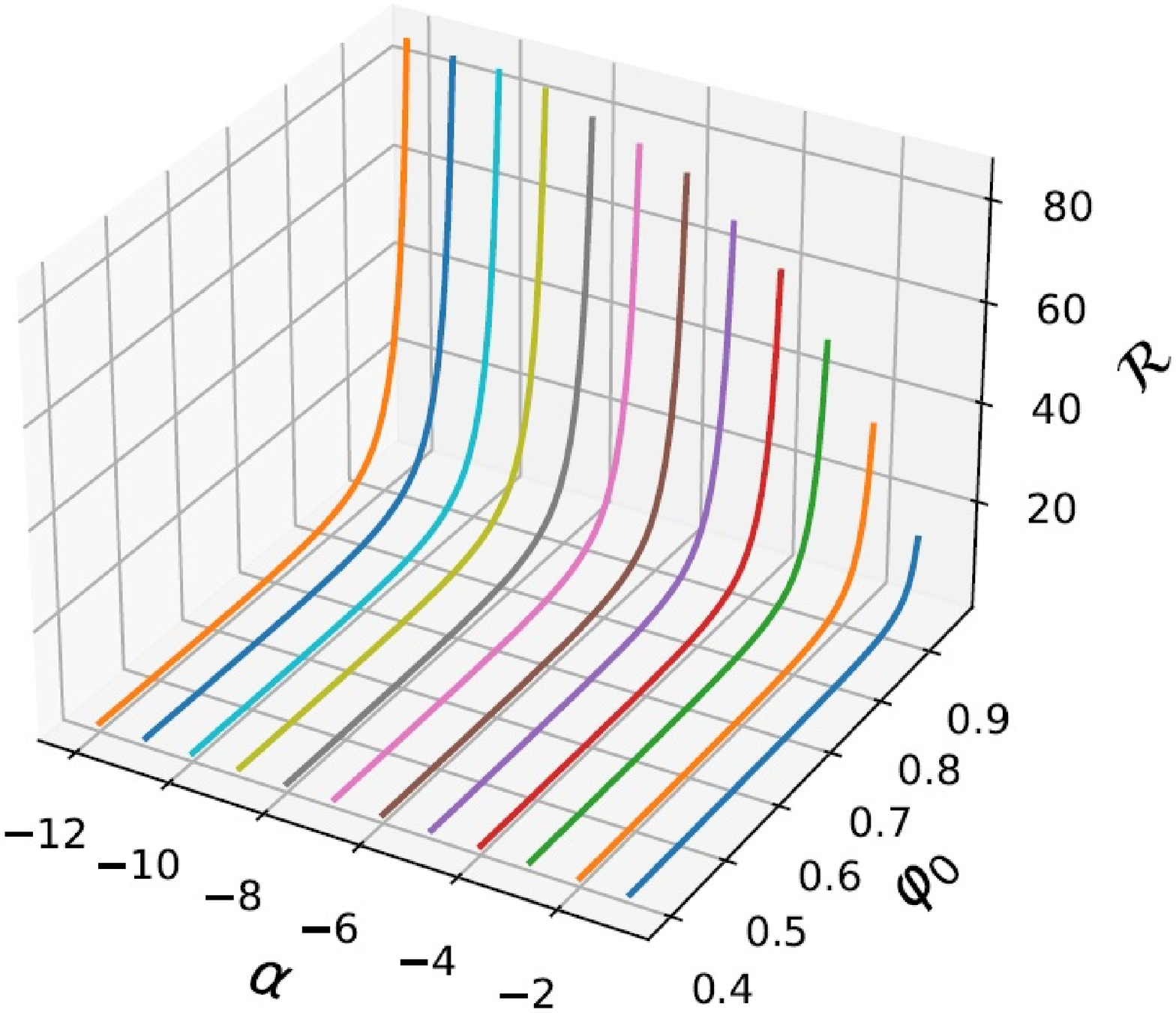}
   \caption{The dimensionless curvature $\mathcal{R}$  as a function of  the two parameters, $\f_0$ and $\alpha$,  for Bolt end-point solutions.}
   \label{fig14}
}
\FIGURE[ht]{
   \begin{minipage}{.5\textwidth}
      \centering
      \includegraphics[scale = 0.34]{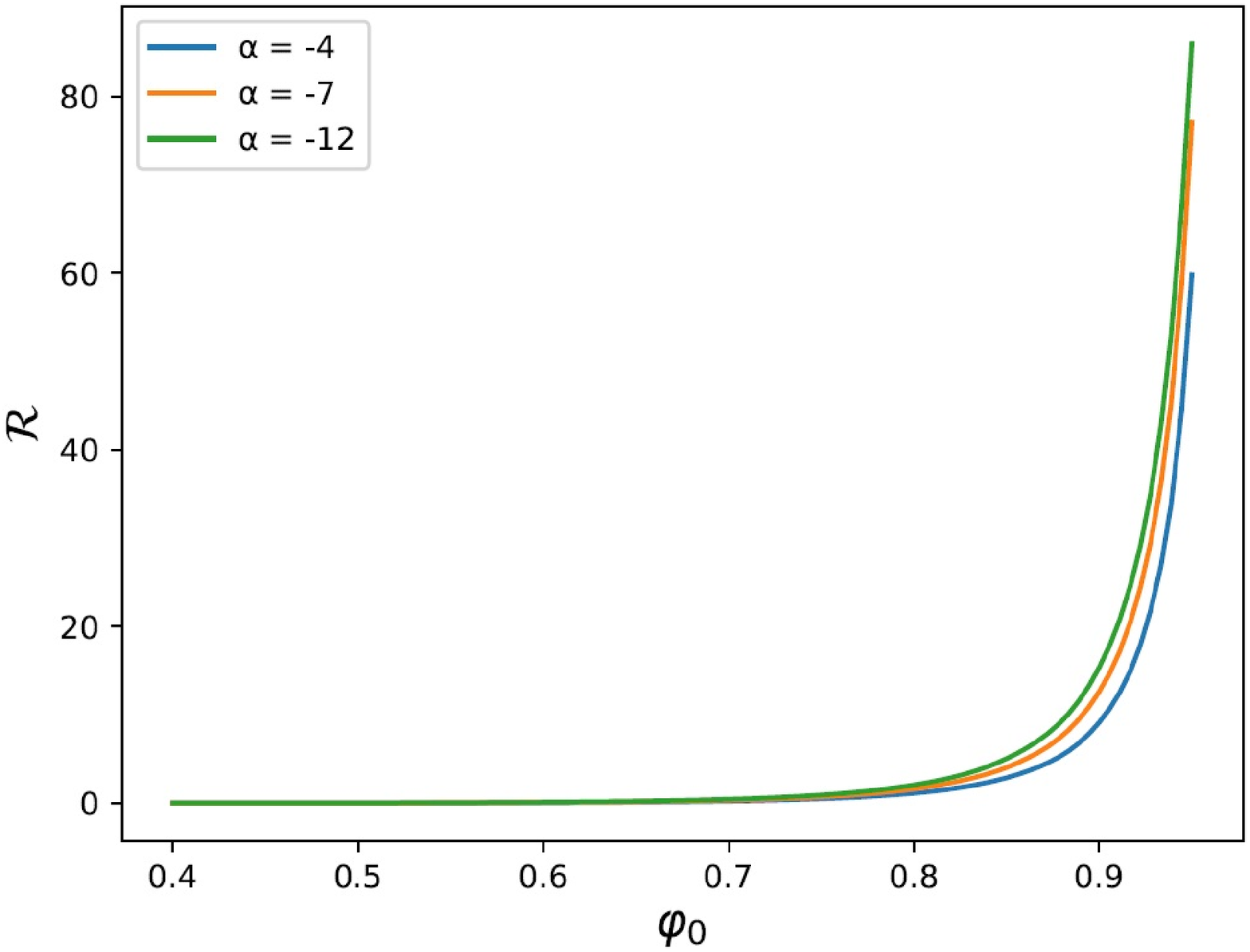}
      \caption{Left: The dimensionless curvature $\mathcal{R}$ as a function of  $\f_0$ for fixed values of $\alpha$ on the left,  for Bolt end-point solutions.
       Right: The dimensionless curvature $\mathcal{R}$ as a function of  $\alpha$ for fixed values of $\f_0$,  for Bolt end-point solutions.}
      \label{fig15}
   \end{minipage}%
   \begin{minipage}{.5\textwidth}
      \centering
\includegraphics[scale = 0.34]{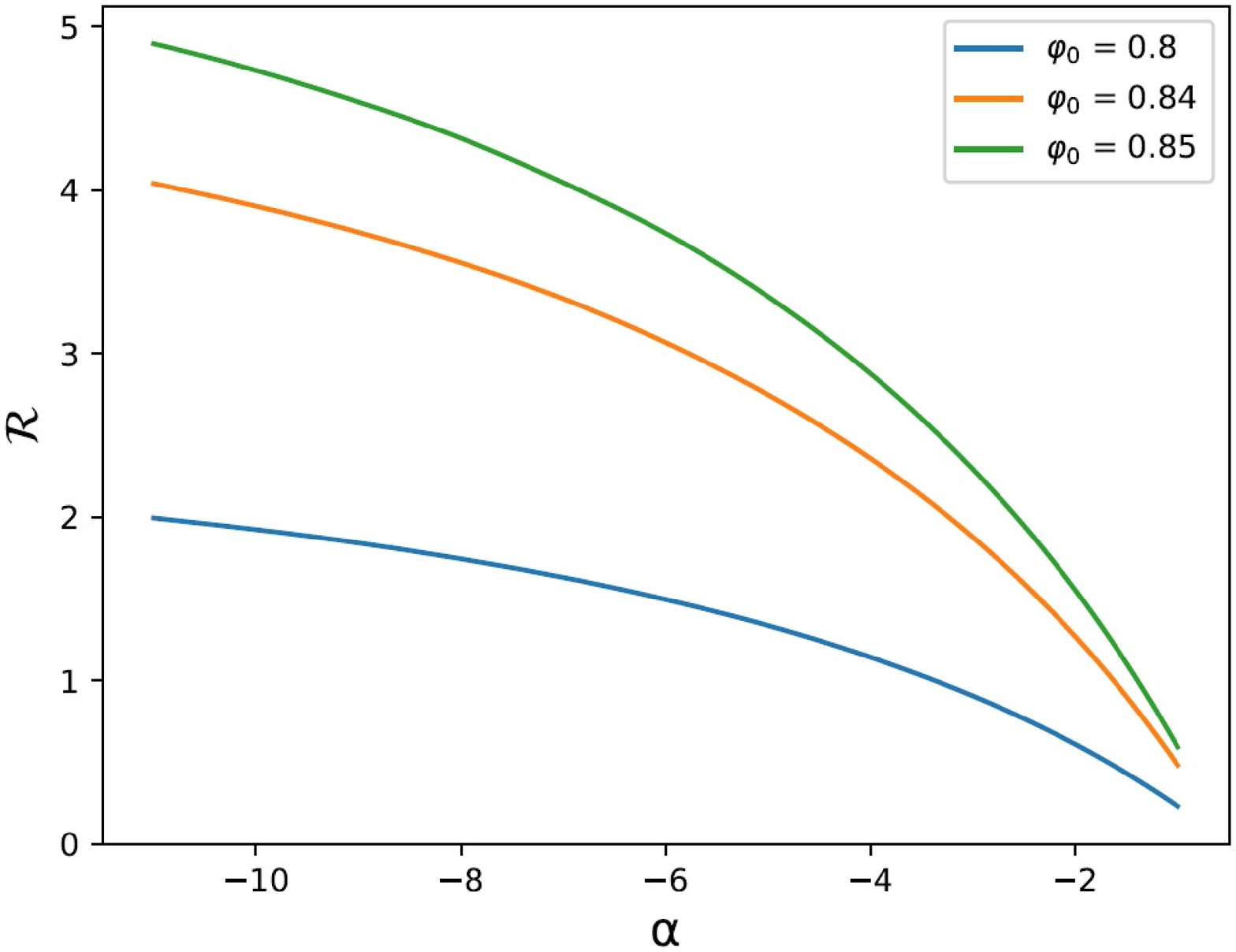}
   \end{minipage}
}

\newpage
\begin{enumerate}
   \item[$\bullet$]  In Figure \ref{fig16} we have plotted the squashing parameter  as a function of  $\alpha$ and $\f_0$.  In the two figures \ref{fig17} we have plotted slices for specific values of the free parameters. We observe that for fixed $\f_0$, the value of the squashing parameter increases with $\alpha$, a behavior similar to what we observed in the case of the Nut IR endpoint. However, since $\alpha$ is bounded now, this implies that $a^2$ is bounded as well. It becomes apparent from the metric (\ref{ga138}) that solutions that asymptote to the Bolt IR endpoint always correspond to 3-spheres with non trivial squashing.
\end{enumerate}

\FIGURE[ht]{
   \includegraphics[scale = 0.6]{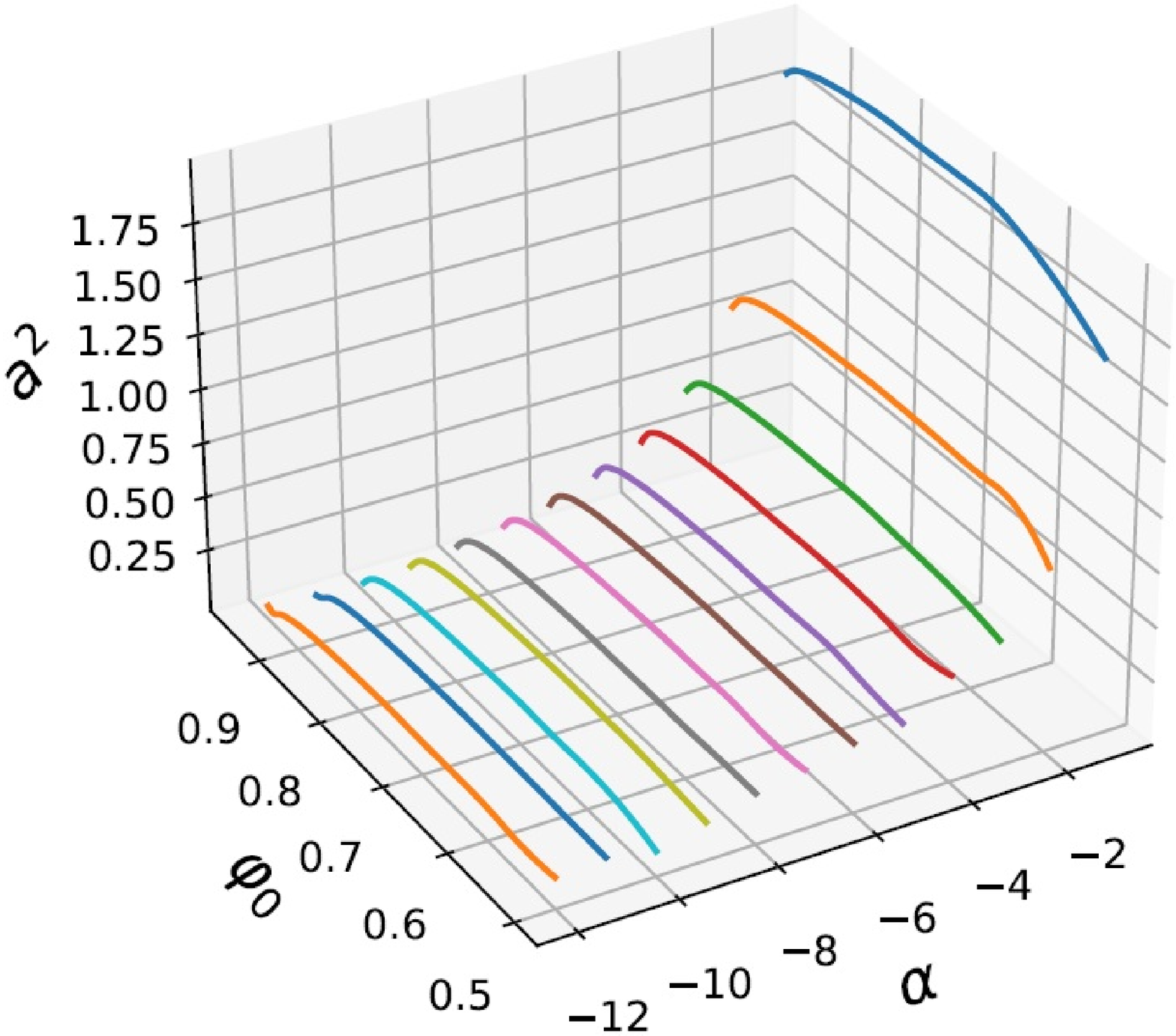}
   \caption{The squashing $a^2$  as a function of  the two parameters, $\f_0$ and $\alpha$,  for Bolt end-point solutions.}
   \label{fig16}
}
\FIGURE[ht]{
   \begin{minipage}{.5\textwidth}
      \centering
      \includegraphics[scale = 0.34]{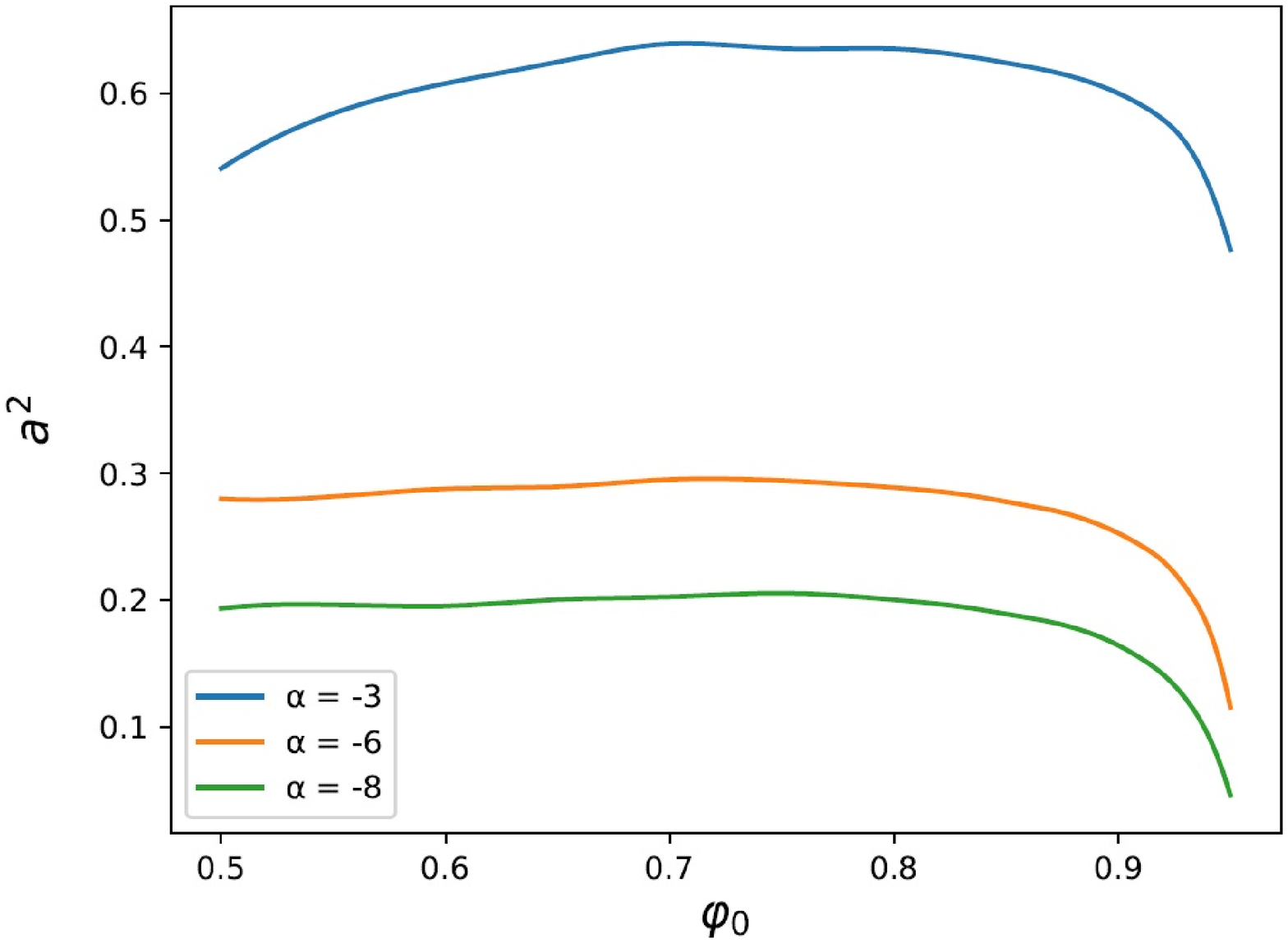}
      \caption{Left: The squashing parameter $a^2$  as a function of  $\f_0$ for certain values of $\alpha$, for Bolt end-point solutions. Right:  The squashing parameter $a^2$ as a function of $\alpha$ for fixed  values of $\f_0$,  for Bolt end-point solutions.}
      \label{fig17}
   \end{minipage}%
   \begin{minipage}{.5\textwidth}
      \centering
\includegraphics[scale = 0.34]{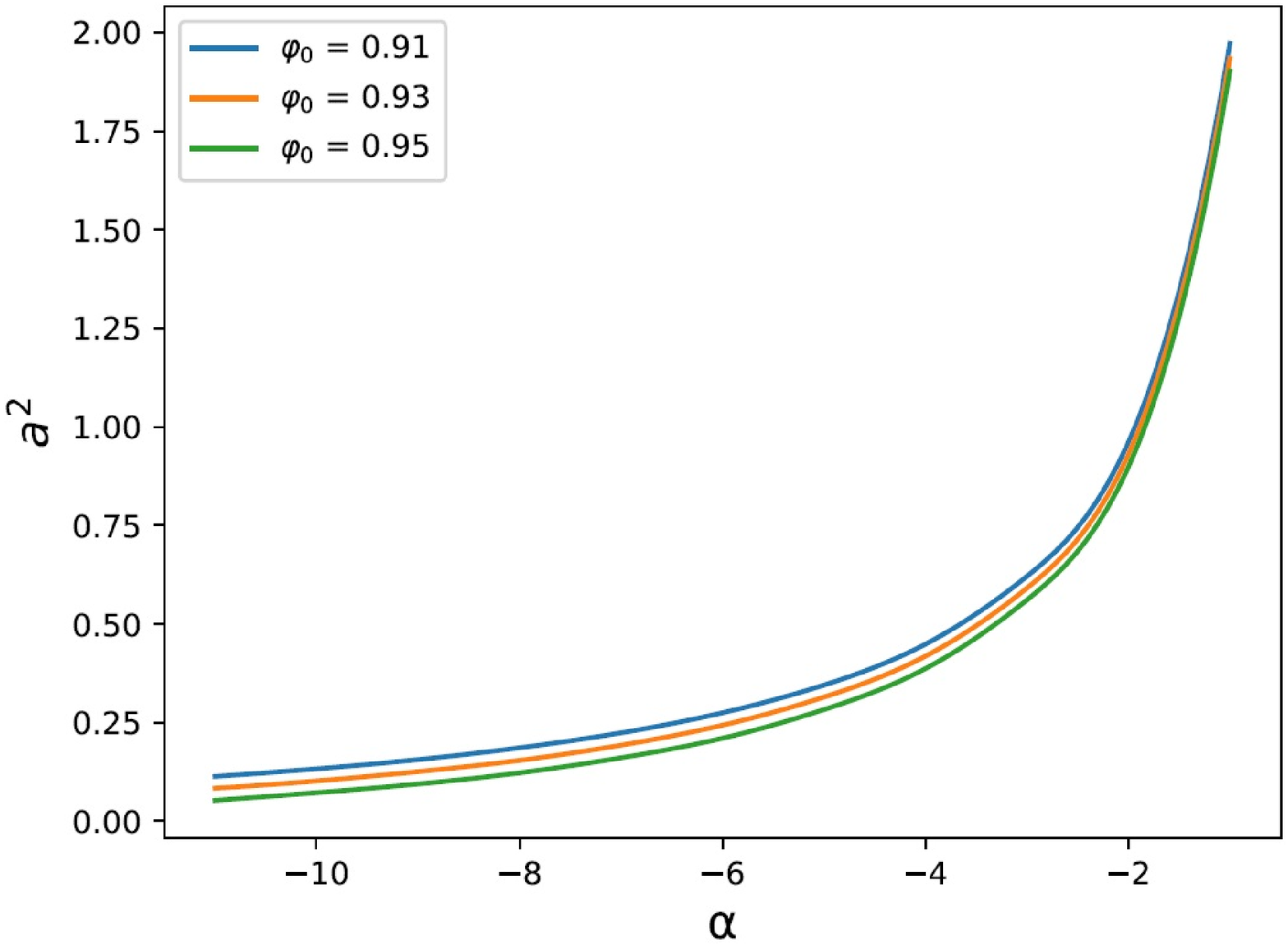}
   \end{minipage}
}

\newpage

\begin{enumerate}
   \item[$\bullet$] In Figure \ref{fig18} we have plotted the integration constant $C_2$  as a function of  the two parameters of the problem, $\f_0$ and $\alpha$. We observe that its behavior is similar to that in Figure \ref{fig19}, which considered the Nut IR endpoint instead of the Bolt. Note, however, that for fixed $\f_0$, $C_2$ increases with increasing $\alpha$, whereas in the previous case it first decreased to a minimum value, and it then continued to increase.
\end{enumerate}

\FIGURE[ht]{
   \includegraphics[scale = 0.59]{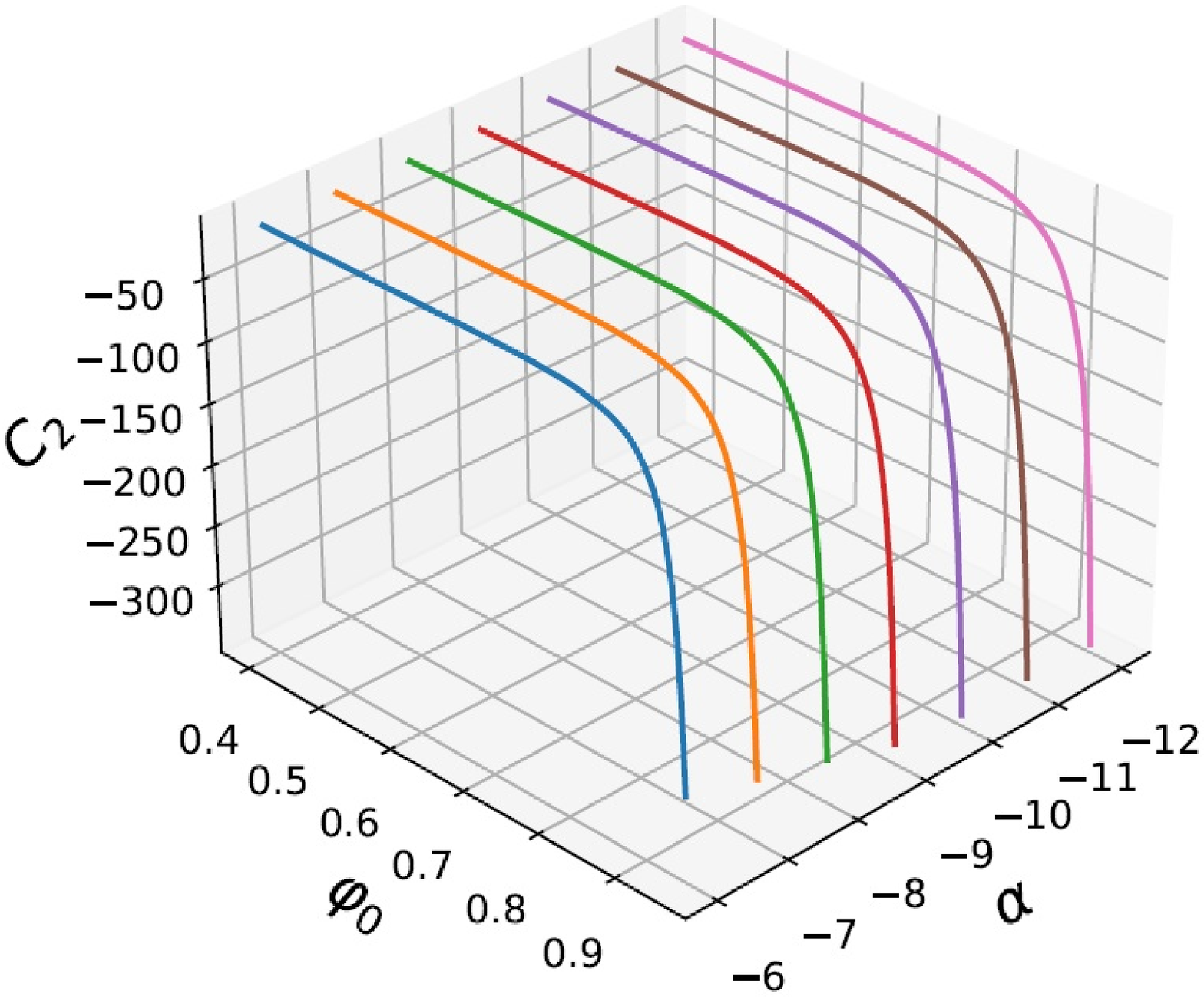}
   \caption{The integration constant $C_2$  as a function of the two parameters, $\f_0$ and $\alpha$,  for Bolt end-point solutions.}
   \label{fig18}
}
\FIGURE[ht]{
   \begin{minipage}{.5\textwidth}
      \centering
      \includegraphics[scale = 0.34]{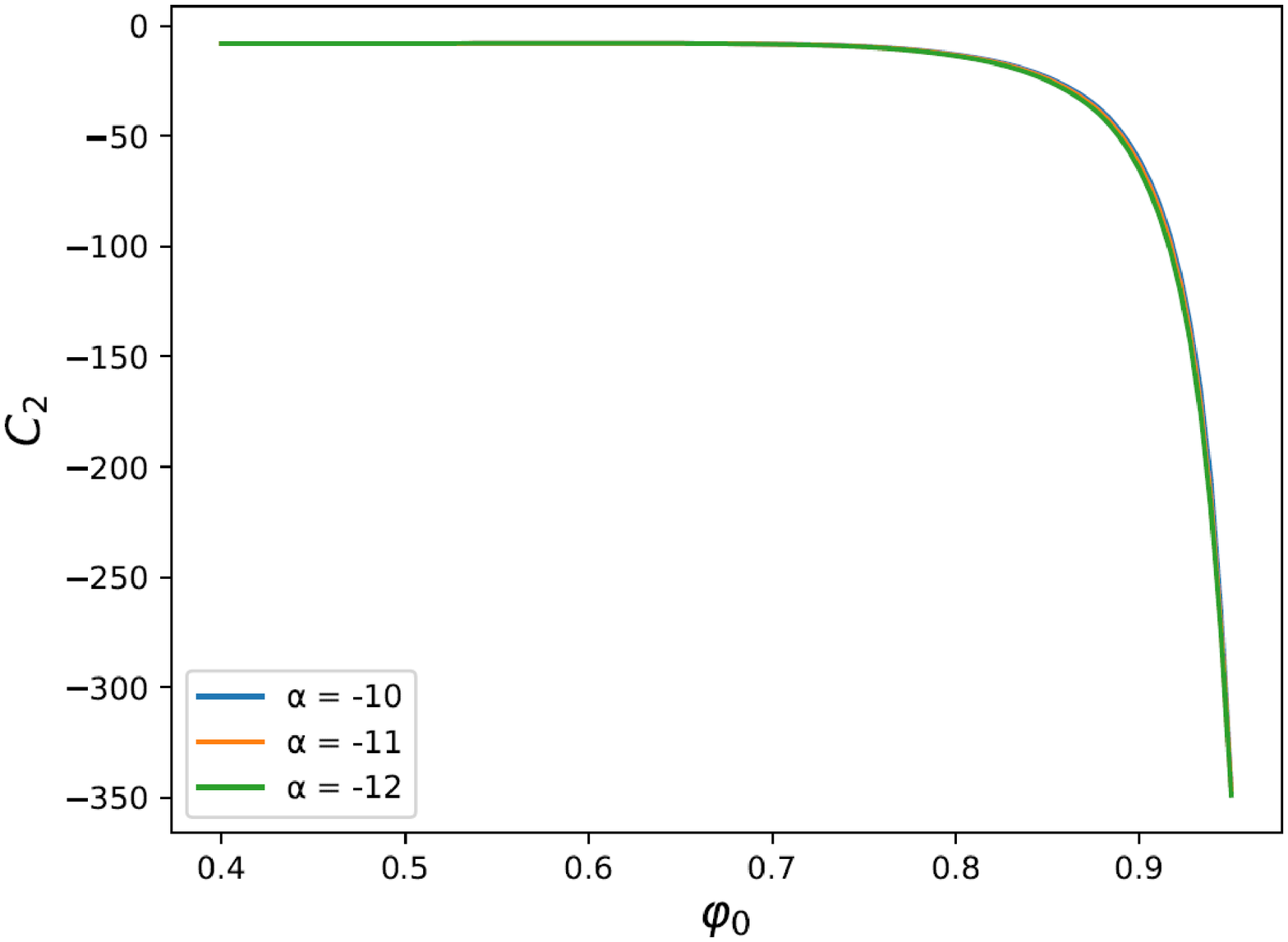}
      \caption{Left: The integration constant $C_2$  as a function of  $\f_0$ for fixed  values of $\alpha$,  for Bolt end-point solutions. Right: the integration constant $C_2$ with respect to $\alpha$ for fixed values of $\f_0$,  for Bolt end-point solutions.}
      \label{fig19}
   \end{minipage}%
   \begin{minipage}{.5\textwidth}
      \centering
\includegraphics[scale = 0.34]{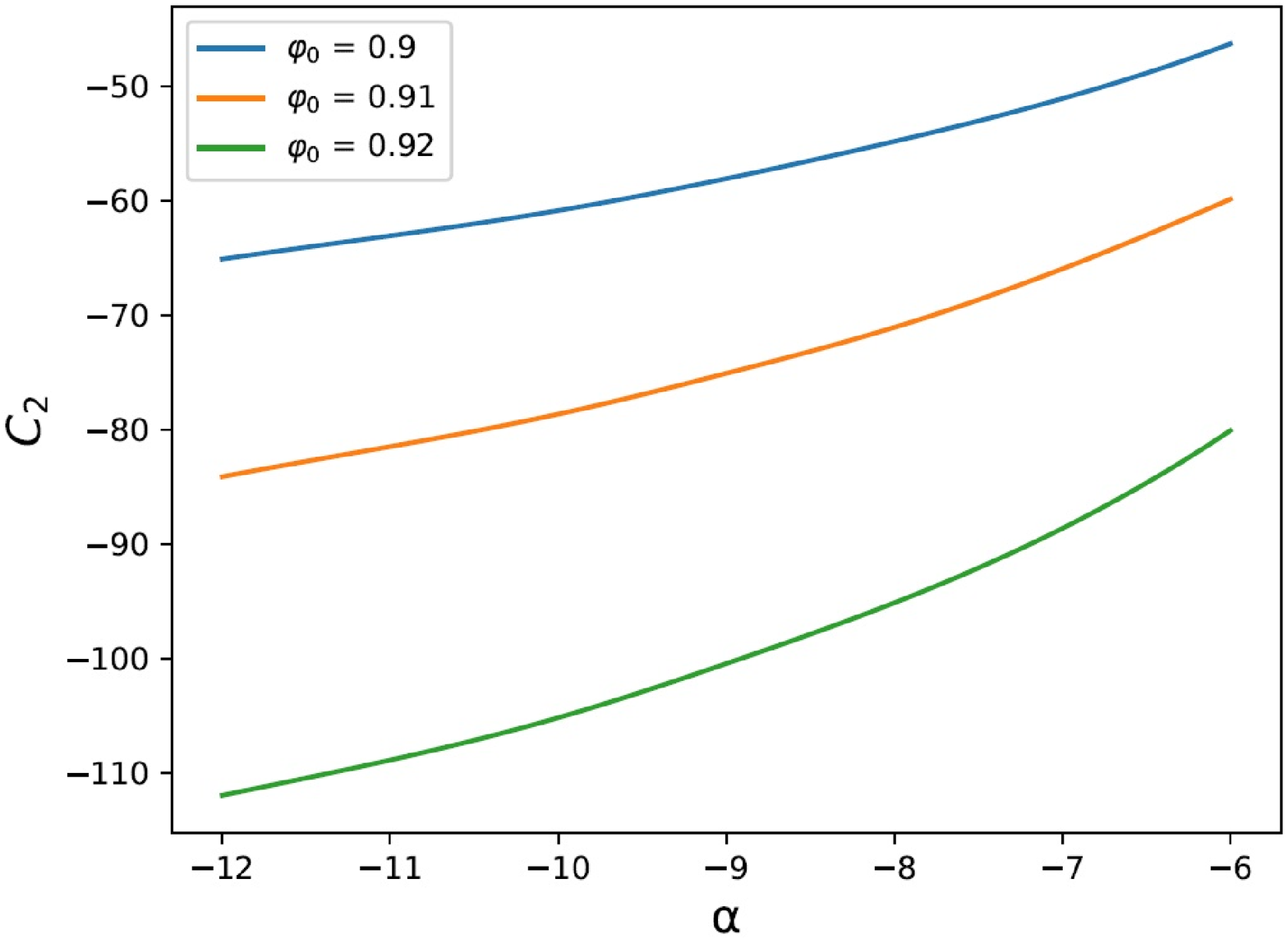}
   \end{minipage}
}

\newpage

\begin{enumerate}
   \item[$\bullet$] In Figure \ref{fig20} we have plotted the integration constant $C_1$  as a function of  the two parameters of the problem, $\f_0$ and $\alpha$. We observe that its behavior is similar to that in Figure \ref{fig21}, which considered the Nut IR endpoint instead of the Bolt.
\end{enumerate}

\FIGURE[ht]{
   \includegraphics[scale = 0.59]{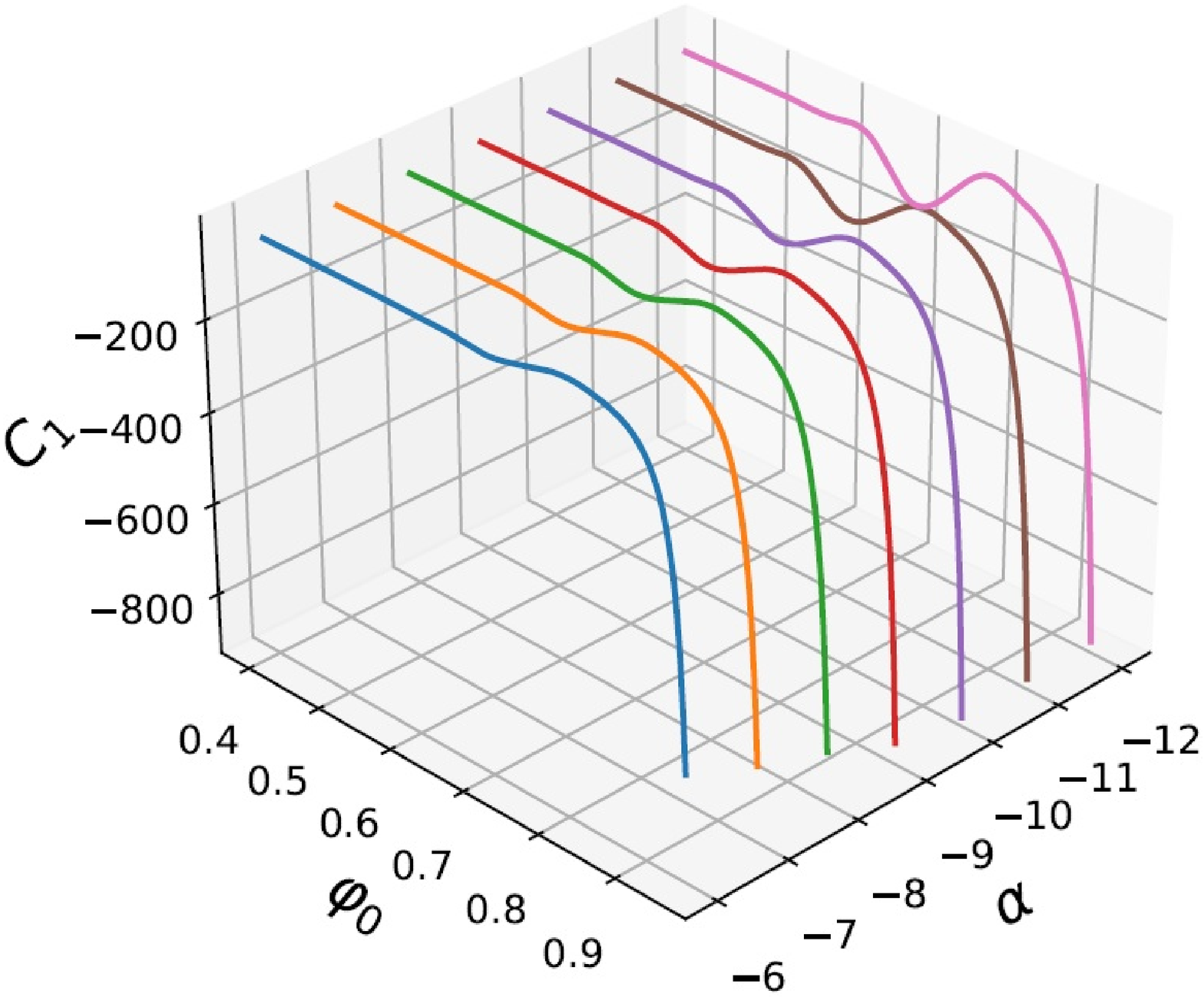}
   \caption{The integration constant $C_1$  as a function of  the two parameters, $\f_0$ and $\alpha$,  for Bolt end-point solutions.}
   \label{fig20}
}
\FIGURE[ht]{
   \begin{minipage}{.5\textwidth}
      \centering
      \includegraphics[scale = 0.33]{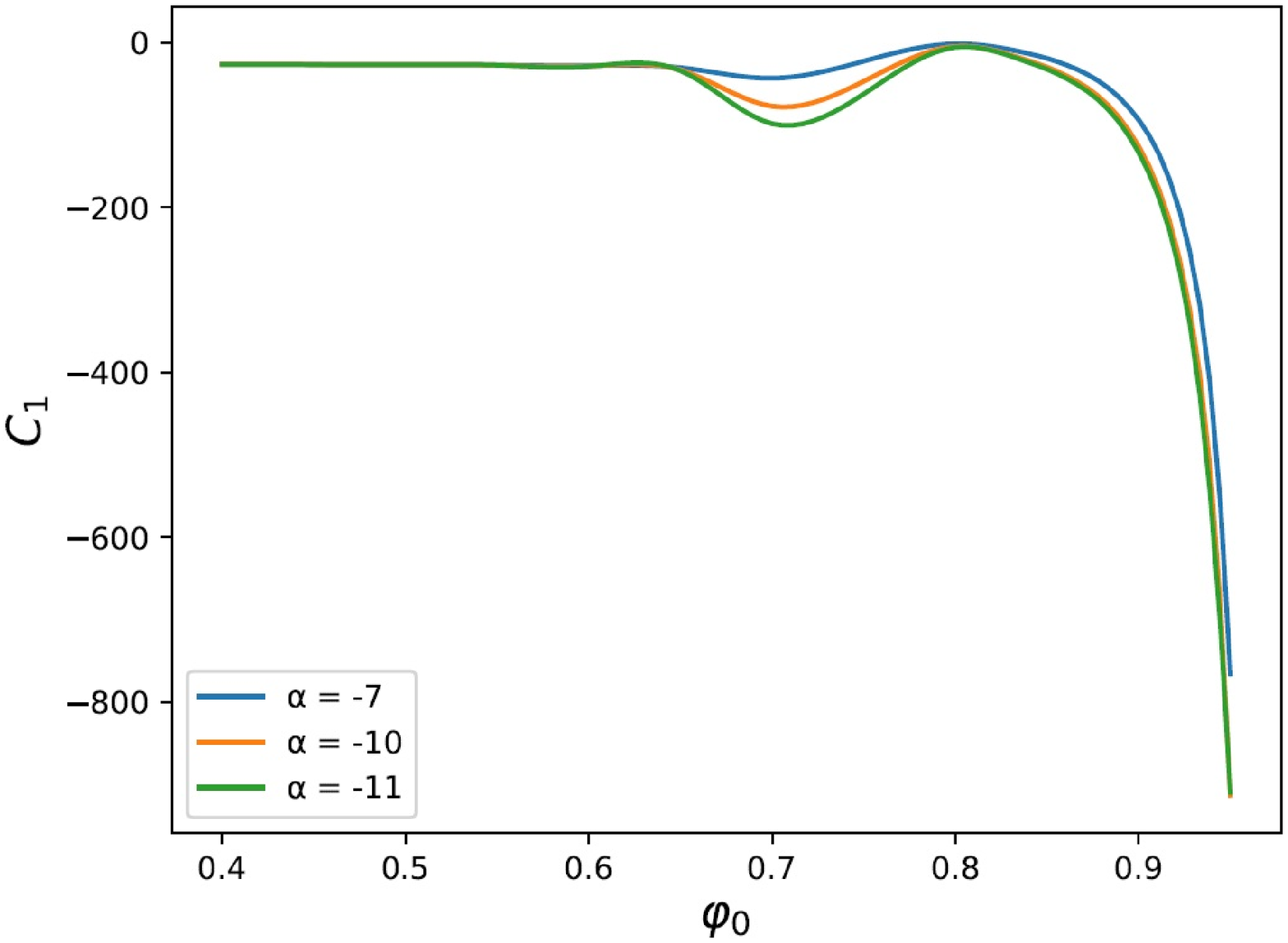}
      \caption{Left: The integration constant $C_1$  as a function of  $\f_0$ for certain values of $\alpha$,  for Bolt end-point solutions. Right: The integration constant $C_1$  as a function of $\alpha$ for fixed values of $\f_0$,  for Bolt end-point solutions.}
      \label{fig21}
   \end{minipage}%
   \begin{minipage}{.5\textwidth}
      \centering
\includegraphics[scale = 0.33]{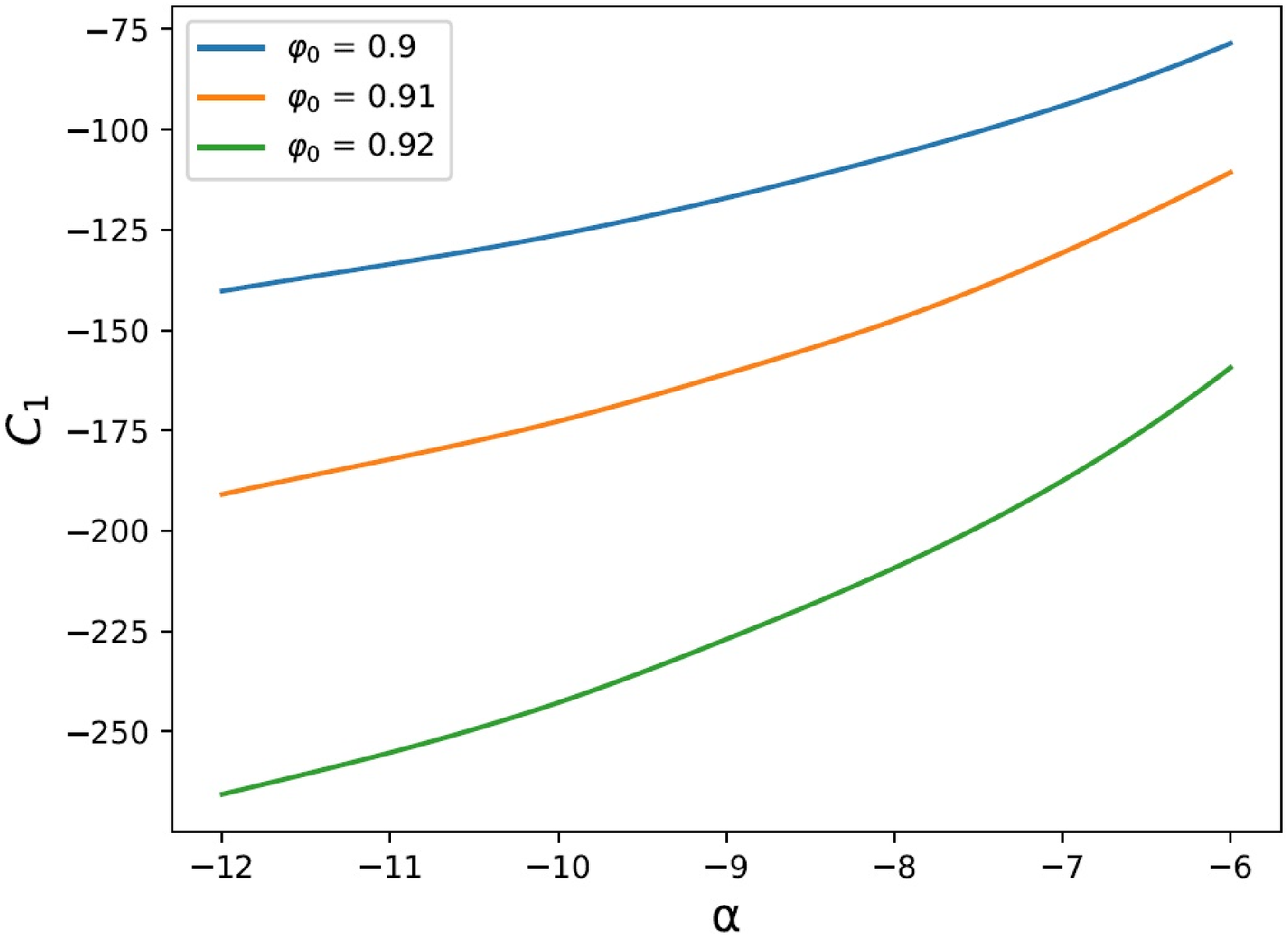}
   \end{minipage}
}
\newpage

Our findings can be summarised as follows: the case of Bolt end-point solutions gives qualitatively similar behaviors as the Nut case, but there is one difference. We cannot obtain all possible values of the squashing parameter $a^2$ but there is an upper bound that can be obtained. This is similar to what is known for the CFT solutions on squashed $S^3$.

\subsection{Synopsis:}

In this section, we have chosen a specific potential, $V(\f)$, given by (\ref{ga153}), and we have displayed full RG flows that start from a UV fixed point, which is located at the right maximum  of $V(\f)$, and end at an IR point $\f_0$, which is located in between the minimum of the potential and the UV fixed point. Depending on the behavior of the superpotentials $S,W_{1,2}$ \footnote{The behavior of $T_{1,2}$ can be extracted from $S,W_{1,2}$ by using the equations of motion (\ref{ga158}),(\ref{ga159}).} near the IR, the integration constants $C_{1,2},\mathcal{R},a$ that appear in the (-)-branch solution (which was discussed in detail in \ref{expansionnearmaxima}) have a specific qualitative behavior, independent of the details of the scalar potential, $V(\f)$. Specifically, their behavior is as follows:

\begin{enumerate}
   \item[$\bullet$] \textbf{The Nut IR endpoint}: In this case, the superpotentials have the following behavior:

   \be
   S \sim \sqrt{\f - \f_0} \ , \ W_{1,2} \sim \left( \f - \f_0 \right)^{-1/2}.
   \label{n3}\ee

   Given a source, $\f_-$, a UV curvature, $R^{uv}$, and a value for the free parameter, $W_{(1)2}$, we can determine the integration constants $\mathcal{R},a^2$. For a fixed value of $W_{(1)2}$, the dimensionless curvature $\mathcal{R}$ is an increasing function of $\f_0$, with the scaling law being $\mathcal{R} \sim \vert \f_- \vert^{-2/\Delta_-}$. Moreover, by fixing $\f_0$ and varying $W_{(1)2}$, we observe that $\mathcal{R}$ is a decreasing function of $W_{(1)2}$. Finally, by choosing the appropriate IR behavior, illustrated in (\ref{n3})
   , we can determine the two parameters $C_{1,2}$ which control the vevs of the problem.

   \item[$\bullet$] \textbf{The Bolt IR endpoint}: In this case, the superpotentials have the following behavior:

   \be
   S \sim \sqrt{\f - \f_0} \ , \ W_1 \sim \left( \f - \f_0 \right)^{-1/2} \ , \ W_2 \sim \sqrt{\f - \f_0}.
   \label{n4}\ee

   Given a source, $\f_-$, a UV curvature, $R^{uv}$, and a value for the free parameter, $T_{(2)2}$, we can determine the integration constants $\mathcal{R},a^2$. For a fixed value of $T_{(2)2}$, the dimensionless curvature $\mathcal{R}$ is an increasing function of $\f_0$, with the scaling law being $\mathcal{R} \sim \vert \f_- \vert^{-2/\Delta_-}$. Moreover, by fixing $\f_0$ and varying $T_{(2)2}$, we observe that $\mathcal{R}$ is an increasing function of $T_{(2)2}$. Finally, by choosing the appropriate IR behavior, illustrated in (\ref{n4}), we can determine the two parameters $C_{1,2}$ which control the vevs of the problem.
\end{enumerate}

%\newpage

\section*{Acknowledgements}\label{ACKNOWL}
\addcontentsline{toc}{section}{Acknowledgements}

\noindent We would like to thank J. Ghosh, F. Nitti, E. Preau and  A. Tersenov  for discussions and help.
We especially thank T. Koutsikos for pointing our some errors in the near boundary expansions of some of the solutions and obtaining the correct coefficients.
  This work was supported in part by the Advanced ERC grant SM-grav, No 669288 as well as the CNRS PICS Project IEA 199430.

\newpage
\appendix
\renewcommand{\theequation}{\thesection.\arabic{equation}}
\addcontentsline{toc}{section}{Appendix\label{app}}
\section*{Appendix}

\section{The geometry of the squashed three-sphere}\label{squashedsphere}

The squashed sphere  metric is parametrized as,
\be
d\Omega_{a}^2=\frac{L^2}{4}[a^2(d\psi+\cos\theta d\phi)^2+(d\theta^2+\sin^2\theta d\phi^2)],
\label{g2a}\ee
where $a\in R^+$ is a dimensionless parameter characterizing the deformation of the 3-sphere. The Euler angles take values as $\theta\in [0,\pi]$, $\phi\in [0,2\pi]$, $\psi\in [0,4\pi]$. The round three-sphere is obtained when $a=1$ and $L$ is its radius.
The volume is,
\be
V_3=2\pi^2 a L^3.
\ee
\label{squasheds3invariants}
The Einstein tensor is,

\be
G_{\psi\psi}={3a^2-4\over 4}\sp G_{\theta\theta}=-{a^2\over 4}\sp G_{\phi\phi}={a^2\over 8}\left[3a^2-5+3(a^2-1)\cos(2\theta)\right],
\ee
\be
G_{\psi\phi}={a\over 4} (3 a^2-4) \cos(\theta).
\ee

The three basic curvature invariants of the squashed sphere metric in (\ref{g2a}) are,
\be
R=2{4-a^2\over L^2} \ , \ R_{\m\n}R^{\m\n}=4{3a^4-8a^2+8\over L^4}\ ,\ R_{\m\n ;\r\s}R^{\m\n ;\r\s}=4{11 a^4-24 a^2+16\over L^4}.
\label{g3}\ee

\section{The FG expansions for the Conformal Fixed Points}\label{PGexpansion}

The aim of this appendix is to express the radial parameter $\epsilon$, appearing in the actions (\ref{ga30.a}),(\ref{ga40.a}), in FG coordinates for each of the cases that appear in sections \ref{taubnut},\ref{taubbolt}. By doing this, we are bringing the two actions into the same coordinate system, and therefore we are able to compare them.

We start of with the metric (\ref{ga11}), and we substitute $r \to \frac{\ell^2}{r}$:

\be
d s^{2}=\frac{\ell^{2}}{r^{2}}\left(\frac{d r^{2}}{\hat{f}(r)}+\frac{\hat{L}^{2}}{4}\left(a^{2} \hat{f}(r)(d \psi+\cos \theta d \phi)^{2}+\hat{g}(r)d\Omega^2\right)\right),
\label{gA1}\ee
where

\be
\hat{f}(r) = \frac{r^2}{\ell^2} f(\frac{\ell^2}{r}) \ , \ \hat{g}(r) = \frac{a^2}{4\ell^2n^2} \left( \ell^4 - \frac{16n^4 r^2}{L^2} \right).
\label{gA2}\ee

In order to go to \textit{FG} coordinates, we need to substitute

\be
\frac{dz}{z} = \frac{dr}{r\sqrt{\hat f(r)}} .
\label{gA3}\ee

All that remains is to examine each case seperately.

\subsection{The Taub-Nut}

In this case the function $\hat{f}$ is,

\be
\hat{f}(r)=\frac{\left(\ell^{2}-4 n^{2} \frac{r}{L}\right)\left(8 \ell^{2} n^{2} \frac{r}{L}\left(2 \frac{r}{L}+1\right)+\ell^{4}-48 n^{4}\left(\frac{r}{L}\right)^{2}\right)}{4 \ell^{4} n^{2} \frac{r}{L}+\ell^{6}}.
\label{gA4}\ee

By using equation (\ref{gA3}) we observe that $z$ can be written in terms of $r$ as follows:

\be
z=\frac{r}{L}\left[1+\frac{r^{2}\left(20 n^{4}-4 n^{2} \ell^{2}\right)}{L^{2} \ell^{4}}+\frac{64 n^{4} r^{3}\left(\ell^{2}-4 n^{2}\right)}{3 L^{3} \ell^{6}}+\mathcal{O}\left(\tilde{r}^{4}\right)\right].
\label{gA5}\ee

By inverting the previous relation, we observe that we can write $\epsilon$ in terms of $z$ as follows:
\be
\epsilon=z+\frac{4 n^{2}\left(\ell^{2}-5 n^{2}\right)}{\ell^{4}} z^{3}+\frac{64 n^{4}\left(4 n^{2}-\ell^{2}\right)}{3 \ell^{6}} z^{4}+\mathcal{O}\left(z^{5}\right).
\label{gA6}\ee

\subsection{The Taub-Bolt}

In this case we have for the function $\hat f$

$$
\hat f(r) = \left( 1 - \frac{r r_0}{\ell^2} \right) \frac{L^2 r_0 \left(r^2 \left(L^2 r_0^2-96 n^4\right)+r \ell ^2 \left(L^2 r_0+16 n^2 r\right)+L^2 \ell ^4\right)-256 n^6 r^3}{L^2 r_0 (L^2 \ell^4 - 16 n^4 r^2)} +
$$
\be
+768n^8r^3 \frac{\ell^2 - r r_0}{L^2 r_0 \ell^4(L^2 \ell^4 - 16 n^4 r^2)},
\label{gA7}\ee
and by using equation (\ref{gA3}) we find for z that
$$
z = {\cal O}\left(r^5\right)+ \left( \frac{r}{L} \right)^4  \frac{\left(16 L^2 n^2 \left(r_0\right){}^2 \left(\ell ^2-6 n^2\right)+L^4 \left(r_0\right){}^4+256 n^6 \left(\ell ^2-3 n^2\right)\right)}{6r_0 \ell ^6} -
$$
\be -\frac{4 n^2 r^3 \left(\ell ^2-5 n^2\right)}{L^3 \ell ^4}+ \frac{r}{L} .\ee

Upon inverting, we obtain for $\epsilon$ the following:
$$
\epsilon =   {\cal O} \left(z^4 \right) + \frac{ \left(16 L^2 n^2 \left(r_0\right){}^2 \left(\ell ^2-6 n^2\right)+L^4 \left(r_0\right){}^4+256 n^6 \left(\ell ^2-3 n^2\right)\right)}{6 L r_0 \ell ^6} z^3  + $$
\be   \frac{4 n^2 \left(\ell ^2-5 n^2\right)}{\ell ^4} z^2 + z .\label{gA8}\ee

\section{Curvature Invariants}\label{Curvatureinvariants}

The aim of this appendix is to express the curvature invariants in terms of the first order formalism functions defined in (\ref{ga46}) - (\ref{ga50}). Recall that for a general metric in a (3+1)-dimensional Einstein-Dilaton (ED) theory the action has the form (\ref{ga1}),
which upon varying with respect to $g,\f$ gives the following equations of motion:

\be
\nabla^\mu \nabla_\mu \f - V'(\f) = 0,
\label{g178}\ee
\be
R_{\mu\nu} - \frac{1}{2} g_{\mu\nu} \left( R - \frac{1}{2} (\partial \f)^2 - V(\f) \right) - \frac{1}{2} \partial_\mu \f \partial_\nu \f = 0.
\label{g179}\ee

We can now construct some of the curvature invariants by using the equations of motion.
By tracing (\ref{g179}) with $g^{\mu \nu}$ we obtain
\be
R = {1\over 2}(\pa \f)^2+2V(\f)= \frac{1}{2} \dot{\f}^2 + 2 V(\f) = \frac{S^2}{2} + 2 V(\f).
\label{g180.a}\ee

Therefore, near an  end point of the flow, $S\to 0$, the Ricci scalar takes the form
\be
R \approx 2V(\f),
\label{g180}\ee
and since $V(\f)$ is analytic everywhere, $R$ is regular when $S \to 0$.

Solving $(\ref{g179})$ for $R^{\mu \nu}$ we obtain
$$
R^{\mu\nu}R_{\mu\nu}=(R-V)^2-{1\over 2}(R-V)(\pa \f)^2+{(\pa\f)^4\over 4},
$$
and using (\ref{g180.a}) we finally obtain
\be
R^{\mu\nu} R_{\mu\nu}  = \frac{1}{4} \left( 3V^2 + \left( S^2 + V \right)^2 \right).
\label{g183}\ee

Therefore, near the end point of the flow, the Ricci square takes the form
\be
R^{\mu\nu} R_{\mu\nu} \approx V^2(\f),
\label{g184}\ee
and since $V(\f)$ is analytic everywhere, $R^{\mu\nu} R_{\mu\nu}$ is regular when $S \to 0$.

Finally, the only curvature invariant that remains that is of interest is the Kretschmann scalar, i.e. the Riemann square:

\be
\mathcal{K} \equiv R_{\alpha \beta; \mu\nu} R^{\alpha \beta; \mu\nu},
\label{g185}\ee
which, upon substituting the metric (\ref{ga3}) gives
\be
\mathcal{K}  =\dot{A_1}^2 \left(8 \dot{A_2}^2+8 \ddot{A_1} + \frac{12}{L^2}e^{2 A_1-4 A_2}\right)+\dot{A_2}^2
\left(16 \ddot{A_2} +\frac{18}{ L^2} e^{2 A_1-4 A_2}-\frac{8}{ L^2} e^{-2 A_2}\right)
\ee
$$
+4 \dot{A_1}^4 -\frac{28}{L^2} e^{2
A_1-4 A_2} \dot{A_2} \dot{A_1} +12 \dot{A_2}^4+4 \ddot{A_1}^2+8
\ddot{A_2}^2+\frac{11}{4L^4} e^{4 A_1-8 A_2}-\frac{6}{ L^4} e^{2 A_1-6 A_2}+\frac{4}{L^4} e^{-4 A_2},
$$
which can be written in the first order formalism as
$$
\mathcal{K} = \frac{1}{16} W_2^2 \left(-4 S W_2'+\frac{9 T_1}{2}-2 T_2\right)+\frac{1}{16} W_1^2 \left(-2 S W_1'+3 T_1+\frac{W_2^2}{2}\right)-\frac{7}{16} T_1 W_1 W_2
$$
\be
+\frac{1}{64} W_1^4+\frac{1}{4} S^2 \left(W_1'\right)^2+\frac{1}{2} S^2 \left(W_2'\right)^2-\frac{3}{8} T_1 T_2+\frac{11 T_1^2}{64}+\frac{1}{4} T_2^2+\frac{3}{64} W_2^4.
\label{g186}\ee

Unlike the previous two, the Kretschmann scalar may or may not blow up near the end point of the flow, depending on the behavior of the functions $W_{1,2}, T_{1,2},S$.

%\newpage

\section{An analysis of the end-points of the flow equations.}\label{extremalpoints}

The aim of this appendix, is to examine the critical points of the first order equations we are solving. This, in turn, shall give us all the possible ways with which an RG flow can start or end. In order to find the
asymptotic behavior of the solutions around such points, we shall consider   points $\f=\f_0$ where $S$ vanishes.
We shall do this by analysing the independent first order equations (\ref{ga51})-(\ref{ga55}),  (\ref{ga54}), which we rewrite here for convenience:

\be
W_2^2+2W_1W_2+T_1-4T_2-4S^2+8V=0,
\label{C1}\ee
\be
4(W_2'-W_1')S+(W_1-W_2)(W_1+2W_2)+4(T_2-T_1)=0,
\label{C2}\ee
\be
SS' - \frac{1}{4} S\left( W_1 + 2 W_2 \right) - V' = 0,
\label{C3}\ee
\be
ST_1'=-\frac{1}{2}(W_1-2W_2)T_1\sp ST_2'= \frac{1}{2} W_2 T_2.
\label{C4}\ee

We are interested in solutions which exhibit the following properties:

\begin{enumerate}
   \item[1)] The function $S$ is always finite and approaches zero as $\f$ approaches the point $\f_0$.
   \item[2)] The curvature invariants are finite. From the analysis done in appendix \ref{Curvatureinvariants}, it becomes clear that the only non-trivial constraint comes from
   the regularity of the Kretschmann scalar.
\end{enumerate}

We have to distinguish  two cases. The first one is the  generic case, and corresponds to a solution around a generic point $\f_0$.  In its the vicinity, the potential $V(\f)$ can be expanded in a regular power series,
\be
V(\f) = V_0 + V_1 (\f - \f_0) + V_2 (\f - \f_0)^2 + \mathcal{O}\left( (\f - \f_0)^3 \right) \sp V_1\not=0.
\label{C5}\ee

The second one corresponds to special points $\f_0$ which are  located at  extrema of the potential. Around such points, $V'(\f_0)=0$  and  the potential $V(\f)$ can be expanded  as,
\be
V(\f) = - \frac{6}{\ell^2} - \frac{\Delta \Delta_-}{2\ell^2} (\f - \f_0)^2 + \mathcal{O} \left( (\f  - \f_0)^3 \right),
\label{C6}\ee
where we have parametrized the expansion in terms of the standard holographic parameters.
In particular, $\Delta_- = 3 - \Delta$ and $\Delta$ is a dimensionless parameter, taking the values $\frac{3}{2} \leq \Delta \leq 3$ when we are at a maximum and $\Delta > 3$ when we are at a minimum, and is associated with the scaling dimension of the perturbing operator around the extremum. $\ell$ is the AdS length scale.

In all the solutions to follow, we shall use the following ansatze for the functions $S,T_1,T_2$\footnote{This behavior was shown to be sufficient to capture all cases in the case of spheres, in \cite{C}.} :
\be
S \approx S_0 x^\gamma \ , \ T_1 \approx C_1 x^{-a} \ , \ T_2 \approx C_2 x^{-b},
\label{C7}\ee
where the constants $\gamma$,$C_1,S_0,C_2$ are subject to the following constraints:
\be
\gamma > 0 \ , \ C_1,S_0,C_2 \neq 0,
\label{C8}\ee
so that $S\to 0$ as $x\to 0$.
We have defined also for convenience
\be
x \equiv \f - \f_0  .
\label{C8.a}\ee

Equation (\ref{C7}) combined with (\ref{C4}) gives the following behavior for the functions $W_1,W_2$:

\be
W_1 \approx 2(a-2b) S_0 x^{\gamma -1} \ , \ W_2 \approx -2b S_0 x^{\gamma-1} .
\label{C9}\ee

As a final remark, we note that we have tacitly assumed that we approach the point $\f_0$ from the right, i.e. $\f > \f_0$. In order to approach it from the left, we simply have to invert the sign of the functions $W_1,W_2$.

\subsection{The generic case, $V_1 \neq 0$}

In this case, the potential has the form (\ref{C5}). Substituting this equation along with (\ref{C7}),(\ref{C9}) into the equations of motion (\ref{C1}) - (\ref{C4}) yields the following leading-order equations:
\be
S_0^2 x^{2\gamma-1} \left( \gamma - \frac{a}{2} + 2b \right) - V_1 +\cdots= 0,
\label{C10}\ee
\be
C_1 x^{-a} - 4 C_2 x^{-b} - 4 x^{2\gamma-2} S_0^2 \left( 2ab -5b^2\right) + 8 V_0+\cdots = 0,
\label{C11}\ee
\be
C_2 x^{-b} - C_1 x^{-a} + (a-b)(a-4b-2\gamma+2) x^{2\gamma-2} S_0^2+\cdots = 0.
\label{C12}\ee

We start by examining equation (\ref{C10}). We observe that since $V_1 \neq 0$, we have two choices:

\begin{enumerate}
   \item[$\bullet$] The first one is to choose $\gamma < \frac{1}{2}$. This implies that the first term is the leading one, and therefore it must vanish on its own.
   \item[$\bullet$] The second case is to choose $\gamma = \frac{1}{2}$. This implies that the two terms have the same order, and therefore they must vanish as a whole.
\end{enumerate}

We proceed to discuss each one in turn.

\subsubsection{$0 < \gamma < \frac{1}{2}$}

As we previously mentioned, this implies that the first term vanishes on its own, therefore we have

\be
\gamma = \frac{a}{2} - 2b .
\label{C13}\ee

Moreover, since $\gamma < \frac{1}{2}$, the three first terms in equation (\ref{C11}) are the leading ones. The two remaining equations therefore become

\be
C_1 x^{-a} - 4 C_2 x^{-b} -4 x^{a-4b-2} S_0^2(2ab-5b^2) + \cdots = 0,
\label{C14}\ee
\be
C_2 x^{-b} - C_1 x^{-a} + 2(a-b) x^{a-4b-2} S_0^2+ \cdots = 0 .
\label{C15}\ee

From the constraint in (\ref{C8}) we observe that in order to avoid setting $C_1 = 0 $ or $C_2 = 0$
we must demand that the first two terms in each equation are either subleading or have the same order as the last one, i.e.

\be
a \leq 2b+1 \ , \ 3b \geq a - 2.
\label{C16}\ee

Thus we have a total of four choices for these two terms:

\begin{enumerate}
   \item[$\bullet$] Both terms are subleading with respect to the last one. This leads to the equations
   \be
   b(2a-5b) = 0 \ , \ a = b,
   \ee
   which are solved by $a = b = 0$. However, this solution is not acceptable since it would imply that $\gamma = 0$.

   \item[$\bullet$] The term $C_1 x^{-a}$ has the same order as the last one. Equation (\ref{C16}) now becomes

   \be
    a = 2b+1 \ , \ b > - 1 \implies \gamma = \frac{1}{2} -b,
   \label{C17}\ee
   while equations (\ref{C14}), (\ref{C15}) become to leading order

   \be
   C_1 = 2(b+1) S_0^2 \ , \ C_1 = 4b S_0^2 (2-b) .
   \ee

   The values of $b$ that satisfy the previous equations are $b = 1, \frac{1}{2}$, which, upon inserting to (\ref{C17}), give values for $\gamma$ that are not positive. Therefore this solution is not acceptable.

   \item[$\bullet$] The term $C_2 x^{-b}$ has the same order as the last one. Equation (\ref{C16}) now becomes

   \be
   3b = a -2 , a < -1 \implies \gamma = \frac{8-a}{6} > \frac{3}{2}\;.
   \ee
   As we have assumed that $\gamma<{1\over 2}$, this solution is not acceptable.

   \item[$\bullet$] Finally, the two terms $C_1 x^{-a}$, $C_2^{-b}$ have the same order as the last one. Equation (\ref{C16}) now becomes

   \be
   a = b = -1,
   \ee
   which, upon insterted to (\ref{C13}), yields a negative value for $\gamma$ that is not acceptable.
\end{enumerate}

We therefore conclude that there are no acceptable solutions for $0 < \gamma < \frac{1}{2}$.

\subsubsection{ $\gamma = \frac{1}{2}$} \label{ircases}

In this case equations (\ref{C1}) - (\ref{C3}) become to leading order

\be
S_0^2 = \frac{V_1}{\frac{1-a}{2} +2b},
\label{C18}\ee
\be
C_1 x^{-a}- 4 C_2 x^{-b} - \frac{4S_0^2}{x}(2ab-5b^2)+ \cdots = 0,
\label{C19}\ee
\be
C_2 x^{-b} - C_1 x^{-a} + (a-b)(a-4b+1) \frac{S_0^2}{x}+ \cdots = 0.
\label{C20}\ee

Similarly to the previous subcase, in order to avoid setting $C_1 = 0$ or $C_2 = 0$ we must demand that the first two terms of the last two equations are either subleading or have the same order as the last one, i.e. we must demand that

\be
a,b \leq 1.
\label{C21}\ee

Therefore, we have a total of four options for these two terms:

\begin{enumerate}
   \item[$\bullet$] Both terms are subleading with respect to the last one. This leads to the following pair of equations:

   \be
   b(2a-5b) = 0 \ , \ (a-b)(a+4b-1) = 0,
   \ee
   which have the following three pairs of solutions:
   \be
   a = 0 , b = 0,
   \ee
   \be
    a = -1 , b = 0,
   \ee
   \be
   a = \frac{5}{3}  \ , \ b = \frac{2}{3}.
   \ee

   Out of the previous three, only the first two satisfy the imposed conditions in (\ref{C21}). Therefore, the two possible solutions are

   \begin{enumerate}
      \item[(1)] The $\f$-Bounce:
      \be
      a = b = 0 \ ,\ \gamma = \frac{1}{2} \ , \  S_0^2 = 2 V_1.
      \label{C22}\ee

      \item[(2)] The \textit{Bolt} IR endpoint:
      \be
      a = -1 \ , \ b = 0 \ , \ \gamma = \frac{1}{2} \ , \ S_0^2 = V_1.
      \label{C23}\ee
   \end{enumerate}

   \item[$\bullet$] The term $C_1 x^{-a}$ has the same order as the last term in (\ref{C19}). This is equivalent to setting $a =1 , b < 1$. Equations (\ref{C19}), (\ref{C20}) now yield

   \be
    C_1 = 2 S_0^2 (1-b)(1-2b) \ , \ C_1 = 4S_0^2( 2b-5b^2),
   \ee
   which have the following pairs of solutions:

   \be
   b = \frac{1}{4} \ , \ C_1 = \frac{3}{4} S_0^2,
   \ee
   \be
   b = \frac{1}{3} \ , \ C_1 = \frac{4}{9} S_0^2.
   \ee

   However, both these solutions are singular. Specifically, by inserting the leading power of the first order formalism functions into equation (\ref{g186}) yields the following:

   \be
    b = \frac{1}{4} : \mathcal{K} \approx \frac{159 S_0^4}{256 x^2},
   \ee
   \be
   b = \frac{1}{3} : \mathcal{K} \approx \frac{32 S_0^4}{108 x^2} .
   \ee

   From the previous two equations it becomes clear that the Kretshmann scalar diverges, and therefore these solutions are singular. Therefore none of these  are of interest.

   \item[$\bullet$] The term $C_2 x^{-b}$ has the same order as the last term in (\ref{C19}). This is equivalent to setting $b = 1, a < 1$. Equations (\ref{C19}),(\ref{C20}) now yield

   \be
    C_2 = (3-a)(a-1) S_0^2 \ , \ C_2 = (5-2a) S_0^2,
    \ee
   which give the following pairs of solutions:
   \be
    a = 2 \ , \ C_2 = S_0^2,
   \ee
   \be
   a = 4 \ , \ C_2 = - 3 S_0^2,
   \ee
   which are both unacceptable, since we assumed that $ a < 1$.

   \item[$\bullet$] Finally, the two terms $C_1 x^{-a}, C_2 x^{-b}$ both have the same order as the last term in (\ref{C19}). This is equivalent to setting $a = b = 1$. Equations (\ref{C19}),(\ref{C20}) now yield

   \be
   C_2 = C_1 = 2 V_1.
   \label{C24}\ee

   Therefore, the only  possible solution is

   \begin{enumerate}
      \item[(3)] The \textit{Nut} IR endpoint:
      \be
      a = b = 1 \ , \ \gamma = \frac{1}{2} \ , \ S_0^2 = \frac{V_1}{2} \ , \ C_1 = C_2 = 2 V_1.
      \label{C25}\ee
   \end{enumerate}
\end{enumerate}

\subsubsection{Conslusion}

Overall, we have found that there are three, regular, asymptotic solutions to the equations of motion, near a generic end-point of the flow. These solutions correspond either to standard IR endpoints or bounces\footnote{Bounces are points where $S$ vanishes and changes sign without the flow ending. At such a point the scalar flow inverts its direction.
 They are singular points of the first order equations but are absolutely regular points of the second order equations. They were studied first in detail in \cite{exotic}.}.
 We collect them here:
  \be
S \approx S_0 x^\gamma \ , \ T_1 \approx C_1 x^{-a} \ , \ T_2 \approx C_2 x^{-b}\sp
W_1 \approx 2(a-2b) S_0 x^{\gamma -1} \ , \ W_2 \approx -2b S_0 x^{\gamma-1} .
\label{C9a}\ee

\begin{enumerate}

\item  The \textit{Bolt} endpoint that appears in the Taub-Bolt solution.
  \be
      a = -1 \ , \ b = 0 \ , \ \gamma = \frac{1}{2} \ , \ S_0^2 = V_1.
      \label{C23a}\ee

It was studied in the context of CFTs defined on a squashed 3-sphere,\cite{Myers},\cite{Hartnoll}, and was analysed in section \ref{conffixedpoints}.
\item
The     \textit{Nut} end-point that appears in the Taub-Nut solution.
 \be
      a = b = 1 \ , \ \gamma = \frac{1}{2} \ , \ S_0^2 = \frac{V_1}{2} \ , \ C_1 = C_2 = 2 V_1.
      \label{C25a}\ee

It was studied in the context of CFTs defined on a squashed 3-sphere,\cite{Myers},\cite{Hartnoll}, and was analysed in section \ref{conffixedpoints}.

 \item The bounce.
  \be
      a = b = 0 \ ,\ \gamma = \frac{1}{2} \ , \  S_0^2 = 2 V_1.
      \label{C22a}\ee

 The bounce solution is not an IR-end-point but a place where the $\f$ flow changes direction.

\end{enumerate}

These solutions shall be studied further in Appendix \ref{IR-endpoints}.

\subsection{Analysis near the extrema of the potential, $V_1 = 0$}

In this case the potential has the form (\ref{C6}). The equations of motion (\ref{C1}) - (\ref{C3}) now become to leading order
\be
S_0^2 \left( \gamma - \frac{a}{2} +2b \right)x^{2\gamma-1} + \frac{\Delta \Delta_-}{\ell^2} x +\cdots= 0,
\label{C31}\ee
\be
C_1 x^{-a} - 4 C_2 x^{-b} - 4 x^{2\gamma-2} S_0^2 (2ab-5b^2) - \frac{48}{\ell^2} +\cdots= 0,
\label{C32}\ee
\be
C_2 x^{-b} - C_1 x^{-a} + (a-b)(2+a-4b-2\gamma) x^{2\gamma-2} S_0^2 +\cdots= 0.
\label{C33}\ee

We start by examining equation (\ref{C31}). Since $\Delta,\Delta_- \neq 0$, the only choice that does not contradict the previous statement or does not coincide with any of the previous cases is to choose

\be
 \gamma = 1.
 \ee

Then, the previous set of equations becomes
\be
S_0^2 = \frac{-\Delta \Delta_-}{\ell^2(1 + 2b - \frac{a}{2})},
\label{C34}\ee
\be
C_1 x^{-a} - 4 C_2 x^{-b} - 4S_0^2 (2ab-5b^2) - \frac{48}{\ell^2} +\cdots= 0,
\label{C35}\ee
\be
C_2 x^{-b} - C_1 x^{-a} + (a-b)(a-4b) S_0^2 +\cdots = 0.
\label{C36}\ee

In order to avoid setting $C_1 = 0$ or $C_2 = 0$, we must demand that the first two terms in (\ref{C35}),(\ref{C36}) must either be subleading or have the same order as the rest. This implies that

\be
 a ,b \leq 0.
\ee

Therefore we have a total of four options:

\begin{enumerate}
   \item[$\bullet$] Both terms $C_1 x^{-a}$, $C_2 x^{-b}$, are subleading with respect to the rest of the terms in (\ref{C35}),(\ref{C36}).
   This leads to

   \be
   \frac{2ab-5b^2}{1 - \frac{a}{2}+2b} = \frac{12}{\Delta \Delta_-} \ , \ (a-b)(a-4b) = 0,
   \label{C37}\ee
   which have the following pairs of solutions:
   \be
    a = b = - \frac{2}{\Delta} \ , \ S_0^2 = \frac{\Delta^2}{\ell^2},
    \label{x1}\ee
   \be
   a = b = - \frac{2}{\Delta_-} \ , \ S_0^2 = \frac{\Delta_-^2}{\ell^2} \ , \ \frac{3}{2} \leq \Delta < 3,
   \label{x2}\ee
   \be a = 4b = - \frac{8}{\sqrt{\Delta \Delta_-}} \ , \ S_0^2 = \frac{-\Delta \Delta_-}{\ell^2} \ , \ \frac{3}{2} \leq \Delta < 3.
   \label{x3}\ee

   However, the last case in (\ref{x3}) implies that $S_0$ is imaginary which is not acceptable.
    Therefore, we remain with two solutions:

   \begin{enumerate}
      \item[(1)] The $(+)$-branch solution \footnote{This is similar to the $(+)$-branch found also in \cite{C}}:
      \be
      \gamma = 1 , a = b = - \frac{2}{\Delta} , S_0^2  =\frac{\Delta^2}{\ell^2}.
      \label{C38}\ee
      \item[(2)] The $(-)$-branch solution \footnote{This is similar to the $(-)$-branch found also in \cite{C}}:
      \be
      \gamma = 1 , a = b = - \frac{2}{\Delta_-} , S_0^2 = \frac{\Delta_-^2}{\ell^2}.
      \label{C39}\ee
   \end{enumerate}

   \item[$\bullet$] The term $C_1 x^{-a}$ has the same order as the last term in (\ref{C35}). This is equivalent to setting $a = 0, b < 0$. This leads to

   \be
   \frac{b^2}{1+2b} = -\frac{2}{\Delta \Delta _-} \ , \ C_1 = \frac{8}{\ell^2},
   \label{C40}\ee
   which give the following pairs of solutions:
   \be
    b = -\frac{1}{1 + \sqrt{\frac{(\Delta-1)(\Delta-2)}{2}}} \ , \ S_0^2 = \frac{2}{\ell^2 b^2} \ , \ \Delta \geq 2,
   \ee
   \be
    b = \frac{1}{-1+\sqrt{\frac{(\Delta-1)(\Delta-2)}{2}}}\ , \ S_0^2 = \frac{2}{\ell^2 b^2} \ , \ 2 \leq \Delta < 3.
   \ee

   Therefore, the two possible solutions are

   \begin{enumerate}
      \item[(3)] The $\nu_+$ solution:
      \be
      \gamma = 1 , a = 0 , b = -\nu_+ \equiv -\frac{1}{\sqrt{\frac{(\Delta-1)(\Delta-2)}{2}} +1}  , S_0^2 = \frac{2}{\ell^2 b^2} , \Delta \geq 2.
      \label{C41}\ee
      \item[(4)] The $\nu_-$ solution:
      \be
      \gamma = 1 , a = 0 , b =- \nu_- \equiv \frac{1}{\sqrt{\frac{(\Delta-1)(\Delta-2)}{2}} -1}  , S_0^2 = \frac{2}{\ell^2 b^2} , 2 \leq \Delta < 3.
      \label{C42}\ee
   \end{enumerate}

   \item[$\bullet$] The term $C_2 x^{-b}$ has the same order as the last term in (\ref{C35}). This is equivalent to setting $b = 0 , a < 0$. This leads to

   \be
   \frac{a^2}{1- \frac{a}{2}} =  -\frac{12}{\Delta \Delta_-} \ , \ C_2 = -a^2 S_0^2.
   \ee

   The previous equation yields the following solutions:
   \be
    a = \frac{3+ \sqrt{9-12\Delta \Delta_-}}{\Delta \Delta_-}  \ , \  S_0^2 = \frac{12}{\ell^2 a^2} \ , \ C_2 = - \frac{12}{\ell^2}\ , \
   \label{C43.a}\ee
   \be
    a = \frac{3- \sqrt{9-12\Delta \Delta_-}}{\Delta \Delta_-}  \ , \  S_0^2 = \frac{12}{\ell^2 a^2} \ , \ C_2 = - \frac{12}{\ell^2}.
   \ee

   However, only the first of these solutions is acceptable. To see why, note that for the second case, we can rewrite $a$ as,
   \be
   a =  \frac{12}{3 + \sqrt{9-12 \Delta \Delta_-} },
   \ee
   which is positive, and therefore contradicts the constraint $ a < 0$. For the first one, we observe that in order for $a$ to be negative, $\Delta \Delta_-$ must be negative. This, in turn, implies that $\Delta_- < 0 \implies \Delta > 3$.

   Therefore the  possible solution is

   \begin{enumerate}
      \item[(5)] The $B_3$ solution:
      \be
      \gamma = 1, b = 0 ,  a =  -\frac{3+ \sqrt{9+12\Delta \Delta_-}}{\Delta \Delta_-}  \ , \  S_0^2 = \frac{12}{\ell^2 a^2} \ , \ C_2 = - \frac{12}{\ell^2}\ , \ \Delta > 3.
      \label{C43}\ee
   \end{enumerate}

 \item[$\bullet$] Finally, both
 terms $C_1 x^{-a},C_2 x^{-b}$ have the same order as the last term in (\ref{C35}). This is equivalent to setting $a = b = 0$. This leads to

   \be
    C_1 = C_2 \ , \ C_1 - 4 C_2 = \frac{48}{\ell^2} \implies C_1 = C_2 = - \frac{16}{\ell^2}.
   \label{CC43}\ee

   However, it can be proved that this solution corresponds to $W_1 = W_2 = 0 , T_1 = T_2 = - \frac{16}{\ell^2} , S = \sqrt{2(V + \frac{6}{\ell^2})} $, which is of no interest.
\end{enumerate}

{In general, the previous solutions can be obtained by assuming the following ansatz for the functions $W_{1,2},S$:
\be
W_{1,2} = \sum_{n=0}^\infty W_{(1,2),n} x^n \ , \ S = x\sum_{n=0} S_n x^n \ .
\label{gb1}\ee

By substituting the previous expansions, along with the potential found in (\ref{C6})} {into the equations of motion } (\ref{num1})-(\ref{num3}){ , we obtain to leading order in $x$ the following:

\be
S_0^2 -\frac{\Delta \Delta_-}{\ell^2} - \frac{1}{4} S_0 (W_{(1)0} + 2 W_{(2)0}) + \ldots = 0,
\ee
\be
(W_{(1)0} + 2 W_{(2)0})(-\frac{48}{\ell^2} + W_{(1)0}^2 + 2 W_{(2)0}^2) + \ldots = 0,
\ee
\be
(W_{(1)0}-2W_{(2)0})(- \frac{48}{\ell^2} + W_{(1)0}^2 + 3 W_{(1)0}W_{(2)0} - W_{(2)0}^2) + \ldots = 0,
\ee
from which we can obtain four classes of solutions: }
\begin{enumerate}
   {\item[1.] The first one corresponds to the choice $W_{(1)0} = W_{(2)0} = \frac{4}{\ell}, S_0 = \frac{\Delta_{\pm}}{\ell}$. These solutions correspond to the (+) and (-) branches that were found in} (\ref{C38}),(\ref{C39}).
   {\item[2.] The second one corresponds to the choice $W_{(1)0} = 4\sqrt{2}, W_{(2)0} = 2 \sqrt{2}, S_0 = -\frac{\sqrt{2}}{\ell b}$, with $b$ being given by }(\ref{C41}),(\ref{C42}) { . These solutions correspond to the $\nu_{\pm}$ solutions that were found in } (\ref{C41}),(\ref{C42}).
   {\item[3.] The third one corresponds to the choice $W_{(1)0} = 4\sqrt{3}, W_{(2)0} = 0, S_0 = - \frac{2\sqrt{3}}{\ell a}$, with $a$ being given by} (\ref{C43.a}). {This solution corresponds to the $B_3$ solution that were found in } (\ref{C43}).
   {\item[4.] Finally, the fourth case corresponds to the choice $W_{(1)0} = W_{(2)0} = 0$, $S_0 = \frac{\sqrt{\Delta \Delta_-}}{\ell}$. This solution corresponds to the solution } (\ref{CC43}).
\end{enumerate}

We have effectively determined all the possible leading asymptotics to the equations of motion for points $\f_0$ near the extrema of the potential. To recapitulate, there are 5 possible asymptotics:

\begin{enumerate}
\item[(1)] The standard $(+)$ branch solution \:
   \be
   a = b = - \frac{2}{\Delta} \sp S_0^2 = \frac{\Delta^2}{\ell^2}.
   \label{g123}\ee
\item[(2)] The standard $(-)$ branch solution :
\be
a = b = - \frac{2}{\Delta_-} \sp S_0^2 = \frac{\Delta^2_-}{\ell^2} \sp \frac{3}{2} \leq \Delta < 3,
\label{g124}\ee
and three non-standard asymptotics:

\item[(3)] The $\nu_+$ solution:
\be
a = 0 \sp b = -\nu_+ \equiv -\frac{ 1}{\sqrt{\frac{(\Delta-1)(\Delta-2)}{2}}+ 1  } \sp S_0^2 = \frac{2}{\ell^2 b^2}  \sp \Delta \geq 2.
\label{g125}\ee
\item[(4)] The $\nu_-$ solution:
\be
a = 0 \sp b = -\nu_- \equiv \frac{1}{\sqrt{\frac{(\Delta-1)(\Delta-2)}{2}}- 1  } \sp S_0^2 = \frac{2}{\ell^2 b^2}  \sp 2 \leq \Delta < 3.
\label{g126}\ee
\item[(5)] The $B_3$ solution:
\be
b = 0 \sp a = \frac{3+\sqrt{9-12\Delta \Delta_-}}{\Delta \Delta_-} \sp S_0^2 = \frac{12}{\ell^2 a^2} ,  \Delta > 3.
\label{g127}\ee
\end{enumerate}

Although we found 5 distinct solutions near extrema, not all of them exist for the same $\Delta$. Specifically, they can be split into two categories: the ones that exist at the minima of the potential, and the ones that exist at the maxima of the potential:

\subsection*{Solutions near minima of the potential}

The solutions are:

\begin{itemize}

\item The $(+)$ branch solution that corresponds to a UV Fixed point perturbed by the vev of an irrlevant operator.

\item The $\nu_+$ solution which as we shall see later on is not asymptotically 3-dimensional.

\item The $B_3$ solution which as we shall see later on is not asymptotically 3-dimensional.

\end{itemize}

\subsection*{Solutions near the maxima of the potential }

The solutions are

\begin{itemize}

\item The $(-)$ branch solution that corresponds to a UV fixed point perturbed by a relevant operator.

\item The $(+)$ branch  solution that corresponds to a UV fixed point perturbed by the vev of a  relevant operator.

\item The $\nu_+$ and  $\nu_-$ solutions. As we shall see, the $\nu_+$ and $\nu_-$ solutions have a near-boundary metric that is \textit{not} 3 dimensional.

\end{itemize}

\subsection{Appendix synopsis}

In this appendix we have analysed the singular points of the flow equations and identified the leading asymptotics of the solutions.
Such points exist when $S=\dot\f$ vanishes.
These solutions can be interpreted as expansions near UV/IR fixed points at which an RG flow can begin/end. Specifically, we can classify them as follows:

\subsection*{IR endpoints}

We have  3 generic cases, namely:

\begin{enumerate}
   \item[(1)]

The \textit{Nut} IR endpoint. This case corresponds to $a =b = 1$ and exists for both the squashed and the regular 3-sphere. Near this point, the functions have the following behavior:

   \be
   S \sim \sqrt{x} \ , \ W_1, W_2 \sim \frac{1}{\sqrt{x}} \ , \ T_1,T_2 \sim \frac{1}{x}.
   \ee

  This asymptotic describes the vanishing  of the whole squashed $S^3$.

   \item[(2)] The \textit{Bolt} IR endpoint. This case corresponds to $a = -1, b = 0$ and exists only for the squashed 3-sphere. Near this point, the functions have the following behavior:

   \be
   S \sim \sqrt{x} \ , \ W_1 \sim \frac{1}{\sqrt{x}} \ , \ W_2 \sim \sqrt{x} \ , \ T_1 \sim x \ , \ T_2 \sim \text{const}.
   \ee

   This asymptotic describes the vanishing of the nontrivial  $S^1$ inside the squashed $S^3$.

   \item[(3)] Finally, the last case is the $\f-$ bounce. This case corresponds to $ a = b = 0 $ and exists for both the squashed and the regular 3-sphere.
   This is not an IR endpoint, rather a bounce in the scalar $\f$, but appears as an end in the first order formalism. The solutions of the second order equations, do not however stop, but continue smoothly further, \cite{exotic}.
   Near this point, the functions have the following behavior:

   \be
   S \sim \sqrt{x} \ , \ W_1,W_2,T_1,T_2 \sim \text{const}.
   \ee

   This case corresponds to a point at which the RG flow inverts its direction in $\f$-space.
\end{enumerate}

\subsection*{UV fixed points}

There are 2 cases that are of interest, namely:

\begin{enumerate}
   \item[(1)] The (-)-branch. This solution exists only around maxima of the potential, and exists for both the squashed and the regular 3-sphere.  Around this point, the functions have the following behavior:

   \be
   S \sim x \ , \ W_1,W_2 \sim \text{const} \ , \ T_1 ,T_2 \sim x^{2/\Delta_-}.
   \ee

   This behavior describes RG flows that start from a maxima of the potential (UV fixed point).
   \item[(2)] The (+)-branch. This solution exists around maxima as well as minima of the potential, and exists for both the squashed and the regular 3-sphere.  Around this point, the functions have the following behavior:

   \be
   S \sim x \ , \ W_1,W_2 \sim \text{const} \ , \ T_1 ,T_2 \sim x^{2/\Delta}.
   \ee

   This behavior describes RG flows that either start from a maximum of the potential (UV fixed point) or arrive at the  minimum. In both cases, we have an AdS boundary.

\end{enumerate}

\subsection*{Exotic solutions}

We have also found the three asymptotics $\nu_{\pm},B_3$. The $\nu_+$ and $\nu_-$ solutions exist around maxima of the potential, while $\nu_+$ and $B_3$ exist around minima of the potential.
For all such solutions, the structure of the asymptotics corresponds to a boundary, however, this is not a standard FG boundary.

\newpage

\section{Perturbation theory around generic flow endpoints} \label{IR-endpoints}

In this appendix, we shall set up the perturbation theory around the leading-order solutions found in the previous appendix, in order to justify whether these correspond to bona-fide asymptotic solutions.
The relevant point $\f_0$ is a generic point, not situated at an extremum of the potential.
 As we found in \ref{ircases}, there are three possibilities:

\begin{enumerate}
\item[(1)] $\f$-Bounce: This corresponds to the case
$$
a = b = 0,
$$
and desribes a flow that inverts its direction when it reaches the point $\f_0$.
\item[(2)] Bolt-IR endpoint: This corresponds to the case
$$
a = -1 \ , \ b = 0,
$$
and describes the squashing of the circle $S^1$.
\item[(3)] Nut-IR endpoint: This corresponds to the case
$$
a = b = 1,
$$
and describes the squashing of the three sphere.
\end{enumerate}

We shall proceed to discuss each case in turn.

\subsection{$\f$-Bounce:}

We parametrize the series solutions as

\be
S = \sqrt{x} \sum_{n=1}^{\infty}S_n x^{n\over 2},
\label{C26}\ee
\be
W_1 = \frac{1}{\sqrt{x}} \sum_{n=0}^{\infty} W_{(1)n}x^{n\over 2}\sp
W_2 = \frac{1}{\sqrt{x}} \sum_{n=0}^{\infty} W_{(2)n}x^{n\over 2},
\label{C27}\ee
\be
T_1 = \frac{1}{x}  \sum_{n=0}^{\infty} T_{(1)n}x^{n\over 2}\sp
T_2 = \frac{1}{x} \sum_{n=0}^{\infty} T_{(2)n}x^{n\over 2}.
\label{C29}\ee

By substituting these expansions into the equations of motion (\ref{ga51}),(\ref{ga53}) - (\ref{ga54}) and solving recursively for the first few coefficients, we find :

\be
S_0^2 = 2 V_1 , S_1 = \frac{4 T_{(2)1} - T_{(1)1} - 8 V_0 + 3 W_{(2)1}^2}{12 W_{(2)1}},
\label{g197}\ee
$$ S_2 = -\frac{-5 T_{(1)1}^2+40 T_{(1)1} T_{(2)1}-80 T_{(1)1} V_0-6 T_{(1)} W_{(2)1}^2-80 T_{(2)1}^2}{288 S_0 W_{(2)1}^2}$$
$$ - \frac{320 T_{(2)1} V_0+24 T_{(2)1} W_{(2)1}^2+320 V_0^2+192 V_0 W_{(2)1}^2+288 V_2 W_{(2)1}^2+45 W_{(2)1}^4}{288 S_0 W_{(2)1}^2} ,$$

\be
W_{(1)0} = 0 , W_{(1)1} = \frac{4T_{(2)1} - T_{(1)1} - 8 V_0 - W_{(2)1}^2}{2 W_{(2)1}},
\label{g198}\ee
$$ W_{(1)2} = -\frac{-T_{(1)1}^2+8 T_{(1)1} T_{(2)1}-16 T_{(1)1} V_0+10 T_{(1)1} W_{(2)1}^2-16 T_{(2)1}^2+64 T_{(2)1} V_0}{8 S_0 W_{(2)1}^2} $$
$$ + \frac{8 T_{(2)1} W_{(2)1}^2 +64 V_0^2-16 V_0 W_{(2)1}^2-3 W_{(2)1}^4}{8 S_0 W_{(2)1}^2},$$

\be
W_{(2)0} = 0 , W_{(2)1} = \text{arbitrary} ,  W_{(2)2} = -\frac{-3 T_{(1)1}+4 T_{(2)1}-8 V_0-3 W_{(2)1}^2}{4 S_0},
\label{g199}\ee

\be
T_{(1)0} = 0 , T_{(1)1} = \text{arbitrary} , T_{(1)2} = \frac{T_{(1)1} \left(T_{(1)1}-4 T_{(2)1}+8 V_0+5 W_{(2)1}^2\right)}{2 S_0 W_{(2)1}},
\label{g200}\ee
$$ T_{(1)3} = -\frac{8 T_{(1)1}^2 T_{(2)1}-16 T_{(1)1}^2 V_0-32 T_{(1)1}^2 W_{(2)1}^2-T_{(1)1}^3-16 T_{(1)1} T_{(2)1}^2+64 T_{(1)1} T_{(2)1} V_0}{12 S_0^2 W_{(2)1}^2}$$
$$ + \frac{-80 T_{(1)1} T_{(2)1} W_{(2)1}^2+64 T_{(1)1} V_0^2+136 T_{(1)1} V_0 W_{(2)1}^2+45 T_{(1)1} W_{(2)1}^4}{12 S_0^2 W_{(2)1}^2},$$

\be
T_{(2)0} = 0 , T_{(2)1} = \text{arbitrary} , T_{(2)2} = \frac{T_{(2)1} W_{(2)1}}{S_0} ,
\label{g201}\ee
$$ T_{(2)3} = -\frac{-5 T_{(1)1} T_{(2)1}+8 T_{(2)1}^2-16 T_{(2)1} V_0-9 T_{(2)1} W_{(2)1}^2}{12 S_0^2} .$$

This perturbative solution  has four integration constants, namely $W_{(2)1}$, $T_{(2)1}$, $T_{(1)2}$  and $\f_0$. Therefore, it is the general solution of the first order system.

Since $ S = \dot{\f}$, we have

\be
\dot{\f} = S_0 \sqrt{\f - \f_0} \left( 1 + \mathcal{O}(\sqrt{\f - \f_0}) \right),
\label{g202}\ee
so that
\be
\f - \f_0 \simeq \frac{1}{2} V_1 (u - u_0)^2+\ldots,
\label{g203}\ee
where $u_0$ is an integration constant.

From the defining relations in  (\ref{ga46}),(\ref{ga47}) we observe that we can write the functions $A_1,A_2$ as a power series in u. Specifically, we have

\be
A_1 = A_1^c - \frac{W_{(1)1}}{4} (u - u_0) - \frac{W_{(1)2}S_0}{16} (u-u_0)^2 + \mathcal{O} \left( (u - u_0)^3 \right),
\label{g204}\ee
\be
A_2 = A_2^c - \frac{W_{(2)1}}{4} (u-u_0) - \frac{W_{(2)2} S_0}{16} (u-u_0)^2 + \mathcal{O} \left( (u - u_0)^3 \right),
\label{g205}\ee
where $A_1^c,A_2^c$ are integration constants.

The  Kretschmann scalar
\be
\mathcal{K} = \frac{W_{(2)1}^2 \left(45 T_{(1)1}^2+48 T_{(1)1} (4 V_0-3 T_{(2)1})+16 \left(3 T_{(2)1}^2-4 T_{(2)1} V_0+4 V_0^2\right)\right)}{64 W_{(2)1}^2}
\label{g206}\ee
$$ + \frac{W_{(2)1}^4 (45 T_{(1)1}-24 T_{(2)1}+16 V_0)+3 T_{(1)1} (T_{(1)1}-4 T_{(2)1}+8 V_0)^2+3 W_{(2)1}^6}{64 W_{(2)1}^2} + \mathcal{O}(u-u_0), $$
is regular. \\
The scale factors are therefore,
\be
e^{2A_2} = \frac{4}{L^2 T_{(2)1}} \left(1 + \mathcal{O}(u-u_0) \right),
\label{g207}\ee
\be
e^{2A_1} = \frac{4 T_{(1)1}}{L^2 T_{(2)1}^2} \left(1 + \mathcal{O}(u-u_0) \right),
\label{g208}\ee
and the metric is,
\be
ds^2 \approx du^2 + \frac{4}{T^2_{(2)1}} \left( {T_{(1)1}} \left( d\psi + \cos\theta d\phi \right)^2 + {T_{(2)1}}d\Omega^2 \right).
\label{g209}\ee

Some remarks are in order.

\begin{enumerate}
   \item[$\bullet$] We have $\ddot{\f}(\f = \f_0) = V_1$ and the space-time does not end here. Hence this is a bounce (a change of direction for $\f$).

   \item[$\bullet$] As we can observe from (\ref{g197}), this case can admit two solutions depending on the sign we choose for $S_0$.
   The full solution at that point is obtained by sowing the two solutions on opposite sides of $\f_0$ using both signs of $S_0$.
\end{enumerate}

\subsection{Bolt-like IR endpoint:}

By substituting the same expansions,  (\ref{C26}) - (\ref{C29}) into the equations of motion (\ref{ga51}),(\ref{ga53}) - (\ref{ga54}) and solving recursively, we find:

\be
S_{2n+1} = W_{(1)2n+1} = W_{(2)2n+1} = T_{(1)2n+1} = T_{(2)2n+1} = 0 , n = 0,1,
\ee

\be
S_0^2 = V_1 , S_2 =  \frac{3 V_0 + 6 V_2 - T_{(2)2}}{8 S_0},
\label{g215}\ee
$$ S_4 = \frac{192 V_1^2 + 16 V_1 T_{(1)2} + 47 T_{(2)2}^2 - 154 T_{(2)2} V_0 + 135 V_0^2 + 44 T_{(2)2} V_2}{1152 S_0 V_1} $$
$$ - \frac{ 132 V_0 V_2 + 292 V_2^2}{1152 S_0 V_1},$$
\be
W_{(1)0} = -2 S_0 , W_{(1)2} = \frac{3 T_{(2)2} - V_0 - 2 V_2}{4 S_0},
\label{g216}\ee
$$ W_{(1)4} = -\frac{176 S_0^2 T_{(1)4}-192 V_1^2  +139 T_{(2)2}^2-434 T_{(2)2} V_0+124 T_{(2)2} V_2+243 V_0^2}{576 S_0 V_1}$$
$$ + \frac{84 V_0 V_2+116 V_2^2}{576 S_0 V_1}, $$
\be
W_{(2)0} = 0 , W_{(2)2} = \frac{-T_{(2)2} + 2V_0}{S_0},
\label{g217}\ee
$$ W_{(2)4} = -\frac{-2 S_0^2 T_{(1)4}-8 V_1^2 -3 T_{(2)2}^2+9 T_{(2)2} V_0-2 T_{(2)2} V_2-6 V_0^2+4 V_0 V_2}{8 S_0 V_1}, $$

\be
T_{(1)0} = T_{(1)2} = 0 , T_{(1)4} = \text{arbitrary} ,  T_{(1)6} = \frac{7 T_{(1)4} V_2 - 5 T_{(1)4}T_{(2)3} - 2 T_{(1)4} V_2}{4 V_1},
\label{g218}\ee

\be
T_{(2)0} = 0 , T_{(2)2} = \text{arbitrary} , T_{(2)4} = \frac{(2V_0 - T_{(2)2})T_{(2)2}}{2 V_1},
\label{g219}\ee
$$ T_{(2)6} = -\frac{-S_0^2 T_{(1)4} T_{(2)2}-4 V_1^2 T_{(2)2}+10 T_{(2)2}^2 V_0-4 T_{(2)2}^2 V_2-3 T_{(2)2}^3}{16 V_1^4}$$
$$+ \frac{ T_{(2)2} V_0^2- T_{(2)2} V_0 V_2}{2 V_1^4}.$$

So far, this perturbative solution has three integration constants, namely
$T_{(2)2}$, $T_{(1)4}$, and $\f_0$.

Since $S = \dot{\f}$, we obtain to leading order
\be
\dot{\f} = S_0 \sqrt{\f - \f_0} \left( 1 + \mathcal{O}(\f - \f_0) \right),
\label{g220}\ee
so that
\be
\f - \f_0 = \frac{1}{4} V_1 (u-u_0)^2+\ldots,
\label{g221}\ee
where $u_0$ is an integration constant.

From the defining relations in  (\ref{ga46}),(\ref{ga47}) we obtain the functions $A_1,A_2$:
\be
A_1 = A_1^c + \ln(u-u_0) - \frac{S_0 W_{(1)2}}{16} (u - u_0)^2 + \mathcal{O}\left( (u-u_0)^4 \right),
\label{g222}\ee
\be
A_2 = A_2^c - \frac{S_0 W_{(2)2}}{8} (u - u_0)^2  - \frac{S_0 V_1 W_{(2)4}}{32}(u-u_0)^4 + \mathcal{O}\left( (u-u_0)^6 \right),
\label{g223}\ee
where $A_1,A_2$ are integration constants.

The scale factors $e^{2A_i}$ become
\be
e^{2A_2} = \frac{4}{L^2 T_{(2)2}} \left( 1 + \mathcal{O}\left( (u-u_0)^2 \right) \right).
\label{g224}\ee
\be
e^{2A_1} = \frac{T_{(1)4} V_1}{L^2 T_{(2)2}^2} (u-u_0)^2 \left(  1 + \mathcal{O}\left( (u-u_0)^2 \right) \right).
\label{g225}\ee

The asymptotic   metric is,

\be
ds^2 \approx du^2 + \frac{4}{T_{(2)2}} \left( \frac{T_{(1)4} V_1}{4 T_{(2)2}} (u-u_0)^2\left(d\psi + \cos\theta d\phi \right)^2 + d\Omega^2 \right).
\label{g226}\ee

Since $\psi \in [0,4\pi]$, we must choose

\be
\frac{T_{(1)4} V_1}{T_{(2)2}^2} = \frac{1}{4} \implies T_{(1)4} = \frac{T_{(2)2}^2}{4V_1},
\label{g227}\ee
to obtain
\be
ds^2 \approx du^2 +   \frac{(u-u_0)^2}{4}\left(d\psi + \cos\theta d\phi \right)^2 + \frac{4}{T_{(2)2}}d\Omega^2,
\label{g226a}\ee
in order to avoid the conical singularity and obtain a regular end-point.

Some remarks are in order:

\begin{enumerate}
   \item[$\bullet$] Although we  have $\ddot{\f} \neq 0 $, the space-time does ends here as the $S^1$ fiber shrinks to zero. Hence this is a true end-point of the flow.

   \item[$\bullet$] We have 2 independent integration constants, namely $T_{(2)2}$ and the scalar coupling $\f_0$. These two parameters are in one-to-one correspondence with the sources expected in the problem.

   \item[$\bullet$] We tacitly assumed in (\ref{g215}) that $V_1 > 0$. If it is negative, the solution must be reorganized by taking $x<0$. In general, the solution arrives and stops at $\f_0$ from larger (smaller) values if $V_1>0$ ($V_1<0$).

\end{enumerate}

\subsection{Nut-like IR endpoint}

By substituting the expansions (\ref{C26}) - (\ref{C29}) into the equations of motion (\ref{ga51}),(\ref{ga53}) - (\ref{ga54}) we find the following:

\be
S_0^2 = \frac{V_1}{2} , S_1 = 0 , S_2 = \frac{V_0 + 6 V_2}{12 S_0},
\label{g267}\ee
\be
W_{(1)0} = - 2 S_0 , W_{(1)1} = 0 , W_{(1)2} = \text{arbitrary},
\label{g268}\ee
\be
W_{(2)0} = - 2 S_0 , W_{(2)1} = 0 , W_{(2)2} = \frac{7 V_0 - 6 V_2}{12 S_0} - \frac{W_{(1)2}}{2},
\label{g269}\ee
\be
T_{(1)0} = 2 V_1 , T_{(1)1} = 0 , T_{(1)2} = \frac{8V_0}{3} - 4 S_0 W_{(1)2},
\label{g270}\ee
\be
T_{(2)0} = 2 V_1 , T_{(2)1} = 0 , T_{(2)2} = \frac{3V_0 + 2 V_2}{2} - S_0 W_{(1)2}.
\label{g271}\ee

This solution has two arbitrary integration constants, namely $W_{(1)2}$ and $\f_0$.

Since $S = \dot{\f}$, we obtain to leading order
\be
\f - \f_0 = \frac{1}{8} V_1 (u -u_0)^2+\ldots,
\label{g272}\ee
where $u_0$ is an integration constant.

From the defining relations in  (\ref{ga46}),(\ref{ga47}) we obtain
\be
A_1 = A_1^c + \ln(u-u_0) - \frac{S_0 W_{(1)2}}{8} \left( u -u_0\right)^2 + \mathcal{O} \left( (u-u_0)^4 \right),
\label{g273}\ee
\be
A_2 = A_2^c + \ln(u-u_0) - \frac{S_0 W_{(2)2}}{8} \left(u-u_0 \right)^2 + \mathcal{O} \left( (u-u_0)^4 \right),
\label{g274}\ee
where $A_1^c,A_2^c$ are integration constants.

The  Kretschmann scalar
\be
\mathcal{K} = \frac{1}{96} \left( 113 V_0^2 - 84 V_0 \left(V_2 + 3 S_0 W_{(1)2} \right)   + 36 \left( V_2 + 3 S_0 W_{(1)2} \right)^2 \right) + \mathcal{O}(u-u_0)
\label{g275}\ee
is regular.

The scale factors $e^{2A_i}$ become
\be
e^{2A_2} = \frac{(u-u_0)^2}{4L^2} \left( 1 + \mathcal{O} \left( u-u_0 \right) \right),
\label{g276}\ee
\be
e^{2A_1} = \frac{(u-u_0)^2}{4L^2} \left( 1 + \mathcal{O} \left( u-u_0 \right) \right).
\label{g277}\ee

Finally, the asymptotic metric is,
\be
ds^2 \approx du^2 + \frac{(u-u_0)^2}{4} \left( (d\psi + \cos\theta d\phi)^2 + d\Omega^2 \right).
\label{g278}\ee

Some remarks are in order:

\begin{enumerate}
   \item[$\bullet$] We have $\ddot{\f} = \frac{V_1}{4} \neq 0 $ but the space-time ends here, as the $S^3$ shrinks to zero size . Hence this is a true end-point of the flow.

   \item[$\bullet$] We have two unknown constants, namely $W_{(2)2}$ and the scalar coupling $\f_0$. This the number expected of  sources in this case, with vevs determined uniquely.

   \item[$\bullet$] We tacitly assumed in (\ref{g267}) that $V_1 > 0$. Since $T_{(1)0} = T_{(2)0} = 2 V_1 > 0$, the solution exists only for positive curvature.

   \item[$\bullet$] We expanded the functions $S,W_{1,2},T_{1,2},V$ in powers of $(\f - \f_0)$, which means that we approach $\f_0$ from the right, i.e $\f \geq \f_0$. We can expand the functions in powers of $(\f_0 - \f)$, with the only difference being that $W_1,W_2 \to - W_1 , - W_2$.
\end{enumerate}

\section{Perturbation theory near extrema of the potential}\label{Extremaexpansions}

In this appendix we shall collect analytical expressions for the superpotentials $S$, $W_{1,2}$, $T_{1,2}$ near extrema of the potential, which describe the geometry in the vicinity of the boundary. Consequently, we shall associate these expansions with observables that appear in the UV QFT, such as the UV curvature, $R^{uv}$, the squashing parameter, $a^2$, the vev of the dual operator $\cal{O}$ and the vev of the stress-energy tensor.

Since we are considering expansions around a point $\f_0$ that is located in the vicinity of an extermum of the potential, we can parametrize the potential with AdS data, as in (\ref{C6}), which we replicate here for convenience:

\be
V(\f) = - \frac{6}{\ell^2} - \frac{\Delta \Delta_-}{2\ell^2}(\f-\f_0)^2 + \mathcal{O}\left( (\f - \f_0)^3 \right),
\label{F1}\ee
where $\Delta_- = 3 - \Delta$, with $\Delta$ being associated to the scaling dimension of the perturbing operator $\mathcal{O}$ around the extremum. In particular, as it was mentioned in appendix \ref{extremalpoints}, it takes the values $\frac{3}{2} \leq \Delta \leq 3$ when we are at a maximum and $\Delta >3 $ when we are at a minimum. $\ell$ is the AdS length scale.

Before we proceed, we shall make a remark: similarly to appendix \ref{extremalpoints}, we define for convenience

\be
x \equiv \f - \f_0.
\label{F2}\ee

The solutions can be split into two categories, and shall be studied as such.
There are no solutions with non-zero curvature corresponding to IR fixed points in the dual theory, like on the round three sphere, \cite{C} .

\subsection*{UV fixed points}

This category is comprised of two cases, namely:

\begin{enumerate}
   \item[(1)] The (-)-branch asymptotics. This solution exists only around maxima of the potential, and exists for both the squashed and the regular 3-sphere. Around this point, the functions have the following behavior:

   \be
   S \sim x \ , \ W_1,W_2 \sim \text{const} \ , \ T_1,T_2 \sim x^{2/\Delta_-}.
   \label{F3}\ee

   This behavior describes RG flows that start from a maximum of the potential (UV fixed point).

   \item[(2)] The (+)-branch asymptotics. This solution exists around maxima as well as minima of the potential, and exists for both the squashed and the regular 3-sphere. Around this point, the functions have the following behavior:

   \be
   S \sim x \ , \ W_1,W_2 \sim \text{const}, T_1,T_2 \sim x^{2/\Delta_+}.
   \label{F4}\ee

   This behavior describes RG flows that either start at a maximum of the potential (UV fixed point) or arrive at a  minimum (UV fixed point). In both cases, we have an AdS boundary.
\end{enumerate}

\subsection*{Exotic Solutions}

This case is comprised by the three asymptotics $\nu_{\pm},B_{3}$. $\nu_-$ and $\nu_+$ exist around maxima of the potential, and $\nu_+$ and $B_3$ exist around minima of the potential.
For all such solutions, the structure of the asymptotics corresponds to a boundary, however, this is not a standard FG boundary.

Before we begin discussing each case in turn, we shall first make some comments regarding perturbative expansions of the leading solutions. Specifically, as we shall see, all the solutions that we study permit continuous deformations of the superpotential functions, as long as they are subleading compared to the leading order expansions. These are non analytic power series expansions and/or are accompanied by a small parameter that justifies their characterisation as subleading. In the following subsection, we shall derive a set of linearised equations which, upon being solved, shall yield the aforementioned subleading expansions.

\subsection{Comments on perturbative expansions}

In order to obtain the aforementioned set of equations, we shall linearise the equations of motion (\ref{ga51}),(\ref{ga53}) - (\ref{ga54}) as follows: we write the superpotential functions as

\be
S = S_{L} + \delta S \ , \ W_1 = W_{1L} + \delta W_{1} \ ,\  W_2 = W_{2L} + \delta W_{2},
\label{F12}\ee
\be
T_{1} = T_{1L} + \delta T_{1} \ , \ T_{2} = T_{2L} + \delta T_{2},
\label{F13}\ee
and substitute them to the equations of motion, keeping terms that are linear to the subleading terms. Here, the functions $S_L,W_{1,2L},T_{1,2L}$ are assumed to satisfy the full non-linear equations. The resulting equations are the following\footnote{We remind the reader that $V(\f)$ is an analytic function of $\f$, and therefore has a regular power series expansion.}:
\be
(S_{L} \delta S)' - \frac{1}{4} S_{L} \left( \delta W_{1} + 2 \delta W_{2} \right) - \frac{1}{4} \delta S_{} \left( W_{1L} + 2 W_{2L} \right) = 0,
\label{F14}\ee
\be
\left( W_{2L} + W_{1L} \right) \delta W_{2} + W_{2L} \delta W_{1} + \frac{1}{2} \delta T_{1} -2 \delta T_{2} - 4 S_{L} \delta S = 0
\label{F15}\ee
$$
\left( W_{2L} - W_{1L}\right)' \delta S + \left( \delta W_{2} - \delta W_{1} \right)' S_{L} + \frac{1}{4} \left( W_{1L} - W_{2L} \right) \left( \delta W_{1} + 2 \delta W_{2} \right),
$$
\be
+ \frac{1}{4}\left( \delta W_{1} - \delta W_{2}\right) \left( W_{1L} + 2 W_{2L} \right) + \delta T_{2} - \delta T_{1} = 0,
\label{F16}\ee
\be
S_{L} \delta T_{2}' + \delta S T_{2L}' = \frac{1}{2} \left( W_{2L} \delta T_{2} + \delta W_{2} T_{2L} \right),
\label{F17}\ee
\be
S_{L} \delta T_{1}' + \delta S T_{1L}' + \frac{1}{2} \left( (W_{1L} - 2 W_{2L}) \delta T_{1} + (\delta W_{1} - 2 \delta W_{2}) T_{1L} \right) = 0.
\label{F18}\ee

Equations (\ref{F14}) - (\ref{F18}) are the required equations. In total, we have 4 first order differential equations and an algebraic one for 5 unknowns: $\delta S, \delta W_{1,2}, \delta T_{1,2}$. The way we are going to solve them is as follows: we shall solve equations (\ref{F14}) - (\ref{F16}) with respect to $\delta W_1,\delta T_{1}, \delta T_2$. After that, we shall substitute the resulting expressions into equations (\ref{F17}),(\ref{F18}). We shall obtain two differential equations for the functions $\delta W_2,\delta S$.

Solving equations (\ref{F14}) - (\ref{F16}) with respect to the functions $\delta W_1, \delta T_1, \delta T_2$ yields the following:

\be
\delta W_1 =-\frac{-4 \delta S' S_L-4 \delta S S_L'+2 \delta W_2S_L+\delta S W_{1L}+2 \delta S W_{2L}}{S_L},
\label{F19}\ee
\be
\delta T_1 =-\frac{-12 \delta S' S_L W_{1L}-4 \delta S W_{1L} S_L'+\delta W_2 S_L W_{1L}-20 \delta S' S_L W_{2L}}{3 S_L}
\label{F20}\ee
$$ - \frac{8 \delta W_2 S_L W_{2L}+16 \delta S' S_L S_L'+16 \delta S'' S_L^2-16 \delta S S_L'^2+16 \delta S S_L S_L''}{3S_L} $$
$$- \frac{8 \delta S S_L^2-12 \delta W_2' S_L^2}{3S_L}+\frac{12 \delta S S_L W_{2L}'+4 \delta S W_{2L} S_L'}{3 S_L}$$
$$  - \frac{7 \delta S W_{1L} W_{2L}+2 \delta S W_{1L}^2+6 \delta S W_{2L}^2}{3S_L} ,$$
\be
\delta T_2 = -\frac{-12 \delta S' S_L W_{1L}-4 \delta S W_{1L} S_L'-5 \delta W_2 S_L W_{1L}-44 \delta S' S_L W_{2L}}{12 S_L}
\label{F21}\ee
$$ - \frac{-12 \delta S S_L W_{2L}'-28 \delta S W_{2L} S_L'+14 \delta W_2 S_L W_{2L}+16 \delta S' S_L S_L'}{12S_L}$$
$$- \frac{32 \delta S S_L^2-12 \delta W_2' S_L^2+13 \delta S W_{1L} W_{2L}+2 \delta S W_{1L}^2+18 \delta S W_{2L}^2}{12S_L}   $$
$$ - \frac{16 \delta S'' S_L^2-16 \delta S S_L'^2+16 \delta S S_L S_L''}{12S_L}. $$

Substituting the previous equations into equations (\ref{F17}),(\ref{F18}) yields the following two equations for the functions $\delta S, \delta W_2$:

\begin{equation}
\begin{aligned}
& \delta S\left(8 W_{1 L}^{\prime} S_L^{\prime}-8 W_{1 L} S_L^{\prime \prime}+8 W_{2 L}^{\prime} S_L^{\prime}+40 W_{2 L} S_L^{\prime \prime}+64 S_L^{\prime} S_L^{\prime \prime}-32 S_L S_L^{\prime \prime \prime}-16 S_L S_L^{\prime}+\right. \\
& +24 S_L W_{2 L}^{\prime \prime}-14 W_{1 L}^{\prime} W_{2 L}-2 W_{1 L} W_{2 L}^{\prime}-4\left(W_{1 L}^2\right)^{\prime}-48 W_{2 L} W_{2 L}^{\prime}+6 T_{1 L}^{\prime}-8 W_{1 L} S_L+16 W_{2 L} S_L+ \\
& +\frac{-2 W_{1 L}^3+12 W_{2 L}^3-3 W_{1 L}^2 W_{2 L}+8 W_{2 L}^2 W_{1 L}+8 S_L^{\prime 2} W_{1 L}+12 T_{1 L} S_L^{\prime}-3 T_{1 L} W_{1 L}-6 T_{1 L} W_{2 L}}{S_L}+ \\
& \left.+\frac{\left.-32 S_L^{\prime 3}-40 W_{2 L} S_L^{\prime 2}+10 S_L^{\prime} W_{1 L} W_{2 L}+8 S_L^{\prime} W_{1 L}^2+4 S_L^{\prime} W_{2 L}^2\right)+}{S_L}\right) \\
& +\delta S_L^{\prime}\left(-8 S_L^{\prime} W_{1 L}+24 S_L W_{1 L}^{\prime}+40 S_L^{\prime} W_{2 L}+64 S_L W_{2 L}^{\prime}+12 T_{1 L}-\right. \\
& \left.\quad-64 S_L S_L^{\prime \prime}-16 S_L^2-18 W_{1 L} W_{2 L}+8 W_{1 L}^2-52 W_{2 L}^2+32 S_L^{\prime 2}\right)+ \\
& \quad+\delta S^{\prime \prime}\left(8 S_L W_{1 L}+72 S_L W_{2 L}-64 S_L S_L^{\prime}\right)-32 S_L^2 \delta S^{\prime \prime \prime}- \\
& -\delta W_2\left(2 S W_{1 L}^{\prime}+16 S_L W_{2 L}^{\prime}+12 T_{1 L}+W_{1 L}^2+6 W_{2 L} W_{1 L}-16 W_{2 L}^2\right)+ \\
& +\delta W_2^{\prime}\left(24 S_L S_L^{\prime}+10 S W_{1 L}-40 S_L W_{2 L}\right)+24 S_L^2 \delta W_2^{\prime \prime}=0
\end{aligned}
\label{F22}\end{equation}

\begin{equation}
   \begin{aligned}
      & \delta S\left(24 S_L W_{2 L}^{\prime \prime}+8\left(S_L^{\prime} W_{1 L}\right)^{\prime}+56\left(S_L^{\prime} W_{2 L}\right)^{\prime}+16 W_{2 L} S_L^{\prime \prime}-26\left(W_{1 L} W_{2 L}\right)^{\prime}-4\left(W_{1 L}^2\right)^{\prime}-42\left(W_{2 L}^2\right)^{\prime}+\right. \\
      & +64 S_L^{\prime} S_L^{\prime \prime}-32 S_L S_L^{\prime \prime \prime}+32 W_{2 L} S_L-32\left(S_L^2\right)^{\prime}+\frac{22 S_L^{\prime} W_{1 L} W_{2 L}-8 W_{1 L} S_L^{\prime 2}-72 W_{2 L} S_L^{\prime 2}}{S_L}+ \\
      & \left.+\frac{8 S_L^{\prime} W_{2 L}^2+4 S_L^{\prime} W_{1 L}^2-32 S_L^{\prime 3}+13 W_{1 L} W_{2 L}^2+2 W_{1 L}^2 W_{2 L}+18 W_{2 L}^3}{S_L}+24 T_{2 L}^{\prime}\right)+ \\
      & +\delta S^{\prime}\left(24 S_L W_{1 L}^{\prime}+8 S_L^{\prime} W_{1 L}+72 S_L^{\prime} W_{2 L}+112 S_L W_{2 L}^{\prime}-38 W_{1 L} W_{2 L}-4 W_{1 L}^2-80 W_{2 L}^2-64 S_L S_L^{\prime \prime}+\right. \\
      & \left.+32 S_L^{\prime 2}-64 S_L^2\right)+\delta S^{\prime \prime}\left(24 S_L W_{1 L}+104 S_L W_{2 L}-64 S_L S_L^{\prime}\right)-32 S_L^2 \delta S^{\prime \prime \prime}+24 S_L^2 \delta W_2^{\prime \prime} \\
      & +\delta W_2\left(10 S_L W_{1 L}^{\prime}-28 S_L W_{2 L}^{\prime}-5 W_{2 L} W_{1 L}+14 W_{2 L}^2-12 T_{2 L}\right)+ \\
      & +S_L \delta W_2^{\prime}\left(10 W_{1 L}-40 W_{2 L}+24 S_L^{\prime}\right)=0
      \end{aligned}
\label{F23}\end{equation}

We have now arrived at two coupled differential equations for the functions $\delta S$,$ \delta W_2$. All that remains is to solve equations (\ref{F22}),(\ref{F23}) with respect to the functions $\delta S$ ,$ \delta W_2$.

We shall now begin to discuss each case in turn.

\subsection{UV fixed points}\label{UVfixedpoints}

Similarly to appendix \ref{IR-endpoints}, we shall attempt to set up the perturbation theory around the leading-order solutions, in order to justify whether these correspond to bona-fide asymptotic solutions.

For reasons that shall become apparent, for this subsection we shall briefly denote $\Delta$ with $\Delta_+$.
We start by calculating the leading-order solutions as follows: we expand the superpotentials in a regular power series with leading powers determined by (\ref{F3}),(\ref{F4}):

\be
S_{L,\pm} = x \sum_{n=0}^\infty S_{n,\pm}\ x^n \ , \ W_{1L,\pm} = \sum_{n=0}^\infty W_{(1)n,\pm}\ x^n \ , \ W_{2L,\pm} = \sum_{n=0}^\infty W_{(2)n,\pm}\ x^n
\label{F5}\ee
\be
T_{1L,\pm} = x^{2/\Delta_\pm} \sum_{n=0}^\infty T_{(1)n,\pm}\ x^n \ , \ T_{2L,\pm} = x^{2/\Delta_\pm} \sum_{n=0}^\infty T_{(2)n,\pm}\ x^n
\label{F6}\ee

Inserting the previous expansions into the equations of motion  (\ref{ga51})-(\ref{ga55}),(\ref{ga54})  and solving recursively for the first few coefficients, we find the following:
\be
S_{L,\pm} = \frac{\Delta_{\pm}}{\ell}x  + \frac{\ell V_3}{\Delta_{\pm}-1}x^2 + \mathcal{O}(x^3)
\label{F9}\ee
\be
W_{1L,\pm} = W_{2L,\pm} = \frac{4}{\ell} + \frac{\Delta_{\pm}}{2\ell} x^2 + \mathcal{O}(x^3)
\label{F10}\ee
\be
T_{1L,\pm} = T_{2L,\pm} = 0
\label{F11}\ee

These expansions correspond to holographic RG flows of field theories defined on flat manifolds, which have been studied extensively. For details, we refer the reader to e.g. \cite{exotic}. Moreover, up to second order no integration constants have entered the expansions. Since we can recursively obtain the higher order coefficients from the lower order ones, we observe that no integration constants appear in the leading order expansions.
We expect that the constants are hidden in the solutions of the linearized equations of motion, (\ref{F19})-(\ref{F21}),(\ref{F22}) - (\ref{F23}). It turns out that the following ansatz for the deformations $\delta S,\delta W_2$, is sufficient to find the general solutions:

\be
\delta S = \mathcal{A} x^\lambda \ , \ \delta W_2 = \mathcal{B} x^g,
\label{F24}\ee
where the constants $\mathcal{A},\lambda,\mathcal{B},g$ must obey the following constraints in order for the corresponding expansions to be subleading:
\be
\lambda \geq 1\ , \ g \geq 0 \ , \ \mathcal{A},\mathcal{B} \neq 0.
\label{F25}\ee

The reasoning behind these constraints is as follows:

\begin{enumerate}
   \item[$\bullet$] Suppose that $\lambda < 1$. Then, the term $\delta S$ is leading compared to $S_{L,\pm}$, since the exponent of $x$ in $\delta S$ is less than that of $S_{L,\pm}$, which is one. This, in turn, implies that we cannot classify $\delta S$ as subleading. \\
   Suppose now that $\lambda = 1$. Then, the subleading term is the same order in $x$ as the leading one, namely $S_{L,\pm}$, and therefore the integration constant appearing can be absorbed into $S_{L,\pm}$ via an appropriate redefinition. \\
   Finally, if $\lambda > 1$, then the exponent of $x$ in the subleading term is greater than the one in $S_{L,\pm}$, and therefore its classification as subleading holds.
   \item[$\bullet$] Similarly for $g$, we observe that since the $W_{2,\pm}$ solution behaves as $W_{2,\pm} \sim x^0$, we must choose $g \geq 0$ in order to avoid the occurence of a non-subleading term (note that if $g < 0$, then $\delta W_2$ diverges, and therefore $W_2$ diverges as well.).
   \item[$\bullet$] Finally, in order for the subleading terms to be non-trivial, we must demand that $\mathcal{A},\mathcal{B}$ are not zero.
\end{enumerate}

Substituting the leading order expansions (\ref{F9}) - (\ref{F11}) along with the ansatze (\ref{F24}) into the linearised equations for $\delta S$,$\delta W_2$ yields to leading order in the integration constants $\mathcal{A},\mathcal{B}$ the following:

\be
-3 \mathcal{B}  x^{1+g}(g - \frac{2}{\Delta_\pm})(2g - \frac{6}{\Delta_\pm}) + 2 \mathcal{A}x^{\lambda}\left(- \frac{120}{\Delta_\pm^3} + 2(\lambda-1)^2(\lambda+1) + \frac{4}{\Delta_\pm^2}(31\lambda-11)\right.
\label{F26}\ee
$$\left.  - \frac{8}{\Delta_\pm}(\lambda-1)(5\lambda+2) \right)+\cdots = 0,$$
\be
3 \mathcal{B}  x^{1+g}(g - \frac{2}{\Delta_\pm})(2g - \frac{6}{\Delta_\pm})  - 2 \mathcal{A}x^{\lambda}\left(- \frac{264}{\Delta_\pm^3} + 2(\lambda-1)^2(\lambda+1) +\frac{4}{\Delta_\pm^2}(61\lambda-17)\right.
\label{F27}\ee
$$ \left. - \frac{8}{\Delta_\pm}(\lambda-1)(8\lambda+5) \right) + \cdots = 0. $$

By adding the two previous equations we arrive at the following:

\be
12 \mathcal{A} x^{\lambda} \left( \lambda + 1 - \frac{3}{\Delta_\pm} \right)\left( \lambda - (1 + \frac{2}{\Delta_\pm}) \right) = 0,
\label{F28}\ee
which leads us to the following cases. As   $\mathcal{A} \not= 0$,

\begin{enumerate}

   \item[$\bullet$] We can choose $\lambda = 1 + \frac{2}{\Delta_\pm}$, which is acceptable iff
   \be
   \frac{2}{\Delta_\pm} > 0 \implies \Delta_\pm > 0.
   \label{F29.a}\ee
   \item[$\bullet$] We can choose $\lambda = \frac{3}{\Delta_{\pm}}-1$, which is acceptable iff
   \be
   \frac{3}{\Delta_\pm} \geq 2.
   \label{F29.b}\ee
\end{enumerate}

Therefore we have two possible choices for the parameter $\lambda$, and therefore we have accounted for two out of the four integration constants. Substituting any one of the two values into equations (\ref{F27}),(\ref{F28}) yields the following equation:

\be
6\mathcal{B} x^{1+g} \left( g - \frac{2}{\Delta_\pm} \right) \left( g - \frac{3}{\Delta_\pm} \right) = 0 ,
\label{F29}\ee
which leads us to the following cases, as  $\mathcal{B} \not= 0$,

\begin{enumerate}

   \item[$\bullet$] We can choose $g = \frac{2}{\Delta_\pm}$, which is acceptable iff
   \be
   \frac{2}{\Delta_\pm} > 0 \implies \Delta_\pm > 0.
   \label{F30.a}\ee
   \item[$\bullet$] We can choose $g = \frac{3}{\Delta_\pm}$, which is acceptable iff
   \be
   \frac{3}{\Delta_\pm} > 0 \implies \Delta_\pm > 0.
   \label{F30.b}\ee
\end{enumerate}

This concludes the analysis of the linearised equations of motion. In total, we have found 4 integration constants, as expected. Before we write the full form of the deformations $\delta S, \delta W_2$, we note that apart from condition (\ref{F29.b}), the rest are satisfied simultaneously, i.e. when $\Delta_{\pm} > 0$. Therefore, for convenience, we shall add them all together, but put the term accompanying the exponent (\ref{F29.b}) in parentheses in order to emphasise this. The full form of the deformations $\delta S, \delta W_2$ is,

\be
\delta S \approx \mathcal{A}_1 x^{\frac{2}{\Delta_\pm}+1} + ( \mathcal{A}_2 x^{\frac{3}{\Delta_\pm}-1} )\ , \ \delta W_2 \approx \mathcal{B}_1 x^{\frac{2}{\Delta_\pm}} + \mathcal{B}_2 x^{\frac{3}{\Delta_\pm}},
\label{F30}\ee
which, upon subsituting to (\ref{F19})-(\ref{F21}), yields the following for the deformations $\delta W_1,\delta T_1,\delta T_2$:

\be
\delta T_1 \approx - \frac{4}{\ell \Delta_\pm} x^{\frac{2}{\Delta_\pm}} \left(4 \mathcal{A}_1 \left(1 - 2 \Delta_\pm \right)  + \mathcal{B}_1 \Delta_\pm \right),
\label{F31}\ee
\be
\delta T_2 \approx - \frac{1}{\ell \Delta_\pm} x^{\frac{2}{\Delta_\pm}} \left( 12 \mathcal{A}_1 \left( 1 - 2 \Delta_\pm \right) + \mathcal{B}_1 \Delta_\pm \right),
\label{F32}\ee
\be
\delta W_1 = \frac{2}{\Delta_\pm} \left( 2\mathcal{A}_1(2\Delta_\pm -1) \right) x^{\frac{2}{\Delta_\pm}} - 2 \mathcal{B}_2 x^{\frac{3}{\Delta_\pm}} + (\Delta_\pm \mathcal{A}_2  x^{\frac{3}{\Delta_\pm}}) .
\label{F33}\ee

For future convenience, we shall redefine the integration constants as follows,

\be
\mathcal{A}_1 = \frac{\Delta_\pm}{2\Delta_\pm -1} \frac{\mathcal{R}_1 + 2 \mathcal{R}_2}{32\ell} \ , \ \mathcal{A}_2 = \frac{C_1 + 2 C_2}{\Delta_\pm \ell} \ , \ \mathcal{B}_1 = \frac{\mathcal{R}_2}{8\ell} \ , \ \mathcal{B}_2 = \frac{C_2}{\ell},
\label{F34}\ee
where the new integration constants now are $\mathcal{R}_{1,2},\mathcal{C}_{1,2}$. The expansions (\ref{F30})-(\ref{F33}) now become

\be
\delta S \approx \frac{\Delta_\pm}{2\Delta_\pm -1} \frac{\mathcal{R}_1 + 2 \mathcal{R}_2}{32\ell} x^{\frac{2}{\Delta_\pm}+1} + ( \frac{C_1 + 2 C_2}{\Delta_\pm \ell} x^{\frac{3}{\Delta_\pm} -1}),
\label{F35}\ee
\be
\delta W_1 \approx \frac{\mathcal{R}_1}{8\ell} x^{\frac{2}{\Delta_\pm}} - \frac{2C_2}{\ell} x^{\frac{3}{\Delta_\pm}} +( \frac{C_1 + 2C_2}{\ell} x^{\frac{3}{\Delta_{\pm}}} ) \ , \ \delta W_2 \approx \frac{\mathcal{R}_2}{8\ell} x^{\frac{2}{\Delta_\pm}} + \frac{C_2}{\ell} x^{\frac{3}{\Delta_\pm}},
\label{F36}\ee
\be
\delta T_1 \approx \frac{\mathcal{R}_1 + \mathcal{R}_2}{2\ell^2} x^{\frac{2}{\Delta_\pm}} \ , \ \delta T_2 \approx \frac{5 \mathcal{R}_2 + 3 \mathcal{R}_1}{8\ell^2} x^{\frac{2}{\Delta_\pm}} .
\label{F37}\ee

Putting everything together, we are now in a position to write down the expressions for $W_{1,2},S,T_{1,2}$ in the vicinity of an extremum of $V$ and up to first order in the integration constants $C_{1,2},\mathcal{R}_{1,2}$. We are thus ready to discuss the (-) and (+) branch seperately.

Finally, we note that we shall revert back to our old notation for $\Delta$, namely $\Delta_+ = \Delta$.

\subsubsection{The (-)-branch}
This branch exists only in the vicinity of maxima of the potential, i.e. when $\Delta$ takes the values $\frac{3}{2} \leq \Delta < 3$ $\implies$  $ 0 < \Delta_- \leq \frac{3}{2}$. From the previous inequality it can easily be seen that $\Delta_-$ satisfies all the constraints (\ref{F29.a}),(\ref{F29.b}),(\ref{F30.a}),(\ref{F30.b}), and therefore none of the integration constants are constrained. The full expansions of the superpotentials are thus:

\be
S_- =  \frac{\Delta_{-}}{\ell}x  + \frac{\ell V_3}{\Delta_{-}-1}x^2 + \mathcal{O}(x^3) + \frac{C_1 + 2 C_2}{\Delta_-\ell} x^{\frac{\Delta}{\Delta_-}}  \left( 1 + \mathcal{O} (x) \right)
\label{F38}\ee
$$ + \frac{\Delta_-(\mathcal{R}_1 + 2 \mathcal{R}_2)}{(2\Delta_- - 1)32\ell} x^{\frac{2}{\Delta_-}+1} \left( 1 + \mathcal{O} (x) \right) ,$$
\be
W_1 = \frac{4}{\ell} + \frac{\Delta_-}{2\ell} x^2 + \mathcal{O} \left( x^3 \right) + \frac{\mathcal{R}_1}{8\ell} x^{\frac{2}{\Delta_-}} \left( 1 + \mathcal{O} (x) \right)+ \frac{C_1}{\ell} x^{\frac{3}{\Delta_-}}  \left( 1 + \mathcal{O} (x) \right),
\label{F39}\ee
\be
W_2 = \frac{4}{\ell} + \frac{\Delta_-}{2\ell} x^2 + \mathcal{O} \left( x^3 \right) + \frac{\mathcal{R}_2}{8\ell} x^{\frac{2}{\Delta_-}} \left( 1 + \mathcal{O} (x) \right) + \frac{C_2}{\ell} x^{\frac{3}{\Delta_-}}  \left( 1 + \mathcal{O} (x) \right),
\label{F40}\ee
\be
T_1 = \frac{\mathcal{R}_1 + \mathcal{R}_2}{2\ell^2} x^{\frac{2}{\Delta_-}} \left( 1 + \mathcal{O} (x) \right),
\label{F41}\ee
\be
T_2 = \frac{5\mathcal{R}_2 +3 \mathcal{R}_1}{8\ell^2} x^{\frac{2}{\Delta_-}} \left( 1 + \mathcal{O} (x) \right),
\label{F42}\ee
where $\mathcal{R}_1,\mathcal{R}_2,C_1,C_2$ are the integration constants.
Since $S = \dot{\f}$, we have to leading order

\be
\f - \f_0 = \f_- \ell^{\Delta_-} e^{\Delta_- \frac{u}{\ell}} \left(1+\mathcal{O}\left(\left(\mathcal{R}_{1}+2 \mathcal{R}_{2}\right)\left|\varphi_{-}\right|^{\frac{2}{\Delta_-}} e^{2 \frac{u}{\ell}}\right)\right)
\label{F43}\ee
$$ + \frac{C_1 + 2 C_2}{\Delta_-(3-2\Delta_-)} \vert \f_- \vert^{\frac{\Delta}{\Delta_-}} \ell^{\Delta}  e^{\Delta \frac{u}{\ell}} \left(1+\mathcal{O}\left(\left(\mathcal{R}_{1}+2 \mathcal{R}_{2}\right)\left|\varphi_{-}\right|^{\frac{2}{\Delta_-}} e^{2 \frac{u}{\ell}}\right)\right) + \ldots ,$$
where $\f_-$ is an integration constant with dimensions $\text{length}^{-\Delta_-} \equiv \text{mass}^{\Delta_-}$. Moreover, since $\frac{3}{2} \leq \Delta < 3$, we observe that as $\f \to \f_0$, $u \to - \infty$. {Comparing the previous expansion, with the expansion }(\ref{fe7}),{we observe that we can identify $\f_-$ as the source of the dual operator $\mathcal{O}$, while its vev is given by

\be
\left\langle \mathcal{O} \right\rangle = \frac{C_1+2C_2}{\Delta_-} \vert \f_- \vert^{\frac{\Delta}{\Delta_-}},
\label{F44}\ee
which is compatible with }(\ref{fe9}).

Continuing, by using the definitions of the functions $W_1,W_2$ in (\ref{ga46}),(\ref{ga47}) we can write the functions $A_1,A_2$ with respect to $u$:

\be
A_{1}=A_{1}^{c}-\frac{u}{\ell}-\frac{\varphi_{-}^{2} \ell^{2\Delta_-}}{16} e^{2\Delta_- \frac{u}{\ell}}-\frac{\mathcal{R}_{1}\left|\varphi_{-}\right|^{\frac{2}{\Delta_-}} \ell^{2}}{196} e^{2 \frac{u}{\ell}}
\label{F45}\ee
$$-\frac{\Delta C_{1}\left|\varphi_{-}\right|^{\frac{3}{\Delta_-}} \ell^{3}}{6(3 - 2\Delta_-)} e^{3 \frac{u}{\ell}}+\ldots,$$
\be
A_{2}=A_{2}^{c}-\frac{u}{\ell}-\frac{\varphi_{-}^{2} \ell^{2\Delta_-}}{16} e^{2\Delta_- \frac{u}{\ell}}-\frac{\mathcal{R}_{2}\left|\varphi_{-}\right|^{\frac{2}{\Delta_-}} \ell^{2}}{196} e^{2 \frac{u}{\ell}}+\ldots
\label{F46}\ee
$$\quad-\frac{\Delta C_{2}\left|\varphi_{-}\right|^{\frac{3}{\Delta_-}} \ell^{3}}{6(3-2\Delta_-)} e^{3 \frac{u}{\ell}}+\ldots,$$
where $A_1^c,A_2^c$ are integration constants. However, these can be determined by looking at the definition of the first order formalism functions $T_1,T_2$. Specifically, to leading order the exponentials $e^{2A_i}$ are,

\be
e^{2A_2} = \frac{32}{L^2(3\mathcal{R}_1 +5 \mathcal{R}_2)} \vert \f_- \vert^{-\frac{2}{\Delta_-}} e^{-2 \frac{u}{\ell}} \left( 1 + \mathcal{O} ( e^{2 \frac{u}{\ell}}) \right)
\label{F47}\ee
\be
e^{2A_1}  =  \frac{128( \mathcal{R}_1 + \mathcal{R}_2 )}{L^2 \left(5 \mathcal{R}_2 + 3\mathcal{R}_1 \right)^2} \vert \f_- \vert^{-\frac{2}{\Delta_-}} e^{-2 \frac{u}{\ell}} \left( 1 + \mathcal{O} ( e^{2 \frac{u}{\ell}}) \right).
\label{F48}\ee

A direct comparison of the two previous equations with equations (\ref{F45}),(\ref{F46}) yields the following for the constants $A_1^c,A_2^c$:

\be
e^{2A_2^c} =  \frac{32}{L^2(3\mathcal{R}_1 +5 \mathcal{R}_2)} \vert \f_- \vert^{-\frac{2}{\Delta_-}}
\label{F49}\ee
\be
e^{2A_1^c} = \frac{128( \mathcal{R}_1 + \mathcal{R}_2 )}{L^2 \left(5 \mathcal{R}_1 + 3\mathcal{R}_2 \right)^2} \vert \f_- \vert^{-\frac{2}{\Delta_-}}.
\label{F50}\ee

Now that we have determined the constants, we can continue our analysis. From equations (\ref{F45}) , (\ref{F46}) we observe that we can write the exponentials $e^{2A_i}$ as follows:

\be
e^{2 A_{2}}=  \frac{32}{L^2(3\mathcal{R}_1 +5 \mathcal{R}_2)} \vert \f_- \vert^{-\frac{2}{\Delta_-}} e^{-2 \frac{u}{\ell}}\left(1-\frac{\varphi_{-}^{2} \ell^{2\Delta_-}}{16} e^{2\Delta_- \frac{u}{\ell}}-\frac{\mathcal{R}_{2}\left|\varphi_{-}\right|^{\frac{2}{\Delta_-} \ell^{2}}}{196} e^{2 \frac{u}{\ell}}\right.
\label{F51}\ee
$$\left.-\frac{\Delta C_{2}\left|\varphi_{-}\right|^{\frac{3}{3-\Delta} \ell^{3}}}{6(3-2 \Delta_-)} e^{3 \frac{u}{\ell}}+\ldots\right)$$
\be
e^{2 A_{1}}= \frac{128( \mathcal{R}_1 + \mathcal{R}_2 )}{L^2 \left(5 \mathcal{R}_2 + 3\mathcal{R}_1 \right)^2}e^{-2 \frac{u}{\ell}}\left(1-\frac{\varphi_{-}^{2} \ell^{2\Delta_-}}{16} e^{2\Delta_- \frac{u}{\ell}}-\frac{\mathcal{R}_{1}\left|\varphi_{-}\right|^{\frac{2}{\Delta_-} \ell^{2}}}{196} e^{2 \frac{u}{\ell}}\right.
\label{F52}\ee
$$\left.-\frac{\Delta C_{1}\left|\varphi_{-}\right|^{\frac{3}{\Delta_-} \ell^{3}}}{6(3-2\Delta_-)} e^{3 \frac{u}{\ell}}+\ldots\right),$$
where the $\ldots$ indicate higher order terms.  We want to write metric in PG coordinates like in (\ref{fe3}), and in order to do that we are going to make the following substitution:
\be
e^{\frac{u}{\ell}} = z.
\label{F53}\ee

The exponentials can be written in terms of z as

\be
\frac{z^2}{\ell^2} L^2 e^{2A_2} =\frac{32}{\vert \ell^{\Delta_-} \f_- \vert^{\frac{2}{\Delta_-}}(3\mathcal{R}_1 +5 \mathcal{R}_2)} \left( 1 - \frac{\f_-^2 \ell^{2\Delta_-}}{16} z^{2\Delta_-} \right.
\label{F54}\ee
$$\left.- \frac{\mathcal{R}_2 \vert \f_- \vert^{\frac{2}{\Delta_-}} \ell^2}{196} z^2 - \frac{\Delta C_2  \vert \f_- \vert^{\frac{3}{\Delta_-}} \ell^3}{6(3-2\Delta_-)} z^3 + \ldots \right)$$
\be
\frac{z^2}{\ell^2} L^2 e^{2A_1} = \frac{128( \mathcal{R}_1 + \mathcal{R}_2 )}{ \left(5 \mathcal{R}_2 + 3\mathcal{R}_1 \right)^2\vert \ell^{\Delta_-} \f_- \vert^{\frac{2}{\Delta_-}}}  \left( 1 - \frac{\f_-^2 \ell^{2\Delta_-}}{16} z^{2\Delta_-} \right.
\label{F55}\ee
$$\left.- \frac{\mathcal{R}_1 \vert \f_- \vert^{\frac{2}{\Delta_-}} \ell^2}{196} z^2 - \frac{\Delta C_1  \vert \f_- \vert^{\frac{3}{\Delta_-}} \ell^3}{6(3-2\Delta_-)} z^3 + \ldots \right).$$

From the previous expansions we immediately identify the coefficients of the term $z^3$, and therefore we can easily find the metric $g_{ij}^{(3)}$:

\be
g_{i j}^{(3)} d x^{i} d x^{j}=- \frac{16\Delta \ell \vert \f_- \vert^{\frac{1}{\Delta_-}}}{3(3-2\Delta_-)(3\mathcal{R}_1 + 5\mathcal{R}_2)}  \left(\frac{4( \mathcal{R}_{1} + \mathcal{R}_2) C_{1}}{3\mathcal{R}_{1}+5\mathcal{R}_{2}}(d \psi+\cos \theta d \phi)^{2}+C_{2} d \Omega^{2}\right),
\label{F56}\ee
and therefore, by virtue of (\ref{fe6}), the vev of the stress energy tensor is,

\be
\left\langle T_{ij} \right\rangle = \frac{3\ell^2}{16\pi G_N} g_{ij}^{(3)}.
\label{F57}\ee

Finally,  the metric near $\f_0$ is,

\be
d s^{2} \approx d u^{2}+\frac{32\vert  \varphi_{-} \vert^{-\frac{2}{\Delta_-}}}{3\mathcal{R}_{1}+5\mathcal{R}_{2}} e^{-2 \frac{u}{\ell}}\left(\frac{4( \mathcal{R}_{1} +\mathcal{R}_2 )}{3\mathcal{R}_{1}+5\mathcal{R}_{2}}(d \psi+\cos \theta d \phi)^{2}+d \Omega^{2}\right).
\label{F58}\ee

We observe that as $\f \to \f_0$, the geometry becomes the same as the one in the boundary of AdS.
Comparing with (\ref{ga1}) we see that the squashing parameter $a$ and the length $L$ can be related to $\mathcal{R}_1,\mathcal{R}_2,\f_-$ as follows:
\be
a^2=\frac{4( \mathcal{R}_{1} + \mathcal{R}_{2})}{(3 \mathcal{R}_{1}+5 \mathcal{R}_{2})}\sp  L^2=\frac{128}{3\mathcal{R}_1 + 5\mathcal{R}_2} \left|\varphi_{-}\right|^{-\frac{2}{\Delta_-}},
\label{asquared}\ee
so that
\be
\frac{\mathcal{R}_{1}}{ \mathcal{R}_{2}}= \frac{5a^2 - 4}{4 -3a^2}.
\ee

We also denote
\be
R^{uv} = \mathcal{R}\vert \f_- \vert^{\frac{2}{\Delta_-}}\to    \mathcal{R}=\frac{\mathcal{R}_1 + 2 \mathcal{R}_2}{8},
\label{dimensionlesscurvature}\ee
where $R^{uv}$ is the scalar curvature of the squashed sphere metric.

We can now rewrite $\mathcal{R}_{1,2}$ as a function of $a,\mathcal{R}$ as,
\be
\mathcal{R}_1=8 \frac{5a^2-4}{4-a^2}\mathcal{R}
\sp \mathcal{R}_2= 8 \frac{4 - 3a^2}{4-a^2}\mathcal{R}.
\ee

This concludes the analysis of the (-) branch. From equations (\ref{F56}) - (\ref{dimensionlesscurvature}), we observe the following:

\begin{enumerate}
   \item[$\bullet$] The integration constants $C_1,C_2$ control the vevs of the stress energy tensor $T_{ij}$ and the dual operator $\mathcal{O}$.
   \item[$\bullet$] The integration constants $\mathcal{R}_1,\mathcal{R}_2$ control the dimensionless curvature $\mathcal{R}$ and the squashing parameter $a^2$.
   \item[$\bullet$] The dual operator $\mathcal{O}$ has a non-zero source that is  controlled by the integration constant $\f_-$.
\end{enumerate}

Therefore, the (-) branch solution corresponds to a UV fixed point perturbed by a relevant operator.

\subsubsection{The (+)-branch}

This branch exists in the vicinity of maxima and minima of the potential, i.e. $\Delta$ can take any value in the range $[\frac{3}{2},\infty) - \left\{ 3\right\}$. From the previous inequality it can easily be seen that $\Delta$ does not satisfy all the constraints (\ref{F29.a}),(\ref{F29.b}),(\ref{F30.a}),(\ref{F30.b}), and therefore one of the integration constants must be set equal to $0$. Specifically, we must set $\mathcal{A}_2$ equal to $0$, which amounts to setting $C_1 = - 2C_2$. The full expansions of the superpotentials are thus:

\be
S = \frac{\Delta}{\ell}x  + \frac{\ell V_3}{\Delta-1}x^2 + \frac{\Delta(\mathcal{R}_1 + 2 \mathcal{R}_2)}{32(2\Delta-1)\ell} x^{\frac{2}{\Delta}+1 } \left( 1 + \mathcal{O} (x) \right),
\label{F59}\ee
\be
W_1 = \frac{4}{\ell} + \frac{\Delta}{2\ell} x^2 + \mathcal{O}( x^3)^3 + \frac{\mathcal{R}_1}{8\ell}x^{\frac{2}{\Delta}} \left( 1 + \mathcal{O} (x) \right)- \frac{2C_2}{\ell} x^{\frac{3}{\Delta}}  \left( 1 + \mathcal{O} (x) \right),
\label{F60}\ee
\be
W_2 = \frac{4}{\ell} + \frac{\Delta}{2\ell} x^2 + \mathcal{O} (x )^3  + \frac{\mathcal{R}_2}{8\ell}x^{\frac{2}{\Delta}} \left( 1 + \mathcal{O} (x) \right) +\frac{C_2}{\ell} x^{\frac{3}{\Delta}}  \left( 1 + \mathcal{O} (x) \right),
\label{F61}\ee
\be
T_1 = \frac{ \mathcal{R}_2 + \mathcal{R}_1}{2\ell^2}x^{\frac{2}{\Delta}} \left( 1 + \mathcal{O} (x) \right),
\label{F62}\ee
\be
T_2 = \frac{ 3\mathcal{R}_1 + 5 \mathcal{R}_2}{8\ell^2} x^{\frac{2}{\Delta}} \left( 1 + \mathcal{O} (x ) \right),
\label{F63}\ee
where $\mathcal{R}_1,\mathcal{R}_2,C_2$ are integration constants.

Since $S = \dot{\f}$, we have to leading order

\be
\f - \f_0 = \f_+ \ell^\Delta e^{\Delta \frac{u}{\ell}} \left( 1 + \mathcal{O} \left( (\mathcal{R}_1 + 2 \mathcal{R}_2) \vert \f_+ \vert^{\frac{2}{\Delta}} e^{2 \frac{u}{\ell}} \right) \right),
\label{F64}\ee
where $\f_+$ is an integration constant that has dimensions $\text{length}^{-\Delta} \equiv \text{mass}^{\Delta} $. Since $\Delta > 0$, we observe that as $\f \to \f_0$ we have $ u \to -\infty$. Comparing the expansion (\ref{F64}), with the expansion in (\ref{fe7}), we observe that the source of the dual operator $\mathcal{O}$ is identically zero, while its vev is given by,

\be
\left\langle \mathcal{O} \right\rangle = (2\Delta -3) \f_+,
\label{F65}\ee
which is compatible with (\ref{fe9}).

Continuing, by using the definitions of the functions $W_1$,$W_2$ in equations\\  (\ref{ga46}),(\ref{ga47}) we observe that we can write the functions $A_1$,$A_2$ in terms of $u$ as follows:

\be
A_1 = A_1^c - \frac{u}{\ell} - \frac{\f_+^2 \ell^{2\Delta}}{16} e^{2\Delta \frac{u}{\ell}} - \frac{\mathcal{R}_1 \vert \f_+ \vert^{\frac{2}{\Delta}} \ell^2}{64} e^{2 \frac{u}{\ell}} + \frac{C_2 \ell^3 \vert \f_+ \vert^{\frac{3}{\Delta}}}{6} e^{3 \frac{u}{\ell}} + \ldots,
\label{F66}\ee
\be
A_2 = A_2^c - \frac{u}{\ell} - \frac{\f_+^2 \ell^{2\Delta}}{16} e^{2\Delta \frac{u}{\ell}} - \frac{\mathcal{R}_2 \vert \f_+ \vert^{\frac{2}{\Delta}} \ell^2}{64} e^{2 \frac{u}{\ell}} -  \frac{C_2 \ell^3 \vert \f_+ \vert^{\frac{3}{\Delta}}}{12} e^{3 \frac{u}{\ell}} + \ldots,
\label{F67}\ee
where $A_1^c$, $A_2^c$ are integration constants. However, these can be determined by looking at the definition of the first order formalism functions $T_1,T_2$. Specifically, to leading order the exponentials $e^{2A_i}$ are

\be
e^{2A_2} = \frac{32}{L^2(3\mathcal{R}_1 +5 \mathcal{R}_2)} \vert \f_+ \vert^{-\frac{2}{\Delta}} e^{-2 \frac{u}{\ell}} \left( 1 + \mathcal{O} ( e^{2 \frac{u}{\ell}}) \right),
\label{F68}\ee
\be
e^{2A_1}  =  \frac{128( \mathcal{R}_1 + \mathcal{R}_2 )}{L^2 \left(5 \mathcal{R}_2 + 3\mathcal{R}_1 \right)^2} \vert \f_+ \vert^{-\frac{2}{\Delta}} e^{-2 \frac{u}{\ell}} \left( 1 + \mathcal{O} ( e^{2 \frac{u}{\ell}}) \right).
\label{F69}\ee

A direct comparison of the previous two equations with equations (\ref{F66}), (\ref{F67}) yields for the integration constants the following:

\be
 e^{2 A_{1}^{c}} =  \frac{128( \mathcal{R}_1 + \mathcal{R}_2 )}{L^2 \left(5 \mathcal{R}_2 + 3\mathcal{R}_1 \right)^2} \vert \f_+ \vert^{-\frac{2}{\Delta}},
 \label{F70}\ee
 \be
 e^{2 A_{2}^{c}} = \frac{32}{L^2(3\mathcal{R}_1 +5 \mathcal{R}_2)} \vert \f_+ \vert^{-\frac{2}{\Delta}}.
\label{F71}\ee

Now that we have determined these integration constants, we can continue our analysis. By looking at equations (\ref{F66}), (\ref{F67}) we observe that we can write the exponentials $e^{2A_i}$ as follows:

\be
e^{2A_1} = e^{2A_1^c} e^{-2 \frac{u}{\ell}} \left( 1 - \frac{\f_+^2 \ell^{2\Delta}}{16} e^{2\Delta \frac{u}{\ell}} - \frac{\mathcal{R}_1 \vert \f_+ \vert^{\frac{2}{\Delta}} \ell^2}{64} e^{2 \frac{u}{\ell}} + \frac{C_2 \ell^3 \vert \f_+ \vert^{\frac{3}{\Delta}}}{6} e^{3 \frac{u}{\ell}} + \ldots \right),
\label{F72}\ee
\be
e^{2A_2} = e^{2A_2^c} e^{-2 \frac{u}{\ell}} \left( 1 - \frac{\f_+^2 \ell^{2\Delta}}{16} e^{2\Delta \frac{u}{\ell}} - \frac{\mathcal{R}_2 \vert \f_+ \vert^{\frac{2}{\Delta}} \ell^2}{64} e^{2 \frac{u}{\ell}} - \frac{C_2 \ell^3 \vert \f_+ \vert^{\frac{3}{\Delta}}}{12} e^{3 \frac{u}{\ell}} + \ldots \right),
\label{F73}\ee
where the $\ldots$ indicate higher order terms. We want to write metric in PG coordinates like in (\ref{fe3}), and in order to do that we are going to make the following substitution:

\be
e^{\frac{u}{\ell}} = z.
\label{F74}\ee

The exponentials can be written in terms of z as,

\be
\frac{z^2}{\ell^2}L^2 e^{2A_1} = \frac{128( \mathcal{R}_1 + \mathcal{R}_2 )}{\left(5 \mathcal{R}_2 + 3\mathcal{R}_1 \right)^2}\left|\ell^{\Delta} \varphi_{+}\right|^{-\frac{2}{\Delta}}  \left( 1 - \frac{\f_+^2 \ell^{2\Delta}}{16} z^{2\Delta} - \frac{\mathcal{R}_1 \vert \f_+ \vert^{\frac{2}{\Delta}} \ell^2}{64} z^2 \right.
\label{F75}\ee
$$ \left. + \frac{C_2 \ell^3 \vert \f_+ \vert^{\frac{3}{\Delta}}}{6} z^3 + \ldots \right) ,$$
\be
\frac{z^2}{\ell^2}L^2 e^{2A_2} = \frac{32}{3\mathcal{R}_1 +5 \mathcal{R}_2}\left|\ell^{\Delta} \varphi_{+}\right|^{-\frac{2}{\Delta}}  \left( 1 - \frac{\f_+^2 \ell^{2\Delta}}{16} z^{2\Delta} - \frac{\mathcal{R}_2 \vert \f_+ \vert^{\frac{2}{\Delta}} \ell^2}{64} z^2 \right.
\label{F76}\ee
$$ \left.  - \frac{C_2 \ell^3 \vert \f_+ \vert^{\frac{3}{\Delta}}}{12} z^3 + \ldots \right) .$$

From the previous expansions we immediately identify the coefficients of the term $z^3$, and therefore we can easily find the metric $g_{ij}^{(3)}$:

\be
g_{ij}^{(3)} dx^i dx^j = \frac{8 C_2 \ell \vert \f_+ \vert^{\frac{1}{\Delta}}}{3(3\mathcal{R}_1 + 5 \mathcal{R}_2)} \left(  \frac{8( \mathcal{R}_2 +  \mathcal{R}_1)}{3\mathcal{R}_1 + 5 \mathcal{R}_2} (d\psi + \cos\theta d\phi)^2 - d\Omega^2 \right),
\label{F77}\ee
and therefore, by virtue of (\ref{fe6}), the vev of the stress-energy tensor is

\be
\left\langle T_{ij} \right\rangle = \frac{3\ell^2}{16\pi G_N} g_{ij}^{(3)}.
\label{F78}\ee

Finally, the metric near $\f_0$ is,

\be
d s^{2} \approx d u^{2}+\frac{32\vert  \varphi_{+} \vert^{-\frac{2}{\Delta}}}{3\mathcal{R}_{1}+5\mathcal{R}_{2}} e^{-2 \frac{u}{\ell}}\left(\frac{4( \mathcal{R}_{1} +\mathcal{R}_2 )}{3\mathcal{R}_{1}+5\mathcal{R}_{2}}(d \psi+\cos \theta d \phi)^{2}+d \Omega^{2}\right).
\label{F79}\ee

We observe that as $\f \to \f_0$, the geometry becomes the same as the one in the boundary of AdS.
Comparing with (\ref{ga1}) we see that the squashing parameter $a$ and the length $L$ can be related to $\mathcal{R}_1,\mathcal{R}_2,\f_+$ as follows:
\be
a^2=\frac{4( \mathcal{R}_{1} + \mathcal{R}_{2})}{(3 \mathcal{R}_{1}+5 \mathcal{R}_{2})}\sp  L^2=\frac{128}{3\mathcal{R}_1 + 5\mathcal{R}_2} \left|\varphi_{+}\right|^{-\frac{2}{\Delta}},
\ee
so that
\be
\frac{\mathcal{R}_{1}}{\mathcal{R}_{2}}= \frac{5a^2 - 4}{4 -3a^2}.
\ee

We also denote
\be
R^{uv} = \mathcal{R}\vert \f_+ \vert^{\frac{2}{\Delta}}\to    \mathcal{R}=\frac{\mathcal{R}_1 + 2 \mathcal{R}_2}{8},
\ee
where $R^{uv}$ is the scalar curvature of the squashed sphere metric.

We can now rewrite $\mathcal{R}_{1,2}$ as a function of $a,\mathcal{R}$ as
\be
\mathcal{R}_1=8 \frac{5a^2-4}{4-a^2}\mathcal{R}
\sp \mathcal{R}_2= 8 \frac{4 - 3a^2}{4-a^2}\mathcal{R}.
\ee

We note that the previous expressions are exactly the same as the ones obtained in (\ref{asquared})-(\ref{dimensionlesscurvature}), if we replace $\f_-$ with $\f_+$.

This concludes the analysis of the (+)-branch. Some remarks are in order:

\begin{enumerate}
   \item[$\bullet$] The integration constants $\mathcal{R}_1,\mathcal{R}_2$ once again control the dimensionless curvature $\mathcal{R}$ and the squashing parameter $a^2$.
   \item[$\bullet$] The integration constant $C_2$ controls only one vev, namely the vev of the stress-energy tensor.
   \item[$\bullet$] Although the source of the dual operator $\mathcal{O}$ vanishes, its vev is non zero and is controlled by the integration constant $\f_+$.
\end{enumerate}

Therefore, the $(+)$ branch solution corresponds to a UV fixed point perturbed by the vev of a relevant operator.

\subsection{Exotic Solutions\label{exot}}

Similarly to the previous subsection, we shall attempt to set up the perturbation theory around the leading-order solutions, in order to justify whether these correspond to bona-fide asymptotic solutions.

We shall proceed as follows: first, we shall investigate the $\nu_+$ and $\nu_-$ solutions, and afterwards the $B_3$ solution. Similarly to the (+) and (-) branches, the $\nu_{\pm}$ solutions can be studied together up to a certain point.

We begin to discuss each one in turn.

\subsubsection{The $\nu_{\pm}$ solutions}

Similarly to the previous subsections, we shall first calculate the leadin-order solutions. To this end, we expand the superpotentials in a regular power series with leading powers determined by (\ref{g125}),(\ref{g126}):

\be
S_{L,\nu_\pm} = x \sum_{n=0}^\infty S_{n,\nu_\pm} x^n \ , \ W_{1L,\nu_\pm} = \sum_{n=0}^\infty W_{(1)n,\nu_\pm} x^n \ , \ W_{2L,\nu_\pm} = \sum_{n=0}^\infty W_{(2)n,\nu_\pm} x^n ,
\label{F80}\ee
\be
T_{1L,\nu_\pm} = \sum_{n=0}^\infty T_{(1)n,\nu_\pm} x^n \ , \ T_{2L,\nu_\pm} = x^{\nu_{\pm}} \sum_{n=0}^\infty \tilde{T}_{(2)n,\nu_\pm} x^n \ , \
\label{F81}\ee
where the constants $\nu_\pm$ where defined in (\ref{g125})-(\ref{g126}), but we rewrite them here for convenience:

\be
\nu_\pm = \pm \frac{1}{\sqrt{\frac{(\Delta-1)(\Delta-2)}{2}}\pm 1}.
\label{F82}\ee

Inserting the previous expansions into the equations of motion (\ref{ga51})-(\ref{ga55}),(\ref{ga54}) and solving recursively for the first few coefficients, we find the following:
\be
S_{L,\nu_{\pm}} =  \frac{\sqrt{2}}{\ell \nu_{\pm}}x + \frac{3\nu_{\pm} \ell V_3}{\sqrt{2}(3-2\nu_\pm)} x^2  +\frac{\frac{4\left(2 \nu_{ \pm}^3+27 \nu_{ \pm}^2+66 \nu_{ \pm}-72\right)}{\ell \nu_{ \pm}\left(12 \nu_{ \pm}^2+33 \nu_{ \pm}-32\right)\left(12+\nu_{ \pm}\right)}+4 V_4 \ell \nu_{ \pm}-\frac{9 V_3^2 \ell^3 \nu_{ \pm}^3}{\left(3-2 \nu_{ \pm}\right)^2}}{2 \sqrt{2}\left(2-\nu_{ \pm}\right)}x^3 + \mathcal{O}(x^4) \ , \
\label{F87}\ee
\be
W_{1L,\nu_\pm} = \frac{4\sqrt{2}}{\ell} + \frac{8 \sqrt{2}\left(\nu_{ \pm}^2+\nu_{ \pm}-2\right)}{\left(12 \nu_{ \pm}^2+33 \nu_{ \pm}-32\right) \ell \nu_{ \pm}} x^2 + \mathcal{O}(x^3) \ , \
\label{F88}\ee
\be
W_{2L,\nu_{\pm}} = \frac{2\sqrt{2}}{\ell} +  \frac{4 \sqrt{2}\left(\nu_{ \pm}^3+14 \nu_{ \pm}^2+56 \nu_{ \pm}-48\right)}{\left(12+\nu_{ \pm}\right)\left(12 \nu_{ \pm}^2+33 \nu_{ \pm}-32\right) \ell \nu_{ \pm}} x^2 + \mathcal{O}(x^3)  \ , \
\label{F89}\ee
\be
T_{1L,\nu_{\pm}} = \frac{8}{\ell^2} + \frac{16\left(2 \nu_{ \pm}-1\right)}{\left(12 \nu_{ \pm}^2+33 \nu_{ \pm}-32\right) \ell^2} x^2 + \mathcal{O}(x^3) \ , \
\label{F90}\ee
\be
T_{2L,\nu_{\pm}} = 0 .
\label{F91}\ee

As the reader can observe from equations (\ref{F87}) - (\ref{F91}), up to second order no integration constants have entered the expansions. Since we can recursively obtain the higher order coefficients from the lower order ones, we conclude that no integration constants appear in the leading order expansions.
We now move  to the linearised equations of motion, (\ref{F19})-(\ref{F21}),(\ref{F22}) - (\ref{F23}). Specifically, we make the following ansatz for the deformations $\delta S,\delta W_2$,

\be
\delta S = \mathcal{A} x^\lambda \ , \ \delta W_2 = \mathcal{B} x^g,
\label{F92}\ee
where the constants $\mathcal{A},\lambda,\mathcal{B},g$, as explained in subsection \ref{UVfixedpoints}, must obey the following constraints in order for the corresponding expansions to be subleading:
\be
\lambda \geq 1\ , \ g \geq 0 \ , \ \mathcal{A},\mathcal{B} \neq 0.
\label{F93}\ee

Substituting (\ref{F92}) along with the expansions (\ref{F87}) - (\ref{F91}) into the linearised equations of motion (\ref{F22})-(\ref{F23}) yields the following to leading order in $\mathcal{A},\mathcal{B}$:

\be
-2 \mathcal{B} x^{1+g} \left( -6\nu_{\pm}^2 -5\nu_{\pm}g + 3g^2 \right) +
\label{F94}\ee
$$  +4\mathcal{A} x^{\lambda} \left[ 6\nu_{\pm}^3 + 2(\lambda-1)^2(\lambda+1)+ \nu_{\pm}^2(11\lambda -17) - \nu_{\pm}(\lambda-1)(11\lambda+3)\right] + \ldots = 0, $$
\be
2\mathcal{B} x^{1+g} (g-\nu_{\pm})(3g-2\nu_{\pm}) -
\label{F95}\ee
$$ -4 \mathcal{A} x^{\lambda} \left[ -26\nu_{\pm}^3 + 2 (\lambda-1)^2(\lambda+1)  + \nu_{\pm}^2(43\lambda-17) - \nu_{\pm}(\lambda-1)(19\lambda+11) \right]+ \ldots = 0$$

By examining equation (\ref{F94}), we observe that we have three choices:

\begin{enumerate}
   \item[$\bullet$] We can choose

   \be
   g+1 < \lambda .
   \label{F96}\ee

   In this case the first term is the leading one, and therefore it must vanish on its own. Equations (\ref{F94}),(\ref{F95}) now yield,

   \be
   \mathcal{B}(-6 \nu_{\pm}^2 - 5 \nu_{\pm}g +3g^2) = 0 \sp \mathcal{B}(g- \nu_{\pm})(3g-2 \nu_{\pm}) = 0 .
   \label{F97}\ee

   The only solution to these equations is $ \mathcal{B} = 0$, which is not acceptable.
   \item[$\bullet$] We can choose

   \be
   g +1 > \lambda.
   \label{F98}\ee

   In this case the second term is the leading one, and therefore it must vanish on its own. Equations (\ref{F94}),(\ref{F95}) now yield,

   \be
   \mathcal{A} \left( 6 \nu_{\pm}^3 + 2(\lambda-1)^2(\lambda+1)+  \nu_{\pm}^2(11\lambda -17) -  \nu_{\pm}(\lambda-1)(11\lambda+3)\right) = 0 \ , \
   \ee
   \be
   \mathcal{A} \left( -26 \nu_{\pm}^3 + 2 (\lambda-1)^2(\lambda+1) +  \nu_{\pm}^2(43\lambda-17) -  \nu_{\pm}(\lambda-1)(19\lambda+11) \right) = 0 .
   \ee

   The previous set of equations has two solutions: the first one is to set $\mathcal{A} = 0$, which is not acceptable. The second one is to set

   \be
   \lambda = -1 +2 \nu_{\pm} .
   \label{F99}\ee

   However, since $\lambda \geq 1$, we must demand that $\nu_{\pm}$ satisfies the condition,

   \be
   \nu_{\pm} \geq 1 .
   \label{F100}\ee

   If $\nu_{\pm}$ is less than one, then we shall set the constant accompanying this expansion equal to 0.
   \item[$\bullet$] The final possibility is

   \be
   g +1 = \lambda .
   \label{F101}\ee

   The two terms are of the same order, and therefore they must vanish as a whole. Equations (\ref{F94}),(\ref{F95}) now become

   \be
   2 \mathcal{A }\left( \nu_{\pm}^2 (11 g-6)+6  \nu_{\pm}^3- \nu_{\pm} g (11 g+14)+\left(2 g^2\right) (g+2)\right)
   \label{F102}\ee
   $$ +\mathcal{B} \left(6  \nu_{\pm}^2+5  \nu_{\pm} g-3 g^2\right) = 0 \ , \ $$
   \be
   2 \mathcal{A} \left((g- \nu_{\pm}) \left(-17  \nu_{\pm} g+26  \nu_{\pm} ( \nu_{\pm}-1)+2 g^2+4 g\right)\right)-\mathcal{B} (g- \nu_{\pm}) (3g-2 \nu_{\pm}) = 0,
   \label{F103}\ee
{or in matrix form,
   \be
   T  \begin{pmatrix}
      2\mathcal{A} \\
      \mathcal{B}
   \end{pmatrix} = 0,
   \ee
   where $T$ is the matrix
   \be
   T = \begin{pmatrix}
      \nu_{\pm}^2 (11 g-6)+6  \nu_{\pm}^3- \nu_{\pm} g (11 g+14)+\left(2 g^2\right) (g+2) & 6  \nu_{\pm}^2+5  \nu_{\pm} g-3 g^2 \\
      (g- \nu_{\pm}) \left(-17  \nu_{\pm} g+26  \nu_{\pm} ( \nu_{\pm}-1)+2 g^2+4 g\right) &- (g- \nu_{\pm}) (3g-2 \nu_{\pm})
   \end{pmatrix}.
   \ee
   In order for our system to have a non trivial solution, we must require the determinant of $T$ to be zero:
   \be
   \det (T) = 0 \implies -24(g - 3 \nu_{\pm}) (g+2- 2 \nu_{\pm}) (g - \nu_{\pm}) (g + \nu_{\pm}) = 0.
   \ee
   The previous equation, along with }(\ref{F102}),(\ref{F103}){, have 4 solutions, namely:}

   \begin{enumerate}
      {\item[1.]
      \be
      g = 3\nu_{\pm}, \mathcal{A} = -\frac{\mathcal{B}}{2(\nu_{\pm}+2)}.
      \label{F104}\ee

      Since $ \nu_{\pm} > 0 \implies g = 3 \nu_{\pm} > 0$, therefore it is acceptable. This solution accounts for the 2nd out of the 4 integration constants.
      \item[2.]
      \be
      g =  \nu_{\pm} , \mathcal{A} = -\frac{\mathcal{B}}{2(\nu_{\pm}-2)}
      \label{F105}\ee
      Since $ \nu_{\pm} > 0 \implies g =  \nu_{\pm} > 0$, therefore it is acceptable. This solution accounts for the 3rd out of the 4 integration constants.
      \item[3.]
      \be
      g = - \nu_{\pm} , \mathcal{A} = -\frac{\mathcal{B}}{6(3 \nu_{\pm}-2)}
      \label{F106}\ee
      Since $ \nu_{\pm}> 0 \implies g < 0$, therefore it is not acceptable. This solution accounts for the 4th out of the 4 integration constants, and we intentionally put it equal to 0.
      \item[4.]
      \be
      g = 2(\nu_{\pm}-1) \ , \ \mathcal{B} = 0 \ , \ \mathcal{A} = \text{uknown}
      \ee
      Since we must have $g \geq 0$, we must demand $\nu_{\pm} \geq 1$. However, since the integration constant $\mathcal{B}$ is zero, we deduce that this case is not acceptable.}
\end{enumerate}
\end{enumerate}

This concludes the analysis of the linearised equations of motion. In total, we have found 4 integration constants, as expected. Moreover, one of them was intentionally put equal to 0. {Before we write the full form of the deformations $\delta S, \delta W_2$, we note that apart from condition} (\ref{F99}){, the rest are satisfied simultaneously, i.e. when $\nu_{\pm} > 0$. Therefore, for convenience, we shall add them all together, but put the term accompanying the exponent }(\ref{F99}) {in parentheses in order to emphasise this.} The full form of the deformations $\delta S, \delta W_2$ is,

\be
\delta S \approx \mathcal{A}_1 x^{1+ \nu_{\pm}}  +\mathcal{A}_2 x^{1+3 \nu_{\pm}} + (\mathcal{A}_3 x^{2\nu_\pm -1}) \ , \ \delta W_2 = 2(2- \nu_{\pm})\mathcal{A}_1 x^{ \nu_{\pm}} -  2( \nu_{\pm}+2)\mathcal{A}_2 x^{3 \nu_{\pm}}.
\label{F107}\ee

Substituting the previous expansions into equations (\ref{F19}) - (\ref{F21}) yields the following for the functions $\delta W_1, \delta T_1 , \delta T_2$:

\be
\delta W_1 \approx 8 \mathcal{A}_2(\nu_{\pm}+2)x^{3\nu_{\pm}},
\label{F108}\ee
\be
\delta T_1 \approx - \frac{8\sqrt{2}}{\ell}(\nu_{\pm}+2)\mathcal{A}_2 x^{3\nu_{\pm}}  + \frac{8\sqrt{2}}{\ell}(2-\nu_{\pm})\mathcal{A}_1 x^{\nu_{\pm}}+ ( - \frac{8\sqrt{2}}{3\ell \nu_\pm} \mathcal{A}_3 x^{2\nu_{\pm}} ),
\label{F109}\ee
\be
\delta T_2 \approx - \frac{8\sqrt{2}}{3\nu_{\pm} \ell} \mathcal{A}_2 x^{2+3\nu_{\pm}} +\frac{8\sqrt{2}}{\ell}(2-\nu_{\pm})\mathcal{A}_1 x^{\nu_{\pm}} + (- \frac{8\sqrt{2}}{3\ell \nu_\pm} \mathcal{A}_3 x^{2\nu_{\pm}}).
\label{F110}\ee

For convenience, we shall redefine the integration constants as follows:

\be
\mathcal{A}_1 = \frac{\ell}{8\sqrt{2}(2-\nu_{\pm})} T_{(2)0} \ , \ \mathcal{A}_2 = \frac{\mathcal{C}}{8(2+\nu_\pm)} \ , \ \mathcal{A}_3 = -\frac{3\ell \nu_{\pm}\mathcal{D}}{8\sqrt{2}},
\label{F111}\ee
where the new integration constants now are $T_{(2)0},\mathcal{C},\mathcal{D}$. The expansions (\ref{F107}) - (\ref{F110}) now become,

\be
\delta S \approx \frac{\ell}{8\sqrt{2}(2-\nu_{\pm})} T_{(2)0} x^{1 + \nu_\pm} + \frac{\mathcal{C}}{8(2+\nu_\pm)} x^{1+3\nu_{\pm}} +( -\frac{3\ell \nu_{\pm}\mathcal{D}}{8\sqrt{2}} x^{2\nu_\pm -1}) ,
\label{F112}\ee
\be
\delta W_2 \approx \frac{\ell}{4\sqrt{2}} T_{(2)0} x^{\nu_\pm} - \frac{\mathcal{C}}{4} x^{3\nu_\pm} \ , \ \delta W_1 \approx \mathcal{C} x^{3\nu_\pm} ,
\label{F113}\ee
\be
\delta T_1 = T_{(2)0} x^{\nu_\pm} - \frac{\sqrt{2}}{\ell} \mathcal{C} x^{3\nu_\pm}  + (  \mathcal{D} x^{2\nu_\pm})  \ ,
\label{F114}\ee
\be
\delta T_2 = T_{(2)0} x^{\nu_\pm} - \frac{\sqrt{2}}{3 \nu_{\pm}\ell(2+\nu_\pm)} \mathcal{C} x^{2+3\nu_\pm} +  ( \mathcal{D} x^{2\nu_\pm}).
\label{F115}\ee

Putting everything together, we are now in a position to write down the expressions for $W_{1,2},T_{1,2},S$ in the vicinity of an extremum of V and up to first order in the integration constants $T_{(2)0},\cal{C},\cal{D}$. We are thus ready to discuss the $\nu_-$ and $\nu_+$ solutions separately.

\subsubsection{The $\nu_{-}$ solution}

This solution exists only when $\Delta$ takes the values $2 \leq \Delta < 3$, i.e. in the vicinity of maxima of the potential. For these values of $\Delta$, the parameter $\nu_-$ takes values in the range $[1,\infty)$, and therefore satisfies the constraint (\ref{F100}). The full expansions of the superpotentials are thus:

$$
S = \frac{\sqrt{2}}{\ell \nu_{-}}x + \frac{3\nu_{-} \ell V_3}{\sqrt{2}(3-2\nu_-)} x^2+\frac{\frac{4\left(2 \nu_{ -}^3+27 \nu_{ -}^2+66 \nu_{ -}-72\right)}{\ell \nu_{ -}\left(12 \nu_{ -}^2+33 \nu_{ -}-32\right)\left(12+\nu_{ -}\right)}+4 V_4 \ell \nu_{ -}-\frac{9 V_3^2 \ell^3 \nu_{ -}^3}{\left(3-2 \nu_{ -}\right)^2}}{2 \sqrt{2}\left(2-\nu_{ -}\right)}x^3 + \mathcal{O}(x^4) +
$$
$$
+ \frac{\mathcal{C}}{8(2+\nu_-)} x^{1+3\nu_{-}}(1+\mathcal{O}(x))   + \frac{\ell}{8\sqrt{2}(2-\nu_{-})} T_{(2)0} x^{1 + \nu_-}(1+\mathcal{O}(x)) -
$$
\be
-\frac{3\ell \nu_{-}\mathcal{D}}{8\sqrt{2}} x^{2\nu_- -1}(1+\mathcal{O}(x)) ,
\label{F116}\ee
\be
W_1 = \frac{4\sqrt{2}}{\ell} + \frac{8 \sqrt{2}\left(\nu_{ -}^2+\nu_{ -}-2\right)}{\left(12 \nu_{ -}^2+33 \nu_{ -}-32\right) \ell \nu_{ -}} x^2 + \mathcal{O}(x^3) + \mathcal{C} x^{3\nu_-} (1 + \mathcal{O}(x)) ,
\label{F117}\ee
$$
W_2 = \frac{2\sqrt{2}}{\ell} + \frac{4 \sqrt{2}\left(\nu_{ -}^3+14 \nu_{ -}^2+56 \nu_{ -}-48\right)}{\left(12+\nu_{ -}\right)\left(12 \nu_{ -}^2+33 \nu_{ -}-32\right) \ell \nu_{ -}}
x^2 + \mathcal{O}(x^3)+
$$
\be
+  \frac{\ell}{4\sqrt{2}} T_{(2)0} x^{\nu_-}(1 + \mathcal{O}(x)) - \frac{\mathcal{C}}{4} x^{3\nu_-}(1 + \mathcal{O}(x)),
\label{F118}\ee
$$
T_1 = \frac{8}{\ell^2} +\frac{16\left(2 \nu_{ -}-1\right)}{\left(12 \nu_{ -}^2+33 \nu_{ -}-32\right) \ell^2} x^2 + \mathcal{O}(x^3) +
$$
\be
+ T_{(2)0} x^{\nu_-}(1 + \mathcal{O}(x)) + \mathcal{D} x^{2\nu_-}(1 + \mathcal{O}(x)) - \frac{\sqrt{2}}{\ell} \mathcal{C} x^{3\nu_-}(1 + \mathcal{O}(x)) ,
\label{F119}\ee
\be
T_2 = T_{(2)0} x^{\nu_-}(1 + \mathcal{O}(x)) + \mathcal{D} x^{2\nu_-}(1 + \mathcal{O}(x)) - \frac{\sqrt{2}}{3 \nu_{-}\ell(2+\nu_-)} \mathcal{C} x^{2+3\nu_-}(1 + \mathcal{O}(x)),
\label{F120}\ee
where $\mathcal{C},T_{(2)0},\mathcal{D}$ are integration constants, and $\nu_-$ is given by,

\be
\nu_- = \frac{1}{1-\sqrt{\frac{(\Delta-1)(\Delta-2)}{2}}} \ , \ 2 \leq \Delta < 3.
\label{F121}\ee

Since $S = \dot{\f}$, we have to leading order,

\be
\f - \f_0 = e^{\frac{\sqrt{2}}{\ell \nu_{\text{-}}} (u-u_0)} \left( 1 + \ldots \right),
\label{F122}\ee
where $u_0$ is an integration constant. Since $\nu_- > 0$, we observe that as $\f \to \f_0$, we have $u \to - \infty$.  Moreover, from the previous equation we observe that we can identify the term $e^{-\frac{\sqrt{2}}{\ell \nu_-} u_0}$ as the free source for the scalar, although this example does not follow the FG expansion.

Continuing, by using the definition of the functions $W_1,W_2$ in equations \\ (\ref{ga46}),(\ref{ga47}), we observe that we can write the functions $A_1,A_2$ in terms of u as follows:

\be
A_1 = A_1^c - \frac{\sqrt{2}}{\ell} (u-u_0) -\frac{\nu_{ -}^2+\nu_{ -}-2}{\left(12 \nu_{ -}^2+33 \nu_{ -}-32\right) \nu_{ -}} e^{\frac{2\sqrt{2}}{\nu_-} \frac{u-u_0}{\ell}} -
\label{F123}\ee
$$ - \frac{\sqrt{2}\ell \mathcal{C}}{24} e^{3\sqrt{2} \frac{u-u_0}{\ell}} + \ldots , $$
\be
A_2 = A_2^c - \frac{u-u_0}{\sqrt{2} \ell} - \frac{\nu_{ -}^3+14 \nu_{ -}^2+56 \nu_{ -}-48}{2\left(12+\nu_{ -}\right)\left(12 \nu_{ -}^2+33 \nu_{ -}-32\right)\nu_{ -}} e^{\frac{2\sqrt{2}}{\nu_{\text{-}}} \frac{u-u_0}{\ell}} +
\label{F124}\ee
$$ + \frac{\sqrt{2}\ell\mathcal{C}}{96} e^{3\sqrt{2} \frac{u-u_0}{\ell}} - \frac{\ell^2 T_{(2)0}}{32} e^{\sqrt{2} \frac{u-u_0}{\ell}} + \ldots , $$
where $A_1^c,A_2^c$ are integration constants that can be determined by substituting the expansions (\ref{F119}),(\ref{F120}), (\ref{F123}) - (\ref{F124}) into equations (\ref{ga48}) - (\ref{ga49}) and solving them to leading order. Specifically, to leading order the exponentials $e^{2A_i}$ are,

\be
e^{2A_2} = \frac{4}{L^2 T_{(2)0}} e^{-\sqrt{2} \frac{u-u_0}{\ell}} \left( 1 + \mathcal{O} \left( e^{\sqrt{2} \frac{u-u_0}{\ell}} \right) \right) ,
\label{F125}\ee
\be
e^{2A_1} = \frac{32}{\ell^2L^2 T_{(2)0}^2}  e^{-2\sqrt{2} \frac{u-u_0}{\ell}} \left( 1 + \mathcal{O} \left( e^{\sqrt{2} \frac{u-u_0}{\ell}} \right) \right) .
\label{F126}\ee

Upon comparing the previous expressions for the exponentials with the ones obtained from (\ref{ga48}) - (\ref{ga49}), we obtain the following:

\be
e^{2A_2^c} = \frac{4}{L^2 T_{(2)0}},
\label{F127}\ee
\be
e^{2A_1^c} = \frac{32}{\ell^2L^2 T_{(2)0}^2}  .
\label{F128}\ee

Finally, the metric near $\f_0$ is,

\be
ds^2 \approx du^2 + \frac{32}{\ell^2 T_{(2)0}^2} e^{-2\sqrt{2} \frac{u-u_0}{\ell}} \left( (d\psi + \cos\theta d\phi)^2 + \frac{\ell^2 T_{(2)0}}{8} e^{\sqrt{2} \frac{u-u_0}{\ell}} d\Omega^2 \right),
\label{F129}\ee
which has a Kretschmann scalar equal to,

\be
\mathcal{K} = \frac{18}{\ell^4},
\label{F130}\ee
which is regular.

As we noted, as $\f \to \f_0$, the coordinate $u$ approaches $-\infty$. Consequently, from (\ref{F129}) we observe that both scale factors of the $S^2$ and the $S^1$ fiber diverge. Therefore, this point is a boundary point, however it is not FG since the scale factors do not diverge at the same rate.

\subsubsection{The $\nu_+$ solution}

This solution exists only when $\Delta >2$, i.e. in the vicinity of maxima as well as minima of the potential. For these values of $\Delta$, the parameter $\nu_+$ takes values in the range $(0,1)$, and therefore it does not satisfy the constraint (\ref{F100}). This, in turn, implies that we must set $\mathcal{D} = 0$ in order to ensure that the expansions found in (\ref{F112}) - (\ref{F115}) are subleading. The full expansions of the superpotentials are thus:

$$
S =  \frac{\sqrt{2}}{\ell \nu_{+}}x + \frac{3\nu_{+} \ell V_3}{\sqrt{2}(3-2\nu_+)} x^2 +\frac{\frac{4\left(2 \nu_{ +}^3+27 \nu_{ +}^2+66 \nu_{ +}-72\right)}{\ell \nu_{ +}\left(12 \nu_{ +}^2+33 \nu_{ +}-32\right)\left(12+\nu_{ +}\right)}+4 V_4 \ell \nu_{ +}-\frac{9 V_3^2 \ell^3 \nu_{ +}^3}{\left(3-2 \nu_{ +}\right)^2}}{2 \sqrt{2}\left(2-\nu_{ +}\right)}x^3 + \mathcal{O}(x^4)  +
$$
\be  +\frac{\mathcal{C}}{8(2+\nu_+)} x^{1+3\nu_{+}}(1+\mathcal{O}(x))  +  + \frac{\ell}{8\sqrt{2}(2-\nu_{+})} T_{(2)0} x^{1 + \nu_+}(1+\mathcal{O}(x))  ,
\label{F131}\ee
\be
W_1 = \frac{4\sqrt{2}}{\ell} + \frac{8 \sqrt{2}\left(\nu_{ +}^2+\nu_{ +}-2\right)}{\left(12 \nu_{ +}^2+33 \nu_{ +}-32\right) \ell \nu_{ +}}
x^2 + \mathcal{O}(x^3) + \mathcal{C} x^{3\nu_+} (1 + \mathcal{O}(x)) ,
\label{F132}\ee
$$
W_2 = \frac{2\sqrt{2}}{\ell} + \frac{4 \sqrt{2}\left(\nu_{ +}^3+14 \nu_{ +}^2+56 \nu_{ +}-48\right)}{\left(12+\nu_{ +}\right)\left(12 \nu_{ +}^2+33 \nu_{ +}-32\right) \ell \nu_{ +}} x^2 + \mathcal{O}(x^3)+
$$
\be
+  \frac{\ell}{4\sqrt{2}} T_{(2)0} x^{\nu_+}(1 + \mathcal{O}(x)) - \frac{\mathcal{C}}{4} x^{3\nu_+}(1 + \mathcal{O}(x)),
\label{F133}\ee
$$
T_1 = \frac{8}{\ell^2} +\frac{16\left(2 \nu_{ +}-1\right)}{\left(12 \nu_{ +}^2+33 \nu_{ +}-32\right) \ell^2} x^2 + \mathcal{O}(x^3) +
$$
\be
+ T_{(2)0} x^{\nu_+}(1 + \mathcal{O}(x)) - \frac{\sqrt{2}}{\ell} \mathcal{C} x^{3\nu_+}(1 + \mathcal{O}(x)) ,
\label{F134}\ee
\be
T_2 = T_{(2)0} x^{\nu_+}(1 + \mathcal{O}(x)) - \frac{\sqrt{2}}{3 \nu_{+}\ell(2+\nu_+)} \mathcal{C} x^{2+3\nu_+}(1 + \mathcal{O}(x)),
\label{F135}\ee
where $\mathcal{C},T_{(2)0}$ are integration constants, and $\nu_+$ is given by,

\be
\nu_+ = \frac{1}{1+\sqrt{\frac{(\Delta-1)(\Delta-2)}{2}}} \ , \ \Delta > 2.
\label{F136}\ee

Since $S = \dot{\f}$, we have to leading order,

\be
\f - \f_0 = e^{\frac{\sqrt{2}}{\nu_+} \frac{u-u_0}{\ell}} \left( 1 + \ldots \right),
\ee
where $u_0$ is an integration constant. Note that as $\f \to \f_0$, u approaches $- \infty$. { Moreover, from the previous equation we observe that we can identify the term $e^{-\frac{\sqrt{2}}{\ell \nu_+} u_0}$ as the free source for the scalar as in the previous case.
}

Continuing, by using the definition of the functions $W_1,W_2$ in equations \\ (\ref{ga46}),(\ref{ga47}), we observe that we can write the functions $A_1,A_2$ in terms of u as follows:

\be
A_1 = A_1^c - \frac{\sqrt{2}}{\ell} (u-u_0) -\frac{\nu_{+}^2+\nu_{+}-2}{\left(12 \nu_{+}^2+33 \nu_{+}-32\right) \nu_{+}} e^{\frac{2\sqrt{2}}{\nu_{+}} \frac{u-u_0}{\ell}} -
\label{F137}\ee
$$ - \frac{\sqrt{2}\ell \mathcal{C}}{24} e^{3\sqrt{2} \frac{u-u_0}{\ell}} + \ldots , $$
\be
A_2 = A_2^c - \frac{u-u_0}{\sqrt{2} \ell} - \frac{\nu_{+}^3+14 \nu_{+}^2+56 \nu_{+}-48}{2\left(12+\nu_{+}\right)\left(12 \nu_{+}^2+33 \nu_{+}-32\right)\nu_{+}} e^{\frac{2\sqrt{2}}{\nu_{+}} \frac{u-u_0}{\ell}} +
\label{F138}\ee
$$ + \frac{\sqrt{2}\ell\mathcal{C}}{96} e^{3\sqrt{2} \frac{u-u_0}{\ell}} - \frac{\ell^2 T_{(2)0}}{32} e^{\sqrt{2} \frac{u-u_0}{\ell}} + \ldots , $$
where $A_1^c,A_2^c$ are integration constants that can be determined by substituting the expansions (\ref{F134}),(\ref{F135}), (\ref{F137}) - (\ref{F138}) into equations (\ref{ga48}) - (\ref{ga49}) and solving them to leading order. Specifically, to leading order the exponentials $e^{2A_i}$ are,

\be
e^{2A_2} = \frac{4}{L^2 T_{(2)0}} e^{-\sqrt{2} \frac{u-u_0}{\ell}} \left( 1 + \mathcal{O} \left( e^{\sqrt{2} \frac{u-u_0}{\ell}} \right) \right) ,
\label{F139}\ee
\be
e^{2A_1} = \frac{32}{\ell^2L^2 T_{(2)0}^2}  e^{-2\sqrt{2} \frac{u-u_0}{\ell}} \left( 1 + \mathcal{O} \left( e^{\sqrt{2} \frac{u-u_0}{\ell}} \right) \right) .
\label{F140}\ee

Upon comparing the previous expressions for the exponentials with the ones obtained from (\ref{ga48}) - (\ref{ga49}), we obtain the following:

\be
e^{2A_2^c} = \frac{4}{L^2 T_{(2)0}},
\label{F141}\ee
\be
e^{2A_1^c} = \frac{32}{\ell^2L^2 T_{(2)0}^2}  .
\label{F142}\ee

Finally, the metric near $\f_0$ is,

\be
ds^2 \approx du^2 + \frac{32}{\ell^2 T_{(2)0}^2} e^{-2\sqrt{2} \frac{u-u_0}{\ell}} \left( (d\psi + \cos\theta d\phi)^2 + \frac{\ell^2 T_{(2)0}}{8} e^{\sqrt{2} \frac{u-u_0}{\ell}} d\Omega^2 \right),
\label{F143}\ee
which has a Kretschmann scalar equal to,

\be
\mathcal{K} = \frac{18}{\ell^4},
\label{F144}\ee
which is regular.

As we noted, as $\f \to \f_0$, the coordinate $u$ approaches $-\infty$. Consequently, from (\ref{F143}) we observe that both scale factors of the $S^2$ and the $S^1$ fiber diverge. Therefore, this point is a boundary point, however it is not FG since the scale factors do not diverge at the same rate.

\subsubsection{The $B_3$ solution}

Similarly to the previous solutions that we studied, we shall first calculate the leading-order solutions. To this end, we expand the superpotentials in a regular power series with leading powers determined by (\ref{g127}):

\be
S_L = x \sum_{n=0}^\infty S_{n} x^n \ , \ W_{1L} = \sum_{n=0}^\infty W_{(1)n} x^n \ , \ W_{2L} = \sum_{n=0}^\infty W_{(2)n} x^n \ , \
\label{F145}\ee
\be
T_{1L} = x^{\tilde{\alpha}} \sum_{n=0}^\infty T_{(1)n} x^n \ , \ T_{2L} = \sum_{n=0}^\infty T_{(2)n} x^n ,
\label{F146}\ee
where the constant $\tilde{\alpha}$ is,

\be
\tilde{\alpha} = - \frac{3 + \sqrt{9 - 12 \Delta \Delta_-}}{\Delta \Delta_-} \  , \  \Delta > 3 .
\label{F147}\ee

Inserting the previous expansions into the equations of motion (\ref{ga51}) - (\ref{ga55}),(\ref{ga54}) we can solve recursively and uniquely for the coefficients of this expansion. We find the following:

\be
S_L = x \left( \frac{2\sqrt{3}}{\ell \tilde{\alpha}} + \frac{\sqrt{3} \ell  \tilde{\alpha} V_3}{ \tilde{\alpha} + 6} x + \mathcal{O}(x^2) \right),
\label{F152}\ee
\be
W_{1L} = - \frac{4\sqrt{3}}{\ell} + \frac{(2-\tilde{\alpha})(4+\tilde{\alpha})(12 \tilde{\alpha}^{-2} + \Delta \Delta_-)}{2\sqrt{3}(\tilde{\alpha}-4)(\tilde{\alpha}+2)\ell} x^2 + \mathcal{O}(x^3) ,
\label{F153}\ee
\be
W_{2L} =  \frac{2}{\sqrt{3}\ell} \frac{12 \tilde{\alpha}^{-2} +\Delta \Delta_-}{\tilde{\alpha}-4} x^2 + \mathcal{O}(x^3) ,
\label{F154}\ee
\be
T_{1L} = 0 ,
\label{F155}\ee
\be
T_{2L} = - \frac{12}{\ell^2} + \frac{12\tilde{\alpha}^{-2}+\Delta \Delta_-}{\ell^2(4\tilde{\alpha}^{-1}-1)}x^2 + \mathcal{O}(x^3) .
\label{F156}\ee

As the reader can observe from equations (\ref{F152}) - (\ref{F156}), up to second order no integration constants have entered the expansions. Since we can recursively obtain the higher order coefficients from the lower order ones uniquely, we conclude that no integration constants appear in the leading order expansions.
 To find the integration constants, we shall make use of the linearised equations of motion, (\ref{F19})-(\ref{F21}),(\ref{F22}) - (\ref{F23}). Specifically, we make the following ansatz for the deformations $\delta S,\delta W_2$,

\be
\delta S = \mathcal{A} x^\lambda \ , \ \delta W_2 = \mathcal{B} x^g,
\label{F157}\ee
where the constants $\mathcal{A},\lambda,\mathcal{B},g$ must obey the following constraints in order for the corresponding expansions to be subleading:
\be
\lambda \geq 1\ , \ g \geq 0 \ , \ \mathcal{A},\mathcal{B} \neq 0.
\label{F158}\ee

Substituting (\ref{F92}) along with the expansions (\ref{F87}) - (\ref{F91}) into the linearised equations of motion (\ref{F22})-(\ref{F23}) yields the following:

\be
- \mathcal{B} x^{1+g} (g-\tilde{\alpha})(6g+\tilde{\alpha}) + 4 \mathcal{A} x^{\lambda} \left( -\tilde{\alpha}^3 -2\tilde{\alpha}^2(1+\lambda) \right.
\label{F159}\ee
$$\left.  + 2(\lambda-1)^2(\lambda+1) + \tilde{\alpha} (\lambda-1)^2 \right) =0, $$
\be
\mathcal{B} x^{1+g} (3\tilde{\alpha}^2 - 5\tilde{\alpha}g + 6g^2) - 4 \mathcal{A} x^{\lambda} \left( -\tilde{\alpha} + \tilde{\alpha}^2(\lambda-1) + \lambda \tilde{\alpha} (3\lambda-2) + 2 (\lambda+1)(\lambda-1)^2 \right) = 0.
\label{F160}\ee

By examining equation (\ref{F161}), we observe that we have three choices:

\begin{enumerate}
   \item[$\bullet$] We can choose
   \be
   1+g  < \lambda .
   \label{F161}\ee

   In this case the first term is the leading one, and therefore it must vanish on its own. Equations (\ref{F159}) , (\ref{F160}) now yield,

   \be
   \mathcal{B}(g-\tilde{\alpha})(6g+\tilde{\alpha}) = 0 \sp \mathcal{B} (3\tilde{\alpha}^2 - 5\tilde{\alpha}g + 6g^2) = 0.
   \ee

   The only solution to these equations is $\mathcal{B} = 0$, which is not acceptable.
   \item[$\bullet$] We can choose
   \be
   \lambda < 1+g .
   \label{F162}\ee

   In this case the second term is the leading one, and therefore it must vanish on its own. Equations (\ref{F159}) , (\ref{F160}) now yield,
   \be
   \mathcal{A } \left( -\tilde{\alpha}^3 -2\tilde{\alpha}^2(1+\lambda)  + 2(\lambda-1)^2(\lambda+1) + \tilde{\alpha} (\lambda-1)^2 \right) =0,
   \ee
   \be
   \mathcal{A} \left( -\tilde{\alpha} + \tilde{\alpha}^2(\lambda-1) + \lambda \tilde{\alpha} (3\lambda-2) + 2 (\lambda+1)(\lambda-1)^2 \right) = 0.
   \ee

   There are 3 possible solutions to the previous two equations:

   \begin{enumerate}
      \item[1.] $ \mathcal{A}  = 0$, which is not acceptable.
      \item[2.] $\lambda = 1-\tilde{\alpha}$. However, since $\tilde{\alpha} > 0 \implies \lambda = 1- \tilde{\alpha} < 1$, which is not acceptable. This solution accounts for 1 of the 4 integration constants, and we intentionally set it equal to 0.
      \item[3.] $\lambda = -(1 + \frac{\tilde{\alpha}}{2})$. However, since $ \tilde{\alpha} > 0 \implies \lambda = -(1 + \frac{\tilde{\alpha}}{2}) < -1 < 1$, which is not acceptable. This solution accounts for the 2nd of the 4 integration constants, and we intentionally set it equal to 0.
   \end{enumerate}

   \item[$\bullet$] Finally we can choose
   \be
   \lambda = 1 + g.
   \ee

   In this case the two terms are of the same order, and therefore they must vanish as a whole. Equations (\ref{F159}),(\ref{F160}) now become,
   \be
   \mathcal{A} (\tilde{\alpha}-g) \left(-4 \tilde{\alpha}^2-4 \tilde{\alpha}(3 g+4)-8 g (g+2)\right)+\mathcal{B}(\tilde{\alpha}-g) (\tilde{\alpha} +6  g) =0,
   \ee
   \be
   \mathcal{B} \left(3 \tilde{\alpha}^2-5 \tilde{\alpha} g+6 g^2\right)-4 \mathcal{A} g (\tilde{\alpha}+g) (\tilde{\alpha}+2 g+4) = 0,
   \ee
   or in matrix form,
   \be
   T \begin{pmatrix}
      \mathcal{A} \\
      \mathcal{B}
   \end{pmatrix} = 0,
   \ee
   where $T$ is the matrix
   \be
   \begin{pmatrix}
      (\tilde{\alpha}-g) \left(-4 \tilde{\alpha}^2-4 \tilde{\alpha}(3 g+4)-8 g (g+2)\right) & (\tilde{\alpha}-g) (\tilde{\alpha} +6  g) \\
      -4g (\tilde{\alpha}+g) (\tilde{\alpha}+2 g+4) & 3 \tilde{\alpha}^2-5 \tilde{\alpha} g+6 g^2
   \end{pmatrix}.
   \ee
   In order for our system to have a non trivial solution, we must require the determinant of $T$ to be zero:
   \be
   \det(T) = 0 \implies -12 \tilde{\alpha} (\tilde{\alpha} - 2 g) (\tilde{\alpha} - g) (\tilde{\alpha}+ g) (4 + \tilde{\alpha} + 2 g) = 0.
   \ee
   The previous system of equations has 4 solutions, namely:
   \begin{enumerate}
       \item[1.]
      \be
      g = \tilde{\alpha} \sp \mathcal{A } =  \frac{\mathcal{B}}{2(3\tilde{\alpha}+4)}.
      \ee
      Since $\tilde{\alpha} > 0 \implies g = \tilde{\alpha} > 0$, which is acceptable. This solution accounts for the 3rd out of the 4 integration constants.
      \item[2.]
      \be
      g =  \frac{\tilde{\alpha}}{2} \sp \mathcal{A} =  \frac{\mathcal{B}}{3(\tilde{\alpha}+2)}.
      \ee
      Since $ \tilde{\alpha} > 0 \implies g =  \frac{\tilde{\alpha}}{2} > 0$, which is acceptable. This solution accounts for the 4th integration constant.
      \item[3.]
      \be
      g = - \tilde{\alpha} \ , \ \mathcal{B} =0 \ , \ \mathcal{A} = \text{unknown}.
      \ee
      Since $\tilde{\alpha} > 0 \implies g < 0$. Moreover, the integration constant $\mathcal{B}$ must be zero. From the previous two we deduce that this case is not acceptable.
      \item[4.]
      \be
      g = -2 - \frac{\tilde{\alpha}}{2} \ ,\ \mathcal{B} = 0 \ , \ \mathcal{A} = \text{unknown}.
      \ee
      Since $\tilde{\alpha} > 0 \implies g < 0$. Moreover, the integration constant $\mathcal{B}$ must be zero. From the previous two we deduce that this case is not acceptable.
   \end{enumerate}
\end{enumerate}

We thus conclude that we have two possibilities, namely $ g= \tilde{\alpha}$ and $g =  \frac{\tilde{\alpha}}{2}$. Therefore, the general form of the functions $\delta S , \delta W_2$ is,

\be
\delta S \approx  \mathcal{A}_1 x^{1+\tilde{\alpha}} + \mathcal{A}_2 x^{1+ \frac{\tilde{\alpha}}{2}} \sp \delta W_2 \approx 2(3\tilde{\alpha}+4)\mathcal{A}_1 x^{\tilde{\alpha}} + 3(\tilde{\alpha}+2) \mathcal{A}_2 x^{\frac{\tilde{\alpha}}{2}}.
\label{F163}\ee

Substituting the previous expansions into equations (\ref{F19}) - (\ref{F21}) yields the following for the functions $\delta W_1, \delta T_1, \delta T_2$:

\be
\delta W_1 \approx  -(6\tilde{\alpha}+8) \mathcal{A}_1 x^{\tilde{\alpha}} - 2 \mathcal{A}_2(\tilde{\alpha}+2) x^{ \frac{\tilde{\alpha}}{2}},
\label{F164}\ee
\be
\delta T_1 \approx  \frac{8\sqrt{3}}{\ell}\mathcal{A}_1 (4 + 3\tilde{\alpha}) x^{\tilde{\alpha}} +\frac{48-36 \tilde{\alpha}+16 \tilde{\alpha}^2+11 \tilde{\alpha}^3}{\sqrt{3} \ell \tilde{\alpha}(2+\alpha)(-4+\alpha)} \mathcal{A}_2 x^{2+ \frac{\tilde{\alpha}}{2}},
\label{F165}\ee
\be
\delta T_2 \approx -\frac{2\sqrt{3}}{\ell} \mathcal{A}_1(4+3\tilde{\alpha}) x^{\tilde{\alpha}} - \frac{6\sqrt{3}}{\ell} \mathcal{A}_2 (\tilde{\alpha}+2) x^{ \frac{\tilde{\alpha}}{2}}.
\label{F166}\ee

For convenience, we shall redefine the integration constants as follows:

\be
\mathcal{A}_1 \equiv   \frac{\ell}{8\sqrt{3}(4+3\tilde{\alpha})}\tilde{T}_{(1)0} \sp \mathcal{A}_2 \equiv \frac{\mathcal{C}}{3(\tilde{\alpha} +2)},
\ee
where the new integration constants now are $T_{(1)0},\mathcal{C}$. The expansions (\ref{F163}) - (\ref{F166}) now become,

\be
\delta S \approx \frac{\ell}{8\sqrt{3}(4+3\tilde{\alpha})}\tilde{T}_{(1)0} x^{1+\tilde{\alpha}} +  \frac{\mathcal{C}}{3(\tilde{\alpha} +2)} x^{1+ \frac{\tilde{\alpha}}{2}} ,
\label{F167}\ee
\be
\delta W_1 = -\frac{\ell\tilde{T}_{(1)0}}{4\sqrt{3}} x^{\tilde{\alpha}} - \frac{2}{3} \mathcal{C} x^{\tilde{\alpha}/2} \ , \ \delta W_2 = \frac{\ell \tilde{T}_{(1)0}}{4\sqrt{3}} x^{\tilde{\alpha}} + \mathcal{C} x^{\frac{\tilde{\alpha}}{2}} ,
\label{F168}\ee
\be
\delta T_1 =\tilde{T}_{(1)0} x^{\tilde{\alpha}}+\frac{48-36 \tilde{\alpha}+16 \tilde{\alpha}^2+11 \tilde{\alpha}^3}{3\sqrt{3} \ell \tilde{\alpha}(2+\tilde{\alpha})^2(-4+\tilde{\alpha})} \mathcal{C} x^{2+\frac{\tilde{\alpha}}{2}} \ , \ \delta T_2 = - \frac{\tilde{T}_{(1)0}}{4} x^{\tilde{\alpha}} - \frac{2\sqrt{3}}{\ell} \mathcal{C} x^{\tilde{\alpha}/2}.
\label{F169}\ee

Putting everything together, we are now in a position to write down the expressions for $W_{1,2},T_{1,2},S$ in the vicinity of an extremum of V and up to first order in the integration constants $T_{(2)0},\cal{C}$. The full expansions are as follows:

\be
S = x \left( \frac{2\sqrt{3}}{\ell \tilde{\alpha}} + \frac{\sqrt{3} \ell  \tilde{\alpha} V_3}{ \tilde{\alpha} + 6} x + \mathcal{O}(x^2) \right) + \frac{\ell}{8\sqrt{3}(4+3\tilde{\alpha})}\tilde{T}_{(1)0} x^{1+\tilde{\alpha}}(1 + \mathcal{O}(x)) +
\label{F170}\ee
$$ +  \frac{\mathcal{C}}{3(\tilde{\mathcal{a}} +2)} x^{1+ \frac{\tilde{\alpha}}{2}}(1+\mathcal{O}(x)),$$
\be
W_1 = - \frac{4\sqrt{3}}{\ell} + \frac{(2-\tilde{\alpha})(4+\tilde{\alpha})(12 \tilde{\alpha}^{-2} + \Delta \Delta_-)}{2\sqrt{3}(\tilde{\alpha}-4)(\tilde{\alpha}+2)\ell} x^2 + \mathcal{O}(x^3) -
\label{F171}\ee
$$ -\frac{\ell\tilde{T}_{(1)0}}{4\sqrt{3}} x^{\tilde{\alpha}}(1 + \mathcal{O}(x)) - \frac{2}{3} \mathcal{C} x^{\tilde{\alpha}/2}(1 + \mathcal{O}(x)), $$
\be
W_2 = \frac{2}{\sqrt{3}\ell} \frac{12 \tilde{\alpha}^{-2} +\Delta \Delta_-}{\tilde{\alpha}-4} x^2 + \mathcal{O}(x^3)+\frac{\ell \tilde{T}_{(1)0}}{4\sqrt{3}} x^{\tilde{\alpha}}(1 + \mathcal{O}(x)) +  \mathcal{C} x^{\tilde{\alpha}/2}(1 + \mathcal{O}(x)),
\label{F172}\ee
\be
T_1 = \tilde{T}_{(1)0} x^{\tilde{\alpha}}(1 + \mathcal{O}(x)) +\frac{48-36 \tilde{\alpha}+16 \tilde{\alpha}^2+11 \tilde{\alpha}^3}{3\sqrt{3} \ell \tilde{\alpha}(2+\tilde{\alpha})^2(-4+\tilde{\alpha})} \mathcal{C} x^{2+\frac{\tilde{\alpha}}{2}}(1 + \mathcal{O}(x)),
\label{F173}\ee
\be
T_2 = - \frac{12}{\ell^2} + \frac{12\tilde{\alpha}^{-2}+\Delta \Delta_-}{\ell^2(4\tilde{\alpha}^{-1}-1)}x^2 + \mathcal{O}(x^3) - \frac{\tilde{T}_{(1)0}}{4} x^{\tilde{\alpha}}(1 + \mathcal{O}(x)) - \frac{2\sqrt{3}}{\ell} \mathcal{C} x^{\tilde{\alpha}/2}(1 + \mathcal{O}(x)),
\label{F174}\ee
where $\tilde{\alpha}$ was defined in (\ref{F147}), but we rewrite it here for convenience:

\be
\tilde{\alpha} =  -\frac{3 + \sqrt{9 + 12 \Delta \Delta_-}}{\Delta \Delta_-} \  , \ \Delta > 3 .
\ee

Since $S = \dot{\f}$, we have to leading order,

\be
\f - \f_0 = e^{\frac{2\sqrt{3}}{\tilde{\alpha}} \frac{u-u_0}{\ell}} \left( 1 + \ldots \right),
\label{F175}\ee
where $u_0$ is an integration constant. Note that as $\f$ approaches $\f_0$, u tends to $-\infty$. From the previous equation we observe that we can identify the term $e^{-\frac{\sqrt{2}\sqrt{3}}{\ell \tilde{\alpha}} u_0}$ as the free source for the scalar as before.

In a similar fashion to the previous solutions, we can obtain expressions for $A_1,A_2$ with respect to u by using the definition of the functions $W_1,W_2$ in equations (\ref{ga46}),(\ref{ga47}) and their expressions from (\ref{F171}),(\ref{F172}):

\be
A_1 = A_1^c + \frac{\sqrt{3}}{\ell} (u-u_0) + \frac{(\tilde{\alpha}-2)(\tilde{\alpha}+4)(12 \tilde{\alpha}^{-2}+\Delta \Delta_-)}{24(\tilde{\alpha}+2)(\tilde{\alpha}-4)} e^{\frac{4\sqrt{3}}{\tilde{\alpha}} \frac{u-u_0}{\ell}} +
\label{F176}\ee
$$+ \frac{\ell^2 \tilde{T}_{(1)0}}{96} e^{2\sqrt{3}\frac{u-u_0}{\ell}}  + \frac{\ell\mathcal{C}}{6\sqrt{3}} e^{\sqrt{3} \frac{u-u_0}{\ell}} + \ldots , $$
\be
A_2 = A_2^c + \frac{\tilde{\alpha}(12\tilde{\alpha}^{-2} + \Delta \Delta_-)}{24} e^{\frac{4\sqrt{3}}{\tilde{\alpha}} \frac{u-u_0}{\ell}} - \frac{\ell^2 \tilde{T}_{(1)0}}{96} e^{2\sqrt{3} \frac{u-u_0}{\ell}} - \frac{\ell \mathcal{C}}{4\sqrt{3}} e^{\sqrt{3} \frac{u-u_0}{\ell}},
\label{F177}\ee
where $A_1^c,A_2^c$ are integration constants that can be determined by substituting the expansions (\ref{F173}),(\ref{F174}), (\ref{F176}) - (\ref{F177}) into equations (\ref{ga48}) - (\ref{ga49}) and solving them to leading order. Specifically, to leading order the exponentials $e^{2A_i}$ are,

\be
e^{2A_2} = - \frac{\ell^2}{3L^2} \left( 1  + \mathcal{O} (e^{\sqrt{3} \frac{u-u_0}{\ell}}) \right),
\label{F178}\ee
\be
e^{2A_1} = \frac{\ell^4 \tilde{T}_{(1)0}}{36L^2} e^{2\sqrt{3} \frac{u-u_0}{\ell}}\left( 1  + \mathcal{O} (e^{\sqrt{3} \frac{u-u_0}{\ell}}) \right).
\label{F179}\ee

Upon comparing the previous expansions for the exponentials with the ones obtained from (\ref{ga48}),(\ref{ga49}), we obtain the following:

\be
e^{2A_2^c} = - \frac{\ell^2}{3L^2},
\label{F180}\ee
\be
e^{2A_1^c} =  \frac{\ell^4 \tilde{T}_{(1)0}}{36L^2}.
\label{F181}\ee

Finally, the metric near $\f_0$ is,

\be
ds^2 = du^2 +\frac{\ell^4 \tilde{T}_{(1)0}}{36}e^{2\sqrt{3} \frac{u-u_0}{\ell}}\left( d\psi + \cos\theta d\phi \right)^2 - \frac{\ell^2}{3} d\Omega^2,
\label{F182}\ee
which has a Kretschmann scalar equal to,

\be
\mathcal{K} = \frac{36}{\ell^4} + \frac{\tilde{T}_{(1)0}^2}{64} e^{4\sqrt{3} \frac{u-u_0}{\ell}},
\label{F183}\ee
which is regular.

As we noted, as $\f \to \f_0$, the coordinate $u$ approaches $-\infty$. Consequently, from (\ref{F182}) we observe that both scale factors of the $S^2$ and the $S^1$ fiber diverge. Therefore, this point is a boundary point, however it is not FG since the scale factors do not diverge at the same rate.

\subsection{Synopsis}

In this appendix we have analysed the behavior of the first order formalism functions around extrema of the potential. Specifically, we have calculated up to second order the leading asymptotics of the solutions and, by linearising the equations of motion, we have obtained the leading powers as well as the coefficients of the subleading expansions. However, we have not examined the possibility of resurgent corrections that involve higher powers of the leading powers.

The aforementioned solutions can be classified depending on whether they are expansions near UV fixed points or whether they are exotic solutions. Specifically, their classification is as follows:

\subsection*{UV fixed points:}

The two cases of interest are the (+) and (-) branches. These cases exist for the regular 3-sphere as well, and were studied extensively in \cite{C}. Through our analysis, we observe that the following hold for the (-) branch:

\begin{enumerate}
   \item[$\bullet$] It exists only near maxima of the potential $V$.
   \item[$\bullet$] It corresponds to a UV fixed point from which RG flows start.
   \item[$\bullet$] The integration constants $C_1,C_2$ control the vevs of the stress energy tensor $T_{ij}$ and the dual operator $\mathcal{O}$.
   \item[$\bullet$] The integration constants $\mathcal{R}_1,\mathcal{R}_2$ control the dimensionless curvature $\mathcal{R}$ and the squashing parameter $a^2$.
   \item[$\bullet$] The dual operator $\mathcal{O}$ has a non-zero source as well as a non-zero vev that are controlled by the integration constants $C_1,C_2$ and $\f_-$.
\end{enumerate}

Therefore, the (-) branch solution corresponds to a UV fixed point perturbed by a relevant operator.

In a similar fashion, we can make the following observations for the (+) branch:

\begin{enumerate}
   \item[$\bullet$] It exists only near maxima as well as minima of the potential $V$.
   \item[$\bullet$] It corresponds to a UV fixed point from which RG flows either start at maxima of the potential or arrive at minima of the potential.
   \item[$\bullet$] The integration constants $\mathcal{R}_1,\mathcal{R}_2$ once again control the dimensionless curvature $\mathcal{R}$ and the squashing parameter $a^2$.
   \item[$\bullet$] The integration constant $C_2$ controls only one vev, namely the vev of the stress-energy tensor.
   \item[$\bullet$] Although the source of the dual operator $\mathcal{O}$ vanishes, its vev is non zero and is controlled by the integration constant $\f_+$.
\end{enumerate}

Therefore, the (+) branch solution corresponds to a UV fixed point perturbed by the vev of a relevant operator.

\subsection*{Exotic Solutions}

This class of solutions is comprised by the three asymptotics $\nu_{\pm},B_{3}$. $\nu_+$,$\nu_-$  exist around maxima of the potential, and $\nu_+$,$B_3$ exist around minima of the potential.
For all such solutions, the structure of the asymptotics corresponds to a boundary, however, this is not a standard FG boundary.

\section{Comments on expectation values}\label{expval}

In the study of solutions around UV fixed points (the (-) and (+) branches, subsection \ref{UVfixedpoints}), we shall be interested in relating the observable quantities of the dual QFT with the integration constants appearing in the first order formalism functions. Specifically, we shall require expressions for the source and the vev of the operator $\mathcal{O}$, as well as for the vev of the stress-energy tensor. The aim of this subsection is to show how we are going to do this.

The general form of the metric we are using is,

\be
ds^2 = du^2 + L^2 \left( e^{2A_1}\left(d\psi + \cos\theta d\phi \right)^2 + e^{2A_2} d\Omega^2 \right).
\label{fe1}\ee

We now make the following coordinate transformation in the metric (\ref{fe1}):

\be
z = e^{\frac{u}{\ell}} \implies du = \frac{\ell dz}{z}.
\label{fe2}\ee

Now the metric takes the form

\be
ds^2 = \frac{\ell^2}{z^2} \left( dz^2 + \frac{z^2}{\ell^2}L^2 e^{2A_1}  \left(d\psi + \cos\theta d\phi \right)^2 + \frac{z^2}{\ell^2}L^2 e^{2A_2} d\Omega^2 \right),
\label{fe3}\ee
or equivalently,

\be
ds^2 = \frac{\ell^2}{z^2} \left( dz^2 + g_{ij}(z,x^i) dx^i dx^j \right),
\label{fe4}\ee
where

\be
g_{ij}(z,x^{i}) = g_{ij}^{(0)}(x^{i}) + g_{ij}^{(2)}(x^{i}) z^2 + g_{ij}^{(3)}(x^{i}) z^3 + \ldots,
\label{fe5}\ee
with the ellipsis containing higher powers of $z$ as well as possible corrections involving $z^{2\Delta}, z^{2\Delta_-}$ if a nontrivial scalar field (see below) is present in the Einstein equations.

The coordinates used in (\ref{fe4}) are known as \textit{Fefferman-Graham} (FG) coordinates. In the FG coordinate system we can identify the sources and vevs of the relevant dual operators  (in our case the stress tensor and scalar). As shown in \cite{Skenderis}, the vev of the stress tensor is given by

\be
\left\langle T_{ij} \right\rangle = \frac{3\ell^2}{16\pi G_N} g_{ij}^{(3)}.
\label{fe6}\ee

The PG expansion for the scalar field $\f$ is,

\be
\f(z,x^{i}) = \f_0 + z^{3-\Delta}\left(\f_{(0)}(x^{i}) + \ldots \right) + z^{\Delta} \left( \f_{(1)}(x^{i}) + \ldots \right)+{\rm resurgent},
\label{fe7}\ee
where the ellipses in the parentheses contain a regular power series expansion in $z$, and the further resurgent corrections involve higher powers of the two leading powers, $z^{\Delta,\Delta_-}$.
 $\f_{(0)}$ is the source and corresponds to the dual QFT coupling
\be
\delta S = \int d^3x \f_{(0)}(x) \mathcal{O}(x),
\label{fe8}\ee
and $\f_{(1)} \sim \left\langle \mathcal{O}\right\rangle$. Specifically, \cite{Skenderis}, the expectation value for the dual operator $\mathcal{O}$ is
given by
\be
\left\langle \mathcal{O}(x) \right\rangle = (2\Delta -3)\f_{(1)}(x).
\label{fe9}\ee

\end{document}